\documentclass{article}
\newif\ifclean
\cleantrue  
\usepackage{graphicx}
\usepackage{amssymb}
\usepackage[letterpaper]{geometry}
\usepackage{subcaption}
\usepackage{amsmath,amssymb,amsthm}
\usepackage{bm}
\usepackage[usenames,dvipsnames]{xcolor}
\usepackage[normalem]{ulem}
\newcommand{\COMMENT}[1]{\textcolor{cyan}{{[ \sc{#1} ]}}} 

\newcommand{\QUESTION}[1]{\textcolor{ForestGreen}{{#1}}}
\newcommand{\red}[1]{\textcolor{red}{{#1}}}

\newlength{\figwidth}
\newlength{\figwidthtwo}
\newlength{\figwidththree}
\newcommand{\aref}[1]{App.\,\ref{#1}}
\newcommand{\fref}[1]{Fig.\,\ref{#1}}
\newcommand{\Fref}[1]{Figure\,\ref{#1}}

\newcommand{\eref}[1]{Eq.\,(\ref{#1})}

\newcommand{\sref}[1]{Sec.\!~\ref{#1}}

\newcommand{\cref}[1]{Ref.\,\cite{#1}}
\newcommand{\crefs}[1]{Refs.\,\cite{#1}}

\newcommand{\ie}{{\it i.e.}\! }

\newcommand{\etal}{{\it et al.}\! }
\newcommand{\apriori}{{\it a priori} }

\newcommand{\Ac}{\mathcal{A}}
\newcommand{\Bc}{\mathcal{B}}
\newcommand{\Ic}{\mathcal{I}}
\newcommand{\Gc}{\mathcal{G}}

\newcommand{\varepsilonb}{\boldsymbol{\varepsilon}}

\newcommand{\eb}{\mathbf{e}}

\newcommand{\nb}{\mathbf{n}}

\newcommand{\pb}{\mathbf{p}}

\newcommand{\Ab}{\mathbf{A}}
\newcommand{\Bb}{\mathbf{B}}
\newcommand{\Cb}{\mathbf{C}}

\newcommand{\Gb}{\mathbf{G}}

\newcommand{\Fb}{\mathbf{F}}

\newcommand{\Eb}{\mathbf{E}}

\newcommand{\Nb}{\mathbf{N}}
\newcommand{\Pb}{\mathbf{P}}
\newcommand{\Sb}{\mathbf{S}}
\newcommand{\Ib}{\mathbf{I}}
\newcommand{\Rb}{\mathbf{R}}
\newcommand{\Qb}{\mathbf{Q}}

\newcommand{\phib}{\boldsymbol{\phi}}

\newcommand{\tr}{\operatorname{tr}}

\newcommand{\sym}{\operatorname{sym}}

\graphicspath{{{./}}}

\newcommand{\caution}{\red{\bf Draft: \today. Do not distribute.}}
\ifclean
\renewcommand{\COMMENT}[1]{{}}
\renewcommand{\QUESTION}[1]{{}}
\else
\fi

\usepackage[utf8]{inputenc}
\usepackage[english]{babel}
\newtheorem{remark}{Remark}

\title{\bf Learning hyperelastic anisotropy from data via a tensor basis neural network}
\author{
J.N. Fuhg$^\dagger$,
N. Bouklas$^\dagger$,
R.E. Jones$^\ddagger$\footnote{corresponding: rjones@sandia.gov}  \\[0.1in]
$^\dagger${\it Cornell University,  921 University Ave,  Ithaca, NY 14853, USA}\\[0.05in]
$^\ddagger${\it Sandia National Laboratories, P.O. Box 969, Livermore, CA 94551, USA}
}
\date{}

\begin{document}
\maketitle

\begin{abstract}
Anisotropy in the mechanical response of materials with microstructure is common and yet is difficult to assess and model.
To construct accurate response models given only stress-strain data, we employ classical representation theory, novel neural network layers, and L1 regularization.
The proposed tensor-basis neural network can discover both the type and orientation of the anisotropy and provide an accurate model of the stress response.
The method is demonstrated with data from hyperelastic materials with off-axis transverse isotropy and orthotropy, as well as materials with less well-defined  symmetries induced by fibers or spherical inclusions.
Both plain feed-forward neural networks and input-convex neural network formulations are developed and tested.
Using the latter, a polyconvex potential can be established, which, by satisfying the growth condition can guarantee the existence of boundary value problem solutions.
\end{abstract}


\section{Introduction} \label{sec:introduction}
Many materials exhibit anisotropic stress response induced by their microstructure, such as embedded fibers or polycrystalline texture.
Crystalline materials are well known to exhibit elastic anisotropy \cite{parry1976elasticity}.
Fiber reinforced composites \cite{nguyen2007modeling}, both engineered \cite{chen1970mechanical,mortazavian2015effects,ma2019mechanical} and biological \cite{yang1998anisotropic,holzapfel2000new,pinsky2005computational,elsheikh2009mechanical,katz2008anisotropic,pandolfi2012fiber,datta2019anisotropy}, are another broad category of materials that exhibit strong anisotropies.
In addition to inherent anisotropies, anisotropy is commonly induced by processing, such as the texture (the biased crystal orientations of a polycrystalline material) generated in worked metals \cite{negahban1989material,bronkhorst1992polycrystalline,beaudoin1994application,svendsen2006application,reese2021using}.

The anisotropy of these materials is difficult to assess directly from mechanical tests \cite{zou2013identification}.
Even with microstructural information and imaging \cite{guild1993microstructural,trimby1999microstructural,cocco2013three,bostanabad2018computational}, such as from computed tomography, the anisotropy observed in the stress response of the material may not exactly correspond to that apparent in the microstructure.
Alternately the microstructure may be too complex to assess symmetry by inspection, as with fiber distributions or semi-structured arrangements of inclusions.

Classical mechanics provides general representations of anisotropic response functions via structure tensors and representation theorems.
Rivlin, Pipkin, Smith, Spencer, Boehler, and co-workers did pioneering work in this field starting in the 1950s \cite{spencer1958finite,spencer1958theory,spencer1962isotropic,pipkin1963material,wineman1964material,smith1964integrity,Smith1965,rivlin1969orthogonal,spencer1971part,spencer1987isotropic,boehler1987representations}.
Particularly notable is Boehler's treatise \cite{boehler1987representations} describing how anisotropy can be represented with a set of tensor generators.
Later, Zheng contributed an often-cited monograph \cite{zheng1994theory} on the application of representation theory to anisotropy, which summarized many of the results to-date.
Although structural tensors have been traditionally used to model symmetries that are fixed in a reference configuration they have also been employed to represent evolving symmetries \cite{jones2006simulating,svendsen2006application,reese2021using}.
Representation of physical stress response functions have additional considerations, such as polyconvexity of the underlying free energy \cite{steigmann2003frame,itskov2004class,kambouchev2007polyconvex,ehret2007polyconvex}, which need to be considered in constructing general representations.

Neural networks (NNs), as universal approximators \cite{hornik1989multilayer}, present a flexible representation for response functions and have been widely employed in constitutive modeling.
Pioneering work by Ghaboussi \cite{ghaboussi1998autoprogressive,jung2006neural} has been extended to enable constitutive modeling in the context of hyperelasticity, viscoelasticity and elastoplasticity \cite{huang2020machine,masi2021thermodynamics,fuhg2021model,vlassis2020geometric,vlassis2021sobolev,holzapfel2021predictive,linka2021constitutive,jones2021neural}.
These works incorporate physical principles and model assumptions to varying degrees as means of obtaining trustworthy surrogates and working in the low-date regime.
In particular, the tensor basis neural network (TBNN) \cite{ling2016machine} was developed based on classical representation theory, which proves that a complete representation can be formed from a finite sum of coefficient functions of scalar invariants, and known tensor basis elements (generators).
A primary benefit of the TBNN approach, compared to more traditional NN-based approaches that learn component-based input-output maps, lies in the simplification of the maps that need to be discovered; the tensor basis elements, which do not need to be learned, carry a significant part of the functional complexity of the representation.

Alternative data driven methods to neural networks exist for the representation of constitutive response functions.
For instance, Flaschel, Kumar, and De Lorenzis \cite{flaschel2021unsupervised,flaschel2022discovering} have developed a sparse regression technique that builds a stress response model from interpretable components.
Utilizing tensor representation theory and building on the TBNN framework, Frankel, Jones and Swiler \cite{frankel2020tensor} and Fuhg and Bouklas \cite{fuhg2021physics} developed Gaussian process (GP) models of hyperelasticity which are particularly well suited to the low data regime.
The latter work highlighted potential benefits of the Gaussian process models compared to NNs.

Knowing the anisotropy of the material response is crucial in maintaining the objectivity of the model when it is deployed in simulations.
Consequently, there has been some work on discovery and representation of anisotropic response functions with machine learning.
In the general context of dynamical systems, Dehmamy \etal \cite{dehmamy2021automatic} devised a Lie algebra convolutional network by working with the generators of symmetries, as opposed to the group.
With this formulation they were able to connect the equivariant properties of the network to conservation laws.
In a more specific setting, Tac \etal \cite{tac2021data} developed a neural network model of hyperelastic soft tissue material with two fiber families.
The model employed a neural network for the isochoric component of the stress response with pre-selected invariant inputs.
Also in the context of material physics, Fuhg and Bouklas \cite{fuhg2021physics} developed a TBGP model of a hyperelastic material with transverse isotropy.
A significant limitation of the applications of TB representations to anisotropic materials to date is the orientation and the type of the symmetry have been assumed to be known.

In this work we develop a TBNN methodology that can discover the type and the orientation of the symmetry of an anisotropic material, and at the same time provide a surrogate for its constitutive response.
In the proposed TBNN symmetries are characterized by a structural tensor whose form (type) and orientation with respect to a canonical frame are learned.
Although it, like other TBNNs, is a proficient representation in a prediction mode, the primary use of the model is to infer the symmetries of a particular material  given samples of its stress response to various deformations.
In \sref{sec:theory} we review the classical theory that the proposed TBNN is based on.
Then, in \sref{sec:method}, we describe the network model and its training methodology in detail.
\sref{sec:data} describes the source models and training data we use to test the model.
\sref{sec:results} demonstrates that the methodology can discover and represent a wide range of anisotropic stress response.
\aref{append:ConverStud} shows that the method can effectively infer material anisotropy in the low data regime, and \aref{append:InputConvex} provides an input-convex variant of the proposed TBNN that satisfies polyconvexity requirements albeit with greater implementation complexity and a larger parameter space (\aref{app:ids} provides tensor identities relevant to the developments of \sref{sec:theory}).
In \sref{sec:conclusion} we conclude with a synopsis of the results and ideas for future work.

\section{Theory} \label{sec:theory}

Hyperelasticity is framed in terms of stress being the derivative of a (free energy) potential $\Psi$ with respect to a suitable deformation measure, for instance the second Piola-Kirchhoff stress $\Sb$ is given by
\begin{equation} \label{eq:eng_conj}
\Sb = \partial_\Eb \Psi \ ,
\end{equation}
where $\Eb$ is the Lagrange strain.
More primitively, the free energy as a function of the deformation gradient $\Fb$ must satisfy symmetry conditions \cite[Sec. 4.3]{ogden1997non} :
\begin{equation} \label{eq:mat_sym}
\Psi(\Fb) = \Psi(\Qb\Fb\Gb) \ ,
\end{equation}
for every $\Qb \in \text{Orth}^+$ and $\Gb \in \mathcal{G} \subseteq \text{Orth}$, where $\Qb$ accounts for the indifference to the current coordinate frame and $\Gb$ embeds material symmetry through invariance to transformations of the reference configuration.
Instead of creating general representations of $\Psi$ for each symmetry group, $\mathcal{G}$, it is customary to introduce a structure tensor, $\Ab$, or a set of structure tensors, $\Ac = \{ \Ab_i \}$, as auxiliary arguments \cite{zhang1990structural,svendsen1994representation}:
\begin{equation} \label{eq:aug_arg}
\Psi = \Psi(\Eb, \Ac) \ ,
\end{equation}
and then appeal to more common isotropic function representation theorems.
This provides an equivalent representation by virtue of Rychlewski’s {\it isotropization} theorem \cite{zhang1990structural}.
A structure tensor characterizes the material symmetry group $\Gc$ through the property group action leaving the structure tensor invariant
\begin{equation}
\Ab = \Gb \boxtimes \Ab  \  \forall \ \Gb \in \Gc \ ,
\end{equation}
where $\boxtimes$ is the Kronecker product ($\Gb \boxtimes \Ab \equiv A_{ij\ldots n} \Gb \eb_i \otimes \Gb \eb_j \otimes \ldots \Gb \eb_n$).
Note that structure tensors are not unique, a set of lower order structure tensors can characterize the same symmetry group as a single higher order tensor \cite{zheng1994theory}.
Employing a set of structure tensors leads to a representation that has symmetries corresponding to the intersection of the respective symmetry groups.

Isotropic function representation theorems \cite[Sec. 37]{gurtin1982introduction} imply the potential $\Psi$ can be expressed in terms of $\Ic = \{ I_k \}$, a set of mutual scalar invariants of $\Eb$ and $\Ac = \{ \Ab_i \}$:
\begin{equation}
\Psi = \Psi(\Eb, \Ac ) = \hat{\Psi}( \Ic) \ .
\end{equation}
Hence the energy conjugacy relation \eqref{eq:eng_conj} gives:
\begin{equation}
\Sb = \partial_\Eb \Psi = 2 \partial_\Cb \Psi
= 2 \sum_i \partial_{\Ic_i} \Psi \ \partial_\Cb \Ic_i
= \sum_i c_i(\Ic) \, \Bb_i
\end{equation}
where $\Bb_i \equiv \partial_\Cb \Ic_i$ are a known tensor basis for the representation of $\Sb$ and $c_i \equiv \partial_{\Ic_i} \Psi$ are the associated coefficient functions.
Here, $\Cb = \Fb^T \Fb = 2 \Eb - \Ib$ is the right Cauchy–Green deformation tensor.
Refer to \aref{app:ids} for the necessary tensor calculus identities.

Given these developments, we assume that the material response can be represented by a potential that is a function of a set of invariants that are possibly redundant.
For example, isotropy with structure tensor $\Ab = \Ib$ is a sub-class of so-called ``complete'' orthotropy, where all symmetric dyads are structure tensors $\Ab_i \in \{ \sym \eb_i \otimes \eb_j\}$.
Here $\eb_i$ is an element of the usual Cartesian basis and $\sym \Ab \equiv 1/2 (\Ab + \Ab^T)$.
For this work we take $\{\Ab_i\}$ to subsume isotropy $\Ab = \sum_i \eb_i \otimes \eb_i$, transverse isotropy $\Ab = \eb_1 \otimes \eb_1$, and orthotropy $\Ab_i \in \{\eb_i \otimes \eb_i , i = 1,3\}$.
For convenience we define
\begin{equation}
\Nb_i = \eb_i \otimes \eb_i \ ,
\end{equation}
so that isotropy $\Ac = \{ \Ib = \sum_i \Nb_i \}$,
transverse isotropy $\Ac = \Nb_1$,
and orthotropy $\Ac = \{ \Nb_1, \Nb_2, \Nb_3 \}$ or, equivalently, $\{ \Nb_1, \Nb_2, \Ib \}$, are in the span of these structure tensors.

These symmetries nest in the sense that each subsumes the higher symmetries.
For isotropy the set of invariants $\Ic_\text{iso} = \{ \tr \Cb, \tr \Cb^2, \tr \Cb^3 \}$ suffice.
The corresponding tensor basis is $\Bc_\text{iso} = \{ \Ib, \Cb, \Cb^2 \}$ and the stress is simply
\begin{equation} \label{eq:stress_iso}
\Sb = ( \partial_{I_1}  \Psi ) \, \Ib + ( 2\partial_{I_2} \Psi ) \, \Cb + ( 3\partial_{I_3} \Psi )\, \Cb^2
\end{equation}
For transverse isotropy, a complete set of invariants is:
\begin{equation}
\Ic_\text{trans} = \Ic_\text{iso} \cup  \{
\tr \Cb \Nb_1, \tr \Cb^2 \Nb_1
\}
\end{equation}
and the corresponding tensor basis is:
\begin{equation}
\Bc_\text{trans} = \Ic_\text{trans} \cup   \{
\sym \Cb \Nb_1,
\sym \Cb^2 \Nb_1
\} \ ,
\end{equation}
given $\Nb_i^n = \Nb_i$ and $\tr \Nb_i = 1$.
The stress for this case is
\begin{equation} \label{eq:stress_trans}
\Sb = \underbrace{( \partial_{I_1}  \Psi ) \, \Ib + ( 2\partial_{I_2} \Psi ) \, \Cb + ( 3\partial_{I_3} \Psi )\, \Cb^2}_{\text{isotropic}}
+
\underbrace{
( \partial_{I_4}  \Psi ) \, \Nb_1
+ ( \partial_{I_5}  \Psi ) \, [  \Cb  \Nb_1 + \Nb_1 \Cb ]}_{\text{anisotropic}}
\end{equation}
With orthotropy the invariants are
\begin{equation}
\Ic_\text{ortho} = \Ic_\text{trans} \cup  \{
\tr \Cb \Nb_2, \tr \Cb^2 \Nb_2
\} \ .
\end{equation}
The corresponding tensor basis is:
\begin{equation}\label{eq:InvOrtho}
\Bc_\text{ortho} = \Bc_\text{trans} \cup \{ \sym  \Cb \Nb_2, \sym \Cb^2 \Nb_2  \} \ ,
\end{equation}
so that the stress representation is:
\begin{eqnarray}\label{eq::OrthoGeneral}
\Sb &=& \underbrace{( \partial_{I_1}  \Psi ) \, \Ib + ( 2\partial_{I_2} \Psi ) \, \Cb + ( 3\partial_{I_3} \Psi )\, \Cb^2}_{\text{isotropic}} \\
&+&
\underbrace{
( \partial_{I_4}  \Psi ) \, \Nb_1
+ ( \partial_{I_5}  \Psi ) \, [  \Cb  \Nb_1 + \Nb_1 \Cb ]
+
( \partial_{I_6}  \Psi ) \, \Nb_2
+ ( \partial_{I_7}  \Psi ) \, [  \Cb  \Nb_2 + \Nb_2 \Cb ]
}_{\text{anisotropic}} \nonumber
\end{eqnarray}
Recall there is no need to employ $\Nb_3$ as a structure tensor since it is in the span of $\{\Ib, \Nb_1, \Nb_2\}$ and $-\nb_i$ leads to the same structure tensors as $\nb_i$.
Alternatively, a polyconvex basis, with the adjugate of $\Cb$ (or $\Cb^{-1}$) substituting for $\Cb^2$, could be employed \cite{steigmann2003frame,itskov2004class,schroder2005variational}, refer to \aref{append:InputConvex} for further developments.
For convenience in the following sections we will use the notation $\Ic_\text{aniso} = \Ic_\text{ortho} \setminus \Ic_\text{iso} $ and $\Bc_\text{aniso} = \Bc_\text{ortho} \setminus \Bc_\text{iso}$.

Given that the orientation of the material anisotropy is not necessarily aligned with the Cartesian lab basis, $\eb_i$, we can express the elemental structure tensors as the rotation of canonically oriented elements:
\begin{equation} \label{eq:rot_G}
\Nb_i(\Rb) = \nb_i \otimes \nb_i
= \Rb \eb_i \otimes \Rb \eb_i
= \Rb \boxtimes ( \eb_i \otimes \eb_i ) \ ,
\end{equation}
where $\nb_i$ is the symmetry axis of $\Nb_i$.
The rotation $\Rb$, has a convenient (tensor basis) Euler-Rodrigues representation:
\begin{equation} \label{eq:rodrigues}
\Rb(\theta\pb) = \exp(\theta \Pb)
= \Ib + (\sin \theta) \Pb + (1-\cos \theta) \Pb^2
\end{equation}
where $\theta\in [0,2\pi]$ is the rotation angle, $\pb$ is the (unit) axis vector, and $\Pb \equiv \varepsilonb \pb$ with $\varepsilonb$ being the third order permutation tensor.
The axial vector $\pb$ can be parameterized in spherical coordinates as
\begin{equation} \label{eq:p_sph_coord}
\pb  = \sin \phi \cos \varphi \, \eb_1 + \sin \phi \sin \varphi \, \eb_2 + \cos \phi \, \eb_3
\end{equation}
with two angles $\phi$ and $\varphi$, or via alternative Rodrigues formulae and representations of $\Rb$.

In the following we will use {\it type} to refer to the symmetry class or degree of symmetry as measured by the size and type of elements in the symmetry group $\Gc$, and {\it orientation} to refer to the orientation of these elements in the reference configuration of the material with respect to the canonical orientations, \ie $\nb_i$ vs. $\eb_i$.

\section{Architecture} \label{sec:method}

\begin{figure}
\centering
\includegraphics[width=0.7\textwidth]{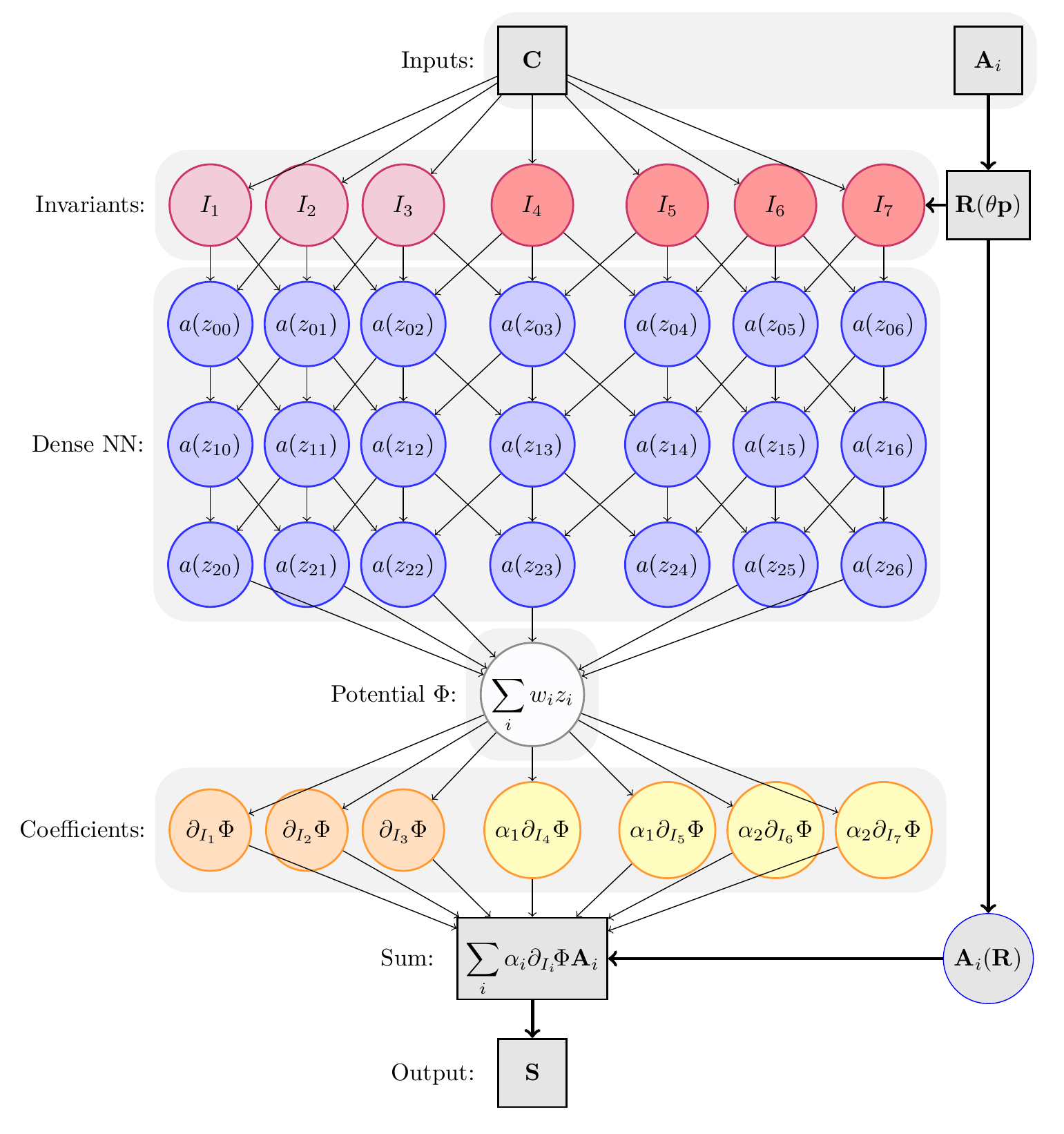}
\caption{Anisotropy discovery TBNN architecture:
isotropic invariants $\Ic_\text{iso}$ (pink);
anisotropic invariants $\Ic_\text{aniso}$ that augment the description and depend on orientation $\theta\pb$ and the associated rotation $\Rb$, (red);
deep densely connected, feed-forward neural network (blue);
elastic potential $\Phi$ (white);
isotropic tensor basis coefficients (orange); and
anisotropic tensor basis coefficients subject to regularization (yellow).
}
\label{fig:nn_architecture}
\end{figure}

The proposed architecture is summarized in the schematic in \fref{fig:nn_architecture}.
It largely resembles a standard TBNN \cite{ling2016machine,jones2018machine} except for the inclusion of a set of elemental structure tensors $\Ac = \{ \Nb_1, \Nb_2 \}$ in their canonical orientation, together with $\Cb$,  in the inputs.
The inclusion of these arguments allows for the construction of the additional invariants that depend on anisotropy and orientation.
Another distinction is the coefficient functions $c_i$ are derived from a potential $\Psi$ formed as the output of a densely connected feed-forward neural network, as opposed to the output of the neural network being the coefficient functions themselves.

A preprocessing layer computes anisotropic invariants $\Ic_\text{aniso} = \{ \tr \Cb \Nb_i, \tr \Cb^2 \Nb_i \}$ and basis elements $\Bc_\text{aniso} = \{\Nb_i, \Cb \Nb_i + \Nb_i \Cb \}$,  where $\Nb_i = \Nb_i(\theta\pb)$ are functions of trainable parameters.
Note the isotropic invariants $\Ic_\text{iso} = \{ \tr \Cb, \tr \Cb^2, \tr \Cb^3 \}$ and basis $\Bc_\text{iso} = \{ \Ib, \Cb, \Cb^2 \}$ could be precomputed and fed in as data since they do not depend on trainable parameters, but we chose to compute them in-line for convenience.
Rotation $\Rb(\phib)$ is formed from axis vector $\phib = \theta \pb$ and the Rodrigues formula \eref{eq:rodrigues}.
The trainable parameters for the rotation of the tensor basis are the rotational angle $\theta \in [0,2 \pi]$ and the unit axis vector $(p_1,p_2,p_3) / \| \pb \| = 1$.
We employed a constrained 3 parameter representation for $\pb$ instead of \eref{eq:p_sph_coord} to avoid the singularities in the 2 parameter spherical coordinate representation.

The invariants $\Ic = \Ic_\text{iso} \cup \Ic_\text{aniso}$ are the inputs to a feed-forward neural network (blue in \fref{fig:nn_architecture}) consisting of input layer with a node for each invariant , $n_{D}-1$ hidden layers of the same width, and an output layer consisting of a single node.
Each layer of the neural network transforms the output of the previous layer through a trainable affine transform followed by the element-wise application of a (pre-selected) non-linear transform:
\begin{equation}\label{eq::NNStand}
\bm{z}_{i+1} = a(\bm{W}_{i} \bm{z}_{i} + \bm{b}_{i}) \ ,
\end{equation}
where $\bm{z}_{i}$ is the input data of the $i$-th layer,  and $\bm{W}_{i} $, $\bm{b}_{i}$ and $a$ are the weights, biases and activation function of the $i^{\text{th}}$ layer.
Since inferring the potential is a regression task the activation function of the output layer is chosen to be identity, \ie the last layer is just a linear mixing of the output of the previous layer.

Once the potential $\Phi = \Phi(\{ I_k \})$ is formed from the output of the last layer of the NN, $z_{n_{D}}$, the  coefficients $c_i$ of the tensor basis representation are obtained from partial derivatives of $\Phi$.
Finally, the tensor basis summation is reformulated as:
\begin{equation}\label{eq::MethodS}
\Sb = 2 \left( \sum_{i=1,3} c_i \Ab_i + \alpha_{1} \sum_{i=4,5} c_i \Ab_i + \alpha_{2} \sum_{i=6,7} c_i \Ab_i\right) \ ,
\end{equation}
where the basis has been selected such that $\Ab_i \equiv \partial_\Cb I_i$, refer to \sref{sec:theory} and the parameters $\alpha_{1}, \alpha_{2}$ are introduced to control the degree of anisotropy.
In order to enable the method to discern the components of $\Bc_\text{aniso}$ needed to represent the data and suppress the involvement of unneeded components, we  promote sparsity in the anisotropic part of the tensor basis representation through $\alpha_i$.
In particular L1 regularization of the $\alpha$-values is employed in the loss:
\begin{equation}
L = \| \Sb - \hat{\Sb} \|_2^2 + \varepsilon  \left( \left| \alpha_{1} \right| + \left| \alpha_{2} \right|  \right) \ ,
\end{equation}
where $\varepsilon$ is a L1 penalty parameter.
Note is it only necessary to penalize the use of the anisotropic components of the basis since the isotropic components are always necessary to represent the response functions of interest.
In the following results we used $\epsilon=1 \times 10^{-4}$.
In general this penalty parameter needs to be tuned; however, data normalization helps to bound values that provide suitable conditioning of the constraint objective relative to the accuracy objective.

\begin{remark}
Alternatively, the coefficients $c_{i}$ could have been directly applied to signal the degree of anisotropy instead of the $\alpha_{i}$, i.e. $(c_{4}+c_{5})$ and $(c_{6}+c_{7})$.
However, regularizing $c_{i}$ means a regularization of a large set of trainable parameters of the neural network since they are dependent on derivatives of the output potential with regards to the input.
Hence, by regularizing only two trainable parameters $\alpha_{1}$ and $\alpha_{2}$ the network can train faster.
Secondly, using batch-wise errors with the batch-size $M$ the magnitude of the sum, e.g. $\sum_{M} (|c_{4}|+|c_{5}|)$, is difficult to judge \apriori.
This means that the regularization parameter $\epsilon$ is also difficult to choose in this case.
For the formulation we propose, the additional $\alpha_i$ are just two trainable parameters, i.e. not network outputs, so this parameterization is not sensitive to the batch-size which makes the regularization process more robust.
\end{remark}

Due to fact that the network architecture needs to fit data through its own derivative with respect to the inputs of the network, we found that a hidden-layer activation function that is at least $C^{1}$ is helpful for the training process.
In this work we applied the $\tanh$ activation function in the hidden layers and a linear function at the final layer.
The constraints $\theta \in [0, 2\pi]$ and $\|\pb\|=1$ were enforced using gradient clipping \cite{zhang2019gradient}.

The newly developed TBNN architecture was implemented in Pytorch \cite{NEURIPS20199015}.
The densely-connected component of the network consisted of 3 hidden layers  with 30 neurons.
The weights and biases of the feed-forward network were initialized using a uniform initialization.
The $6$ additional trainable parameters, $\{ \alpha_{1}, \alpha_{2}, p_{1}, p_{2}, p_{3}, \theta \}$, of the proposed TBNN framework were initialized independently.
In order to not initially favor any particular degree of anisotropy (e.g. $\alpha_{1} \approx 0$ and $\alpha_{2} \approx 1$), the $\alpha$-values were initialized using $\alpha_{1}=\alpha_{2}=0.1$ where the value of $0.1$ is chosen without any fine-tuning but (magnitude-wise) roughly on par with the initial weights of the network.
The values of $p_{i}$ were initialized from a uniform distribution $p_{i} \sim \mathcal{U}[0,1]$ and were then normalized by $p_{i} = p_{i}/\|\pb\|$.
The angle $\theta$ was initialized from a uniform distribution $\theta \sim \mathcal{U}[0,2 \pi]$.
The network parameters were optimized using the Adam optimizer \cite{kingma2014adam} over $10^{6}$ epochs.
All the hyperparameters, such as learning rate and network size, were not tuned extensively since preliminary studies found that the selected parameters produced satisfactory results.
For other applications hyperparameters, such as the L1 penalty $\varepsilon$, would need to be tuned in general; however, data normalization assists in the transferability of hyperparameters from one application to the next.

\section{Data} \label{sec:data}
In order to investigate the robustness of the proposed framework we study its performance on representing five classical hyperelastic material formulations which are summarized in this section.
We also discuss the sampling strategy employed to generate training data from these models.
In this work the sampling strategy and the evaluation of the approach assumes stress data under different loading conditions are accessible, e.g. cases where a micromechanical RVE or other high-fidelity model is available.

\subsection{Models}\label{subSec::Model}
The proposed methodology infers the orientation and type of the material symmetry and provides a model of the stress response.
In the following we divide the material models from which we generate synthetic data into those that can (a) verify or (b) validate the methodology used to solve the inverse problem.
Following the definitions of \cref{tsai1999verification}, we use the {\it verification} process to check the correctness of the inference of material symmetry by using constitutive laws with known symmetries and material orientations.
Then the TBNN formulation designed to solve the inverse problem of discovering anisotropy is {\it validated} on material responses where the symmetries and orientations are \apriori unknown and the proposed approach should find a best fit.

\subsubsection{Material models for verification }\label{sec::ModelVeri}
In this section three hyperelastic laws with anisotropies from known classes: isotropy, transverse isotropy, and orthotropy, are introduced for the purpose of verifying the TBNN.
\aref{app:ids} provides some of the tensor calculus identities necessary to derive the stress response from the potentials.

\paragraph{Isotropic hyperelasticity}
The well-known compressible isotropic neo-Hookean model is described by a strain energy function discussed in \crefs{ciarlet2021mathematical,bonet1998simple}:
\begin{equation}
\Psi_{\text{neo}} = \frac{1}{2} c_1 (I_1 - 3) - c_1 \log J + \frac{1}{2} c_2 (J -1)^2 \ ,
\end{equation}
which yields a second Piola-Kirchhoff stress of the form:
\begin{equation}
\Sb_{\text{neo}} = c_1 (\Ib - \Cb^{-1} ) + c_2 J(J-1) \Cb^{-1}.
\end{equation}
Herein $I_1 = \tr \Cb$ and  $J \equiv \sqrt{\det{\Cb}} = \sqrt{I_3}$.
We choose $c_{1} = \frac{1}{\sqrt{2}}$ and $c_{2} = \frac{10}{3}$ to generate the training data.

\paragraph{Transversely isotropic hyperelasticity}
Bonet and Burton \cite{bonet1998simple} introduced a hyperelastic model of transversely isotropic stress response defined by the strain energy:
\begin{equation}
\Psi = \Psi_{\text{neo}} +  (I_4 - 1) \left( c_0 + c_1 \log J  + c_2 (I_4 -1) \right) - \frac{1}{2} c_0 (I_5 - 1) \ .
\end{equation}
For this model the second Piola-Kirchhoff stress is
\begin{equation}
\Sb = \Sb_{\text{neo}} +  2 c_1 (I_4 -1) \Cb^{-1} + 2 ( c_0 + 2 c_1 \log J + c_2 (I_4 -1) ) \Nb - c_0 (\Cb \Nb + \Nb \Cb) \ .
\end{equation}
Herein $I_4 \equiv \tr \Cb\Nb_1$ and $I_5 \equiv \tr \Cb^2\Nb_1$.
For the results shown in \sref{sec:results} we specify $c_{0} = c_{1} = c_{2} = 1.0$ and $\nb  =(\frac{1}{\sqrt{2}},\frac{1}{\sqrt{2}},0)$.

\paragraph{Orthotropic hyperelasticity}
The orthotropic material we selected a modified version of the strain energy function proposed in \cref{holzapfel2000new} which reads
\begin{equation}
\Psi = c_{1} ( I_{1}- 3 ) + \frac{c_{1}}{c_{2}} (J^{-2 c_{2}}- 1)+ c_{3} \left(\exp( c_{4} (I_{4}-1)^{4})   + \exp(c_{5} (I_{6}-1)^{4})   -2 \right) \ ,
\end{equation}
where $I_4 \equiv \tr \Cb\Nb_1$ and $I_6 \equiv \tr \Cb\Nb_2$.
In this case second Piola-Kirchhoff stress is given by
\begin{equation}
\begin{aligned}
\Sb &= 2 c_{1} \Ib - 2 c_{1} I_{3}^{-c_{2}} \Cb^{-1} + 8 c_{3} c_{4} (I_{4}-1)^{3} \exp(c_{4}(I_{4}-1)^{4}) \Nb_{1} \\
&+ 8 c_{3} c_{5} (I_{6}-1)^{3} \exp(c_{5}(I_{6}-1)^{4}) \Nb_{2} \ .
\end{aligned}
\end{equation}
In this case we choose the following material parameters: $c_{1} = 5.5$, $c_{2}=0.75$, $c_{3} = 5.0$, $c_{4}= 1.5$, $c_{5}=1.5$, and preferred directions:
$\bm{n}_{1} = (\frac{1}{\sqrt{2}},-\frac{1}{\sqrt{2}},0)$ and $\bm{n}_{2} = (-\frac{1}{\sqrt{2}},-\frac{1}{\sqrt{2}},0)$.

\subsubsection{Material models for validation}\label{sec::ModelVali}
The proposed framework is validated on material models that do not have explicitly known anisotropic symmetries and orientations.
For this purpose we obtain stress-strain data from a fiber composite model and an elastic microstructure with inclusions.

\paragraph{Fiber anisotropy}
Distributed fiber models are commonly employed for modeling tissues \cite{zulliger2004strain,nguyen2008nonlinear,holzapfel2010constitutive}.
They are generally characterized by assuming a distribution of fiber orientations $\rho$ and a strain energy $\Psi_\text{fiber}$ for each fiber.
The resulting hyperelastic models can not be explicitly classified in the general orthotropic representation of \eref{eq::OrthoGeneral}.

Following \cref{holzapfel2002nonlinear} and \cref{tonge2013full} and  we assume that the response of a compressible fiber-reinforced material results from the strain energy function composed of a matrix and a fiber contribution:
\begin{equation}
\Psi = c_{1} (I_{1}-3)+ \frac{c_{1}}{c_{2}}(J^{-2 c_{2}} -1 ) + \int_{-\pi}^{\pi} \Psi_{\text{fiber}} (\lambda(\theta)) \rho(\theta) \mathrm{d} \theta.
\end{equation}
The response of the fibers $\Psi_{\text{fiber}}$ is described by
\begin{equation}
\Psi_{\text{fiber}}(\lambda(\theta)) = \frac{k_{1}}{2 k_{2}} \left( \exp \left[ k_{2} ( \lambda^{2}(\theta)-1)^{2} \right] \right) \ ,
\end{equation}
where the stretch projected on a fiber with in-plane orientation $\nb(\theta)$ is
\begin{equation}
\lambda^{2}(\theta) = \Cb : (\nb(\theta) \otimes \nb(\theta)).
\end{equation}
The fiber structure is defined by semicircular von Mises distribution function
\begin{equation}
\rho (\theta) = \frac{\exp(b \cos(2 \theta))}{2 \pi \iota_{0} b} \ ,
\end{equation}
where $\iota_{0}$ is the modified Bessel function of the first kind of order zero
\begin{equation}
\iota_{0}(b) = \frac{1}{\pi} \int_{0}^{\pi} \exp(b \cos(\theta)) d \theta.
\end{equation}
The second Piola-Kirchhoff stress is given by
\begin{equation}
\Sb = 2 c_{1} \Ib - 2 c_{1} I_{3}^{-c_{2}} \Cb^{-1} + \int_{-\pi}^{\pi} \frac{1}{\lambda}
\frac{\partial \Psi_{\text{fiber}}}{\partial \lambda} (\nb(\theta) \otimes \nb(\theta))\rho(\theta) \mathrm{d} \theta
\end{equation}
with
\begin{equation}
\frac{\partial \Psi_{\text{fiber}}}{\partial \lambda} = 2 k_{1}
\lambda (\lambda^{2} - 1) \left( \exp \left[ k_{2} ( \lambda^{2}(\theta)-1)^{2} \right] \right).
\end{equation}

\begin{figure}
\centering
\includegraphics[width=0.5\linewidth]{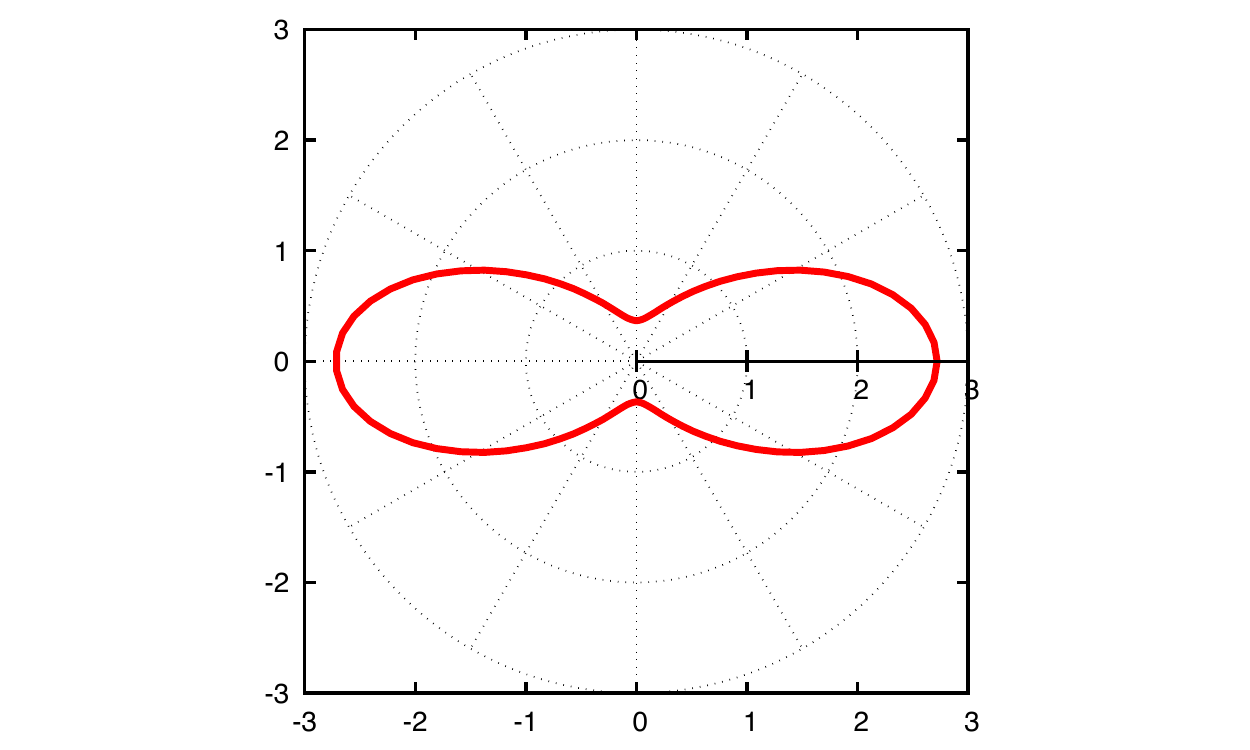}
\caption{In-plane fiber distribution $\rho$.
}
\label{fig:rho}
\end{figure}

Here we specify representative parameters:
$c_{1}=3.0$, $c_{2} = 0.75$, $k_{1} = \frac{4}{15}$, $k_{2}= \frac{170}{15}$  and $b=1$.
The resulting fiber distribution is plotted in \fref{fig:rho}.
The fibers lie in the $\eb_1$-$\eb_2$ plane and the major lobe is in the $\eb_1$ direction.

\begin{figure}
\centering
\includegraphics[scale=0.25]{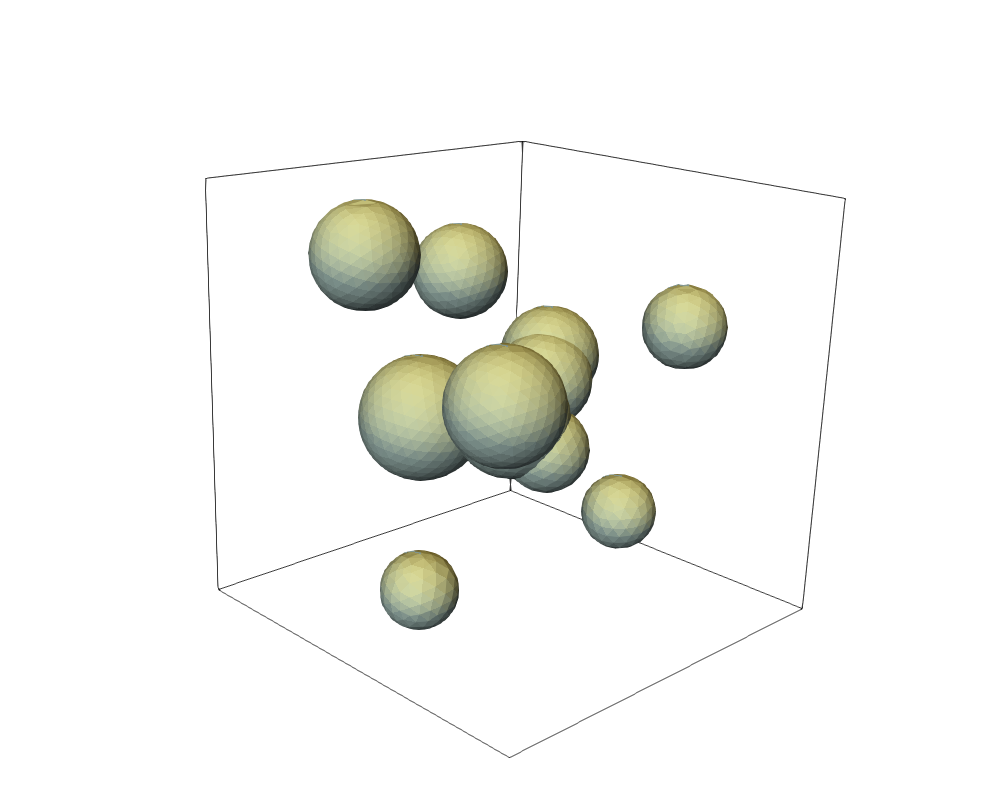}
\caption{Representative volume element with $12$ inclusions.}\label{fig:Incl_RVE}
\end{figure}

\paragraph{Microstructure with inclusions}
Lastly, the framework is also tested on training data from an effective microstructural response, specifically a microstructure with 12 inclusions used as a Representative Volume Element (RVE), see \fref{fig:Incl_RVE}.
The microstructural quantities are denoted by the subscript ``$m$".
Following \cref{fuhg2022local} the stiffness response of the materials in both phases can be characterized by a compressible neoHookean formulation of the form
\begin{equation}
\Psi_{m} =  c_{1} (I_{1} - 3) + \frac{c_{1}}{c_{2}} (J^{-2 c_{2}} - 1).
\end{equation}
The resulting second Piola-Kirchhoff stress reads
\begin{equation}
\Sb_{m} = 2 c_{1} \Ib - 2 I_{3}^{-c_{2}} \Cb^{-1}
\end{equation}
with
\begin{equation}
\begin{aligned}
c_{2} &= \frac{\nu}{1-2\nu}, \qquad c_{1} &= \frac{\mu}{2}.
\end{aligned}
\end{equation}
in terms of the Poisson's ratio $\nu$ and the shear modulus $\mu$.
The shear modulus in both phases is $80$ GPa.
The bulk modulus for the inclusions is $120$ GPa and for the matrix it is $160$ GPa.
The effective response of the microstructure can be obtained using numerical homogenization conditioned on an imposed mean deformation gradient.
The equilibrium equation at the microscale is given by
\begin{equation}
\bm{\nabla}_{m} \cdot  \Pb_{m} = \bm{0} \ ,
\end{equation}
where $\Pb_{m}$ denotes the microscopic first Piola-Kirchhoff stress and $\bm{\nabla}_{m}$ is the microscopic gradient operator with respect to the reference configuration.
The RVE is assumed to have geometrically periodic boundaries.
After solving this boundary value problem, the macroscopic first Piola-Kichhoff stress tensor $\Pb$ is computed via a volume average
\begin{equation}
\Pb = \frac{1}{V} \int_{V} \Pb_{m} \, \mathrm{d} V  \ ,
\end{equation}
over the volume $V$ of the RVE.
The pull-back operation then allows us to obtain the averaged second Piola-Kirchhoff stress tensor
\begin{equation}
\Sb     = \Fb^{-1}  \Pb \ ,
\end{equation}
where $\Fb$ is the applied macroscopic deformation gradient.
For more information we refer to \cref{geers2017homogenization}.

After discretization, the RVE consists of $\approx 25,000$ hexahedral elements.
An author-modified version of the finite element (FE) simulation code provided by Ref. \cite{yaghoobi2019prisms} was used for these computational homogenization simulations.

\subsection{Sampling}\label{sec::Sampling}

In each case, we collect samples over a bounded deformation gradient space around the undeformed configuration ($\Fb = \Ib$)  given by
\begin{equation}\label{eq::DefoSpace}
{F}_{ij} \in \delta_{ij} + [-\lambda, \lambda]
\end{equation}
with $\lambda>0$ and $\delta_{ij}$ being the Kronecker delta.
We employed Latin Hypercube Sampling \cite{stein1987large} to generate $N$ space filling samples in this nine-dimensional bounded space.
The right Cauchy-Green tensor $\Cb = \Fb^{T}\Fb$ of each sample was derived from $\Fb$ and the model evaluated to obtain $\Sb(\Cb)$ and a dataset of the form $\mathcal{D} = \lbrace \Cb_{i}, \Sb_{i} \rbrace_{i=1}^{N}$.
All of the following numerical examples employ $\lambda=0.2$.

For each model $N=2500$ samples were calculated.
Prior to training the TBNN, both the inputs $\Cb$ and outputs $\Sb$ are normalized to lie between $0$ and $1$.

\section{Results} \label{sec:results}

In the following we demonstrate the ability of the proposed approach to: (a) recover both the orientation and the type of anisotropy for material responses where these properties are well classified (cf. \sref{sec::ModelVeri}), and (b) discover best fitting surrogates to data with imprecise symmetries (cf. \sref{sec::ModelVali}).

In order to visualize the proficiency of the trained TBNN surrogates, the ground truth stress as well as the predicted stress of the models are plotted for the loading case
\begin{equation}\label{eq::LoadCase}
\Fb_\text{test} = \Ib + (1-\lambda_{11}) (\eb_{1} \otimes \Eb_{1}), \qquad \lambda_{11} \in [-0.2, 0.2].
\end{equation}
This loading is not part of the training dataset generated with random sampling, refer to \sref{sec::Sampling}.

In addition to the results in this section, the influence of the number of data points on the proposed framework for the material models of \sref{sec::ModelVeri} is reported in \aref{append:ConverStud}.
Also, as a modification of the TBNN approach presented in \sref{sec:method} with a simple feed-forward architecture which does not adhere to polyconvexity requirements \cite{steigmann2003frame}, an alternative input-convex neural network is presented in \aref{append:InputConvex}.
This kind of network enables the learning hyperelastic anisotropy with polyconvex potentials.

\subsection{Recovering anisotropy type and orientation}

\begin{figure}
\begin{subfigure}[b]{0.5\linewidth}
\includegraphics[scale=0.3]{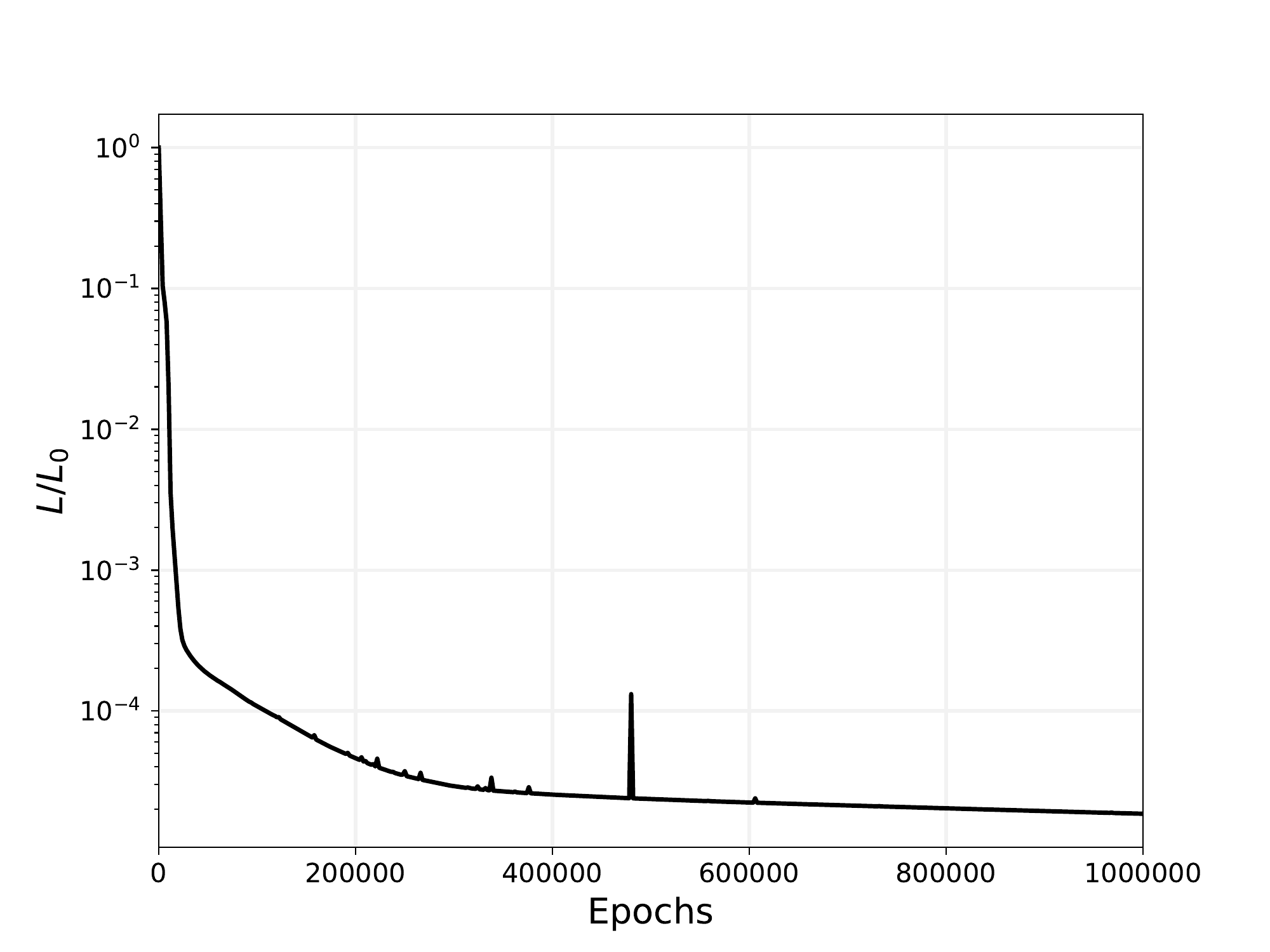}
\caption{Loss}\label{fig:loss_over_epochsIso}
\end{subfigure}
\begin{subfigure}[b]{0.5\linewidth}
\includegraphics[scale=0.3]{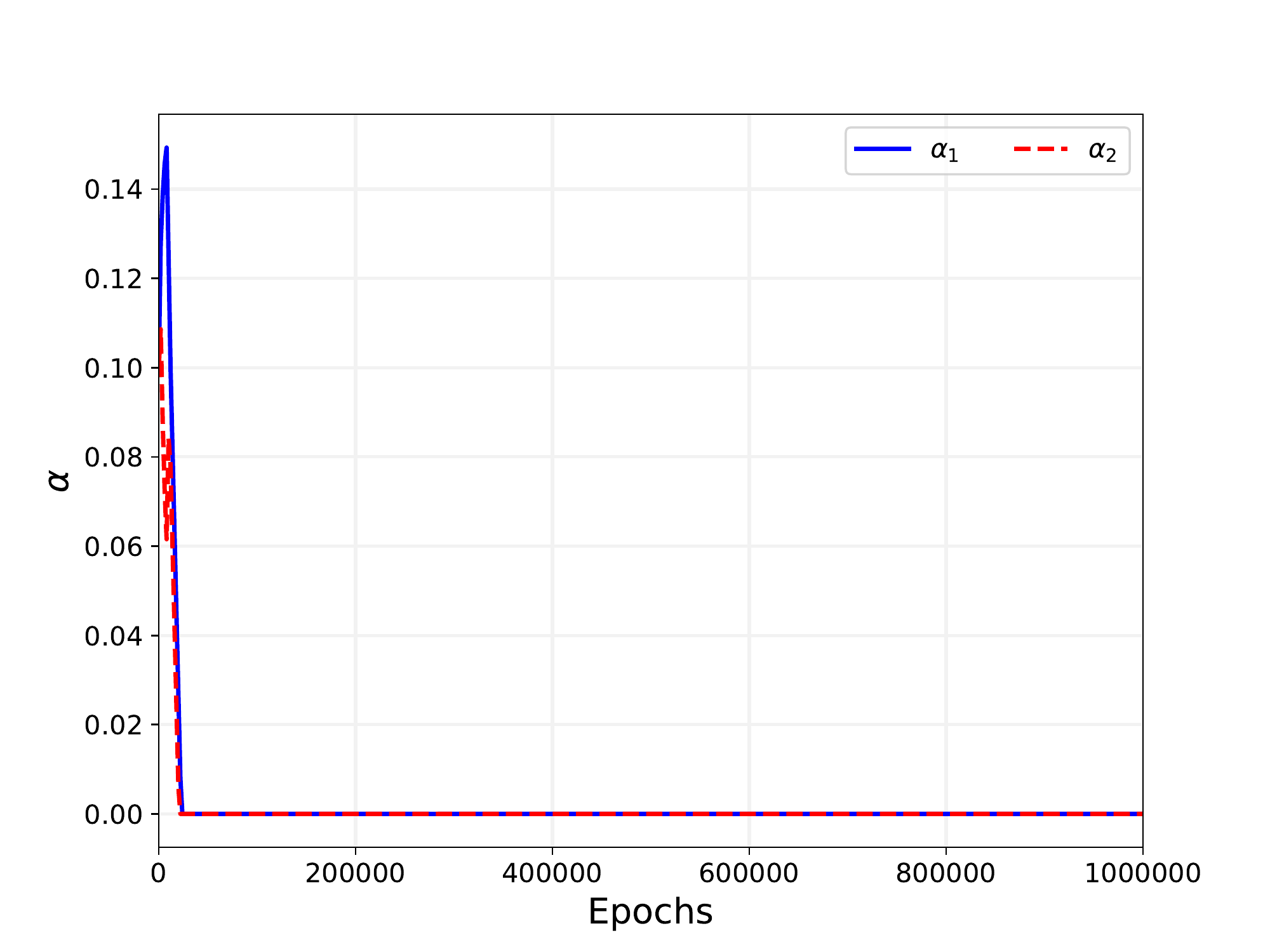}
\caption{Anisotropic coefficients}\label{fig:AniCoeffs_over_epochsIso}
\end{subfigure}
\begin{center}
\begin{subfigure}[b]{0.5\linewidth}
\includegraphics[scale=0.3]{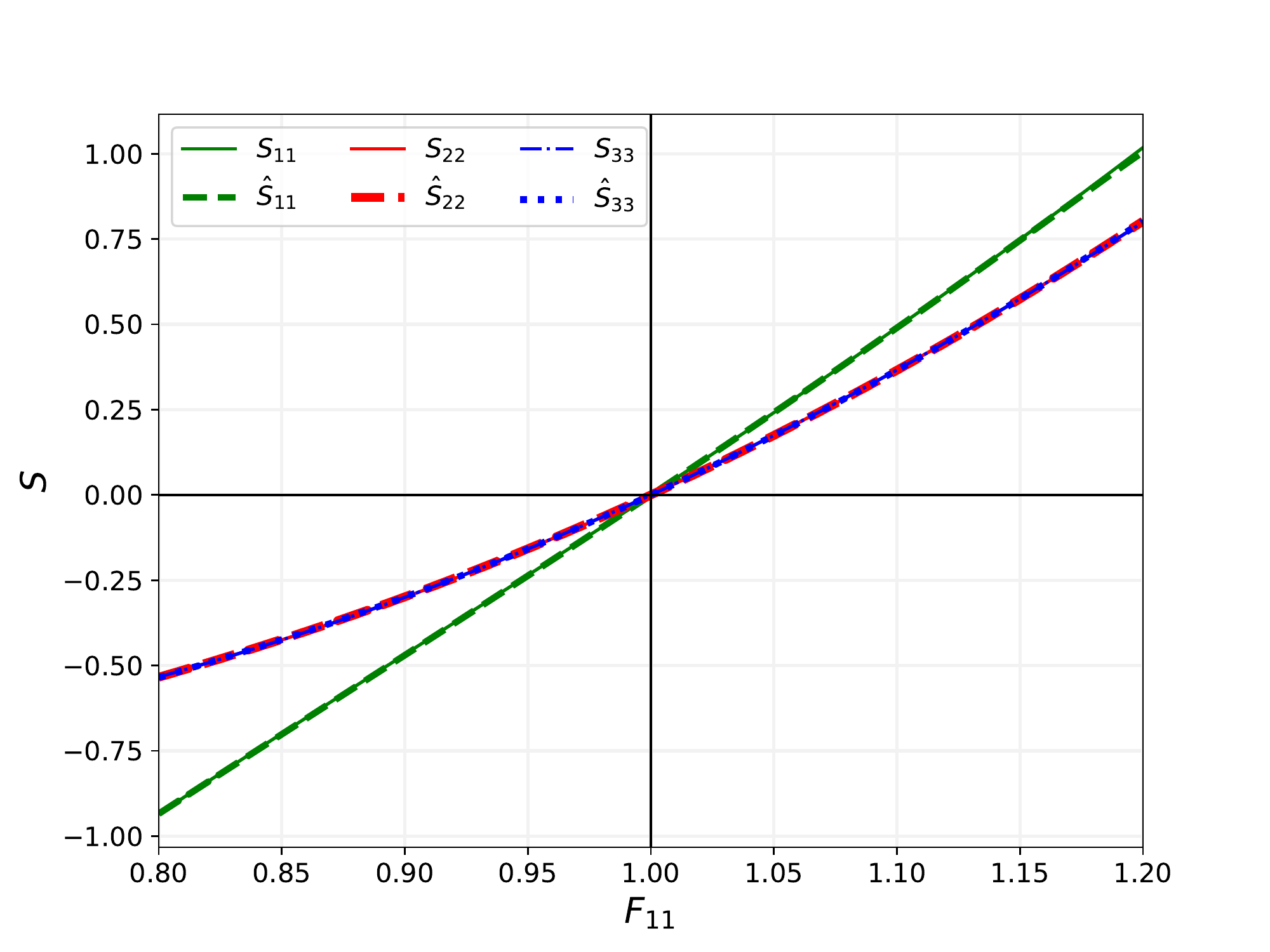}
\caption{Stress comparison}\label{fig:StressIso}
\end{subfigure}
\end{center}
\caption{Results for training the TBNN to data from the \textbf{isotropic} material model of  \sref{sec::ModelVeri}. (a) Training loss over epochs, (b) anisotropic coefficients $\alpha_i$ over epochs, and (c) true (solid lines) and predicted stresses (dashed lines) over $F_{11}$.
}
\end{figure}

We start by demonstrating that the TBNN, as formulated in \sref{sec:method} with a full orthotropic basis, can accurately recover an isotropic response.
\Fref{fig:loss_over_epochsIso} plots the loss over the training process.
We can see that a smooth error convergence is achieved.
For isotropic materials both anisotropic coefficients, $\alpha_{1}$ and $\alpha_{2}$, should converge to zero.
\Fref{fig:AniCoeffs_over_epochsIso} shows that anisotropic coefficients rapidly go to zero over the training.
Hence, this example clearly demonstrates that the method is efficient and effective in recovering isotropic material behavior.
\fref{fig:StressIso} plots the ground truth and the predicted stress for the load case of \eref{eq::LoadCase}.
We can see that although this test data is not part of the training dataset an accurate stress representation is found.

\begin{figure}
\begin{subfigure}[b]{0.5\linewidth}
\includegraphics[scale=0.3]{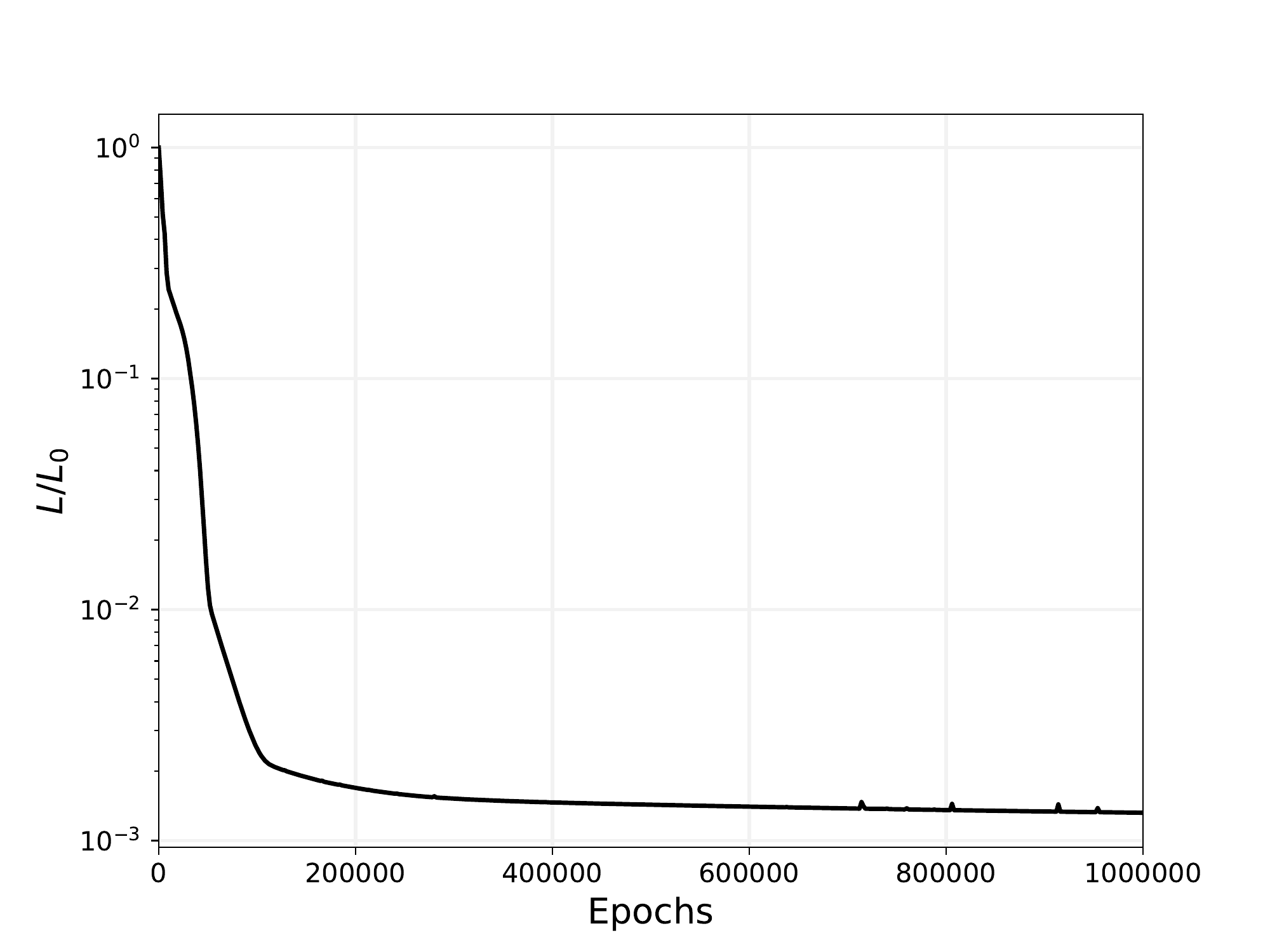}
\caption{Loss}\label{fig:loss_over_epochs}
\end{subfigure}
\begin{subfigure}[b]{0.5\linewidth}
\includegraphics[scale=0.3]{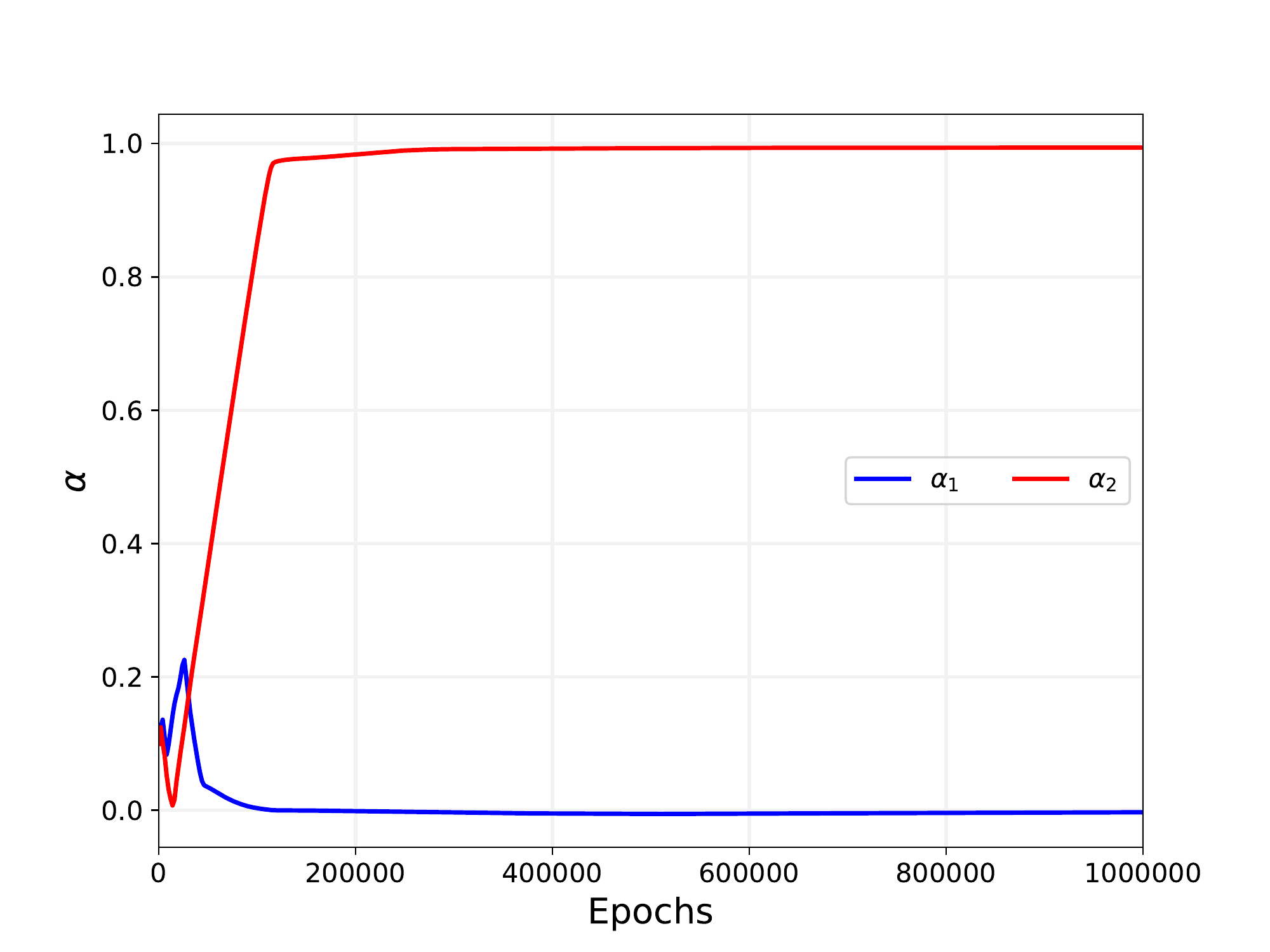}
\caption{Anisotropic coefficients}\label{fig:AniCoeffs_over_epochs}
\end{subfigure}
\begin{center}
\begin{subfigure}[b]{0.5\linewidth}
\includegraphics[scale=0.3]{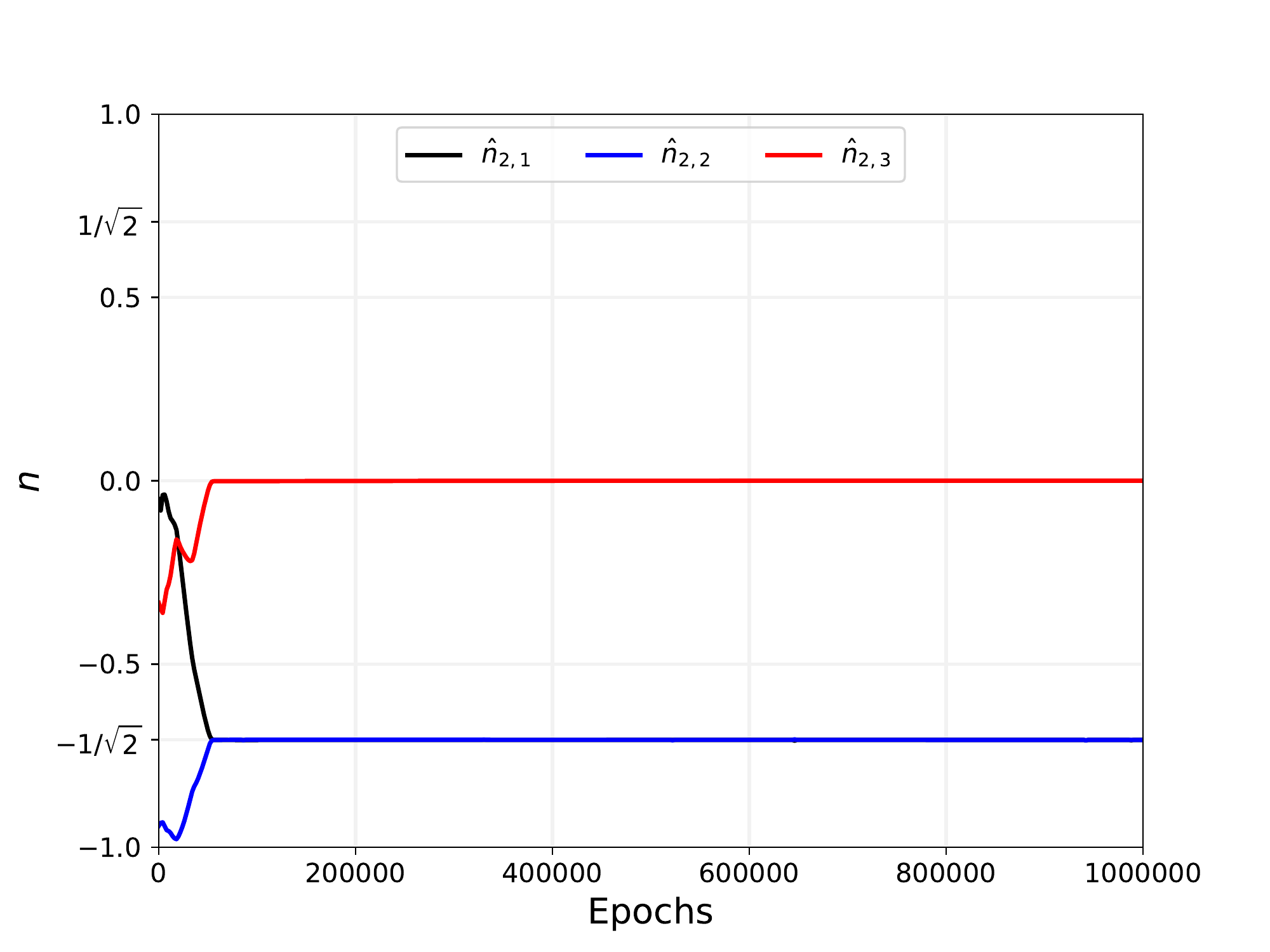}
\caption{Preferred direction}\label{fig:PrefDir_over_epochs}
\end{subfigure}%
\begin{subfigure}[b]{0.5\linewidth}
\includegraphics[scale=0.3]{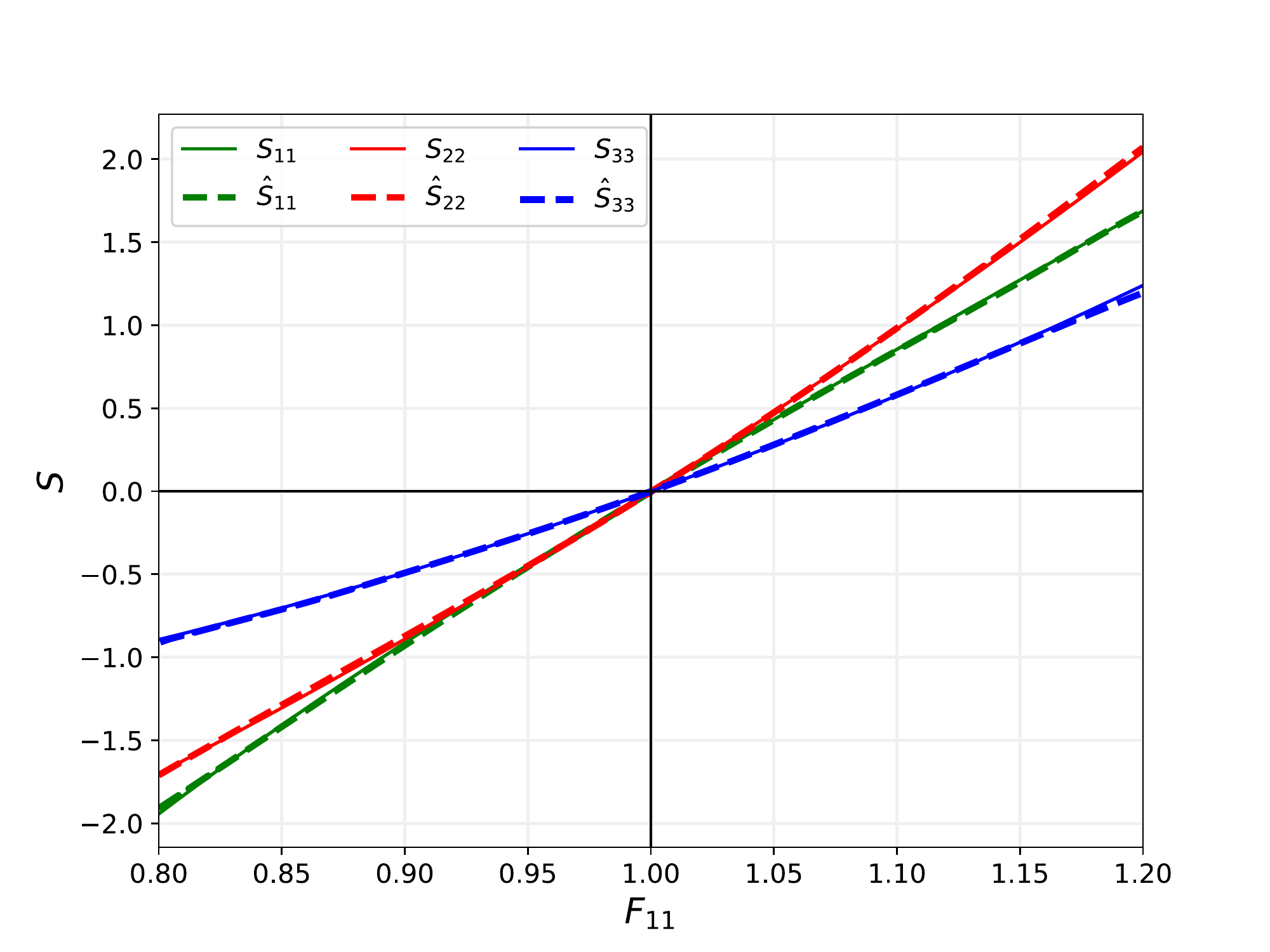}
\caption{Stress comparison}\label{fig:StressTrans}
\end{subfigure}
\end{center}
\caption{Results for training the TBNN to data from the \textbf{transversely isotropic} material model of \sref{sec::ModelVeri} with preferred direction $\nb = [\frac{1}{\sqrt{2}},\frac{1}{\sqrt{2}},0]^{T}$. (a) Training loss over epochs, (b) anisotropic coefficients $\alpha_i$ over epochs, (c) inferred preferred direction associated with $\alpha_{2}$, and (d) true (solid lines) and predicted stresses (dashed lines) over $F_{11}$.}
\end{figure}

Next, we consider the stress-strain data from the transversely isotropic model of \sref{sec::ModelVeri}.
The preferred direction of the data source model is chosen to be $\nb = (\frac{1}{\sqrt{2}},\frac{1}{\sqrt{2}},0).$
\Fref{fig:loss_over_epochs} plots the training loss over the training process.
We can see that, again, the convergence is regular and the error has essentially converged after $\approx 400,000$ epochs.
For this example we expect only one anisotropic coefficient to remain while the other one should tend to zero.
The recovered coefficients over the training process are plotted in \fref{fig:AniCoeffs_over_epochs}, and clearly shows that the presented TBNN framework accurately identifies the degree of anisotropy, with $\alpha_{1} \to 0$ and $\alpha_2 \to 1$.
\Fref{fig:PrefDir_over_epochs} displays evolution of the components of the preferred direction $\bm{n}_{2}$, which converge to the true material orientation (to within a sign).
Hence, this example demonstrates the proposed anisotropy discovery method can accurately recover the response and symmetries of an off-axis transversely isotropic material.
Finally, the diagonal components of the true stress and predicted stress responses for the uniaxial test case are plotted over $F_{11}$ in  \fref{fig:StressTrans} which indicates the validity of the trained surrogate in terms of its prediction accuracy on unseen data.

\begin{figure}
\begin{subfigure}[b]{0.5\linewidth}
\includegraphics[scale=0.3]{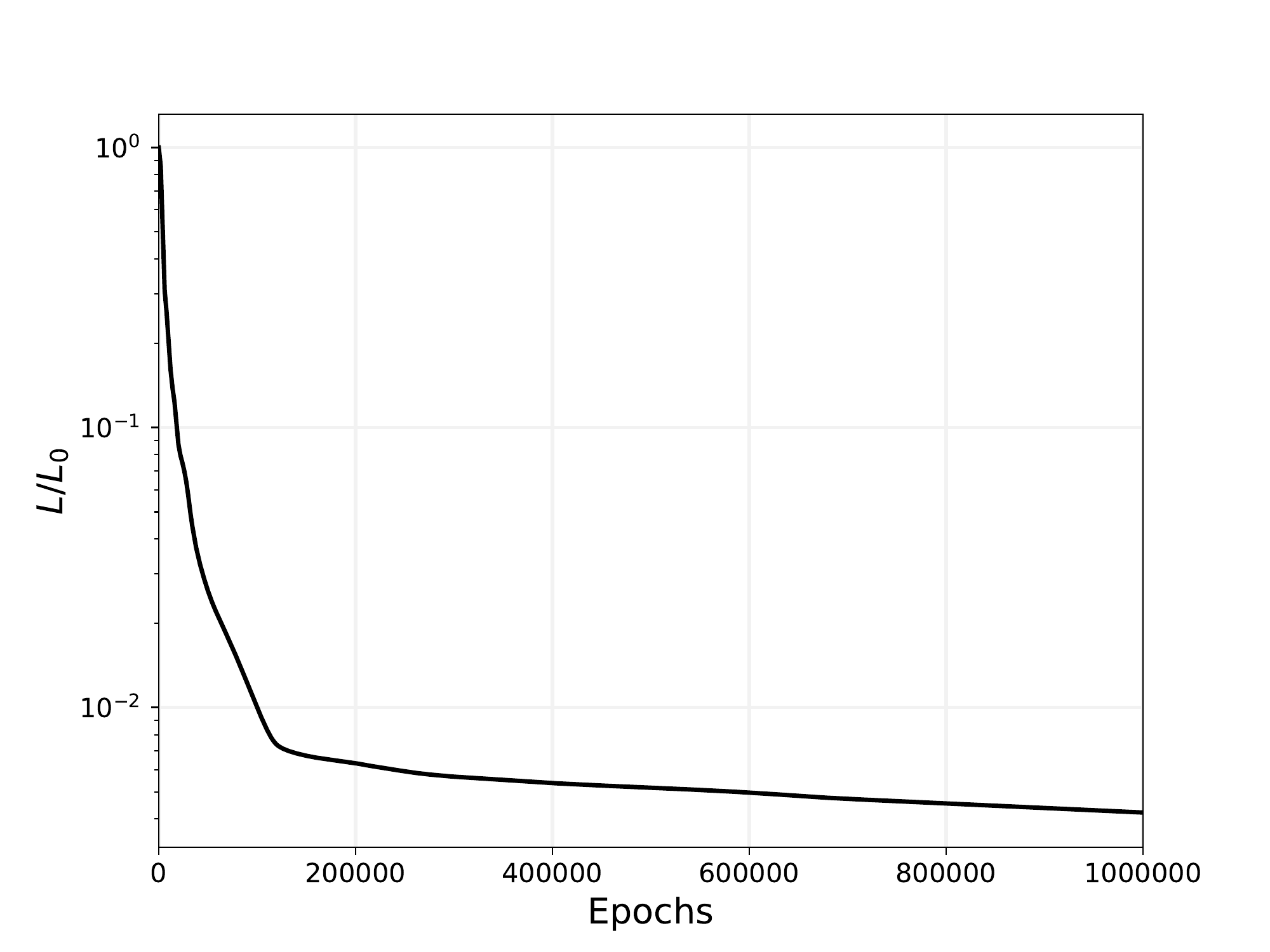}
\caption{}\label{fig:loss_over_epochs_ortho}
\end{subfigure}
\begin{subfigure}[b]{0.5\linewidth}
\includegraphics[scale=0.3]{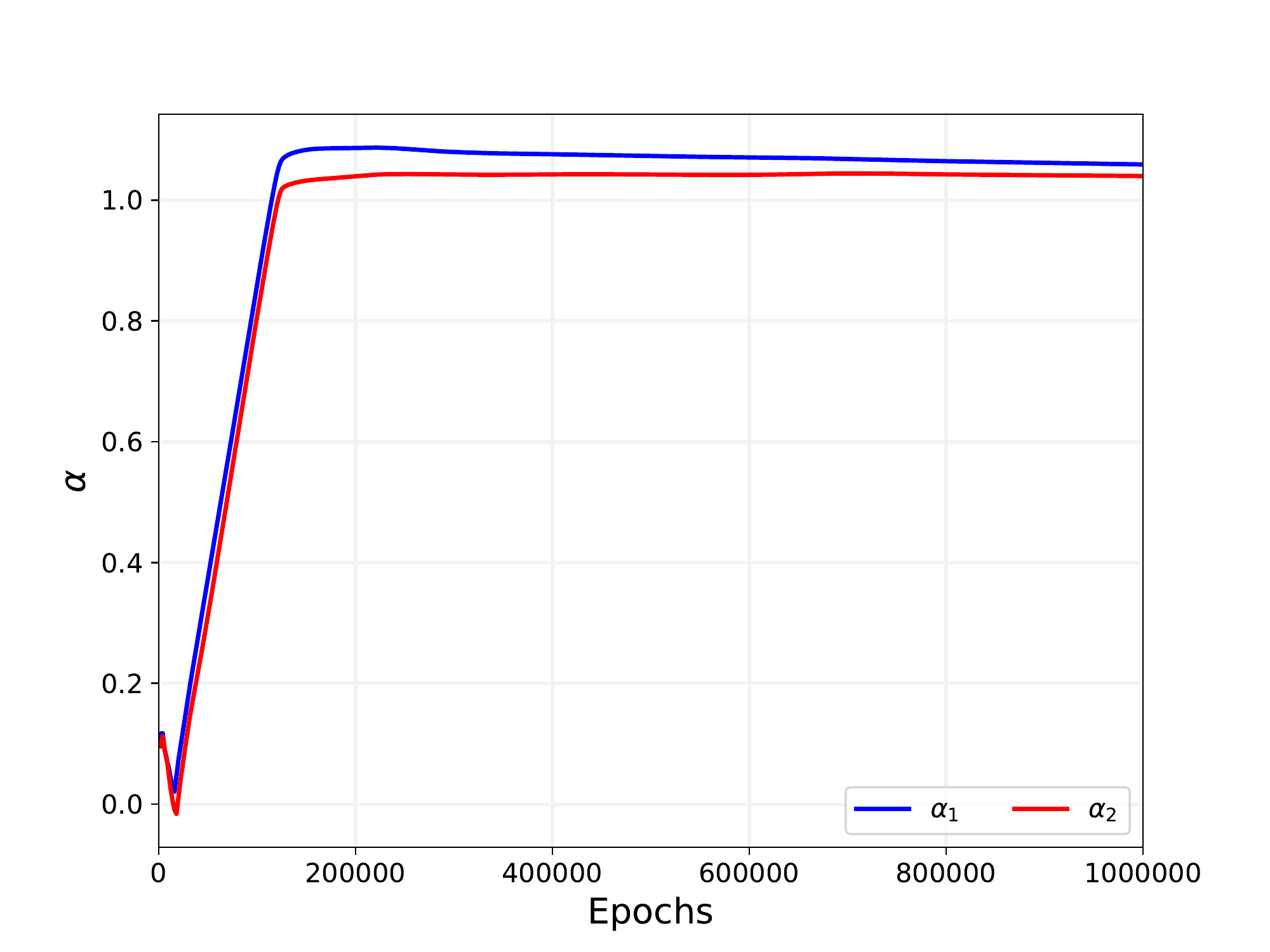}
\caption{}\label{fig:AniCoeffs_over_epochs_ortho}
\end{subfigure}
\begin{subfigure}[b]{0.5\linewidth}
\includegraphics[scale=0.3]{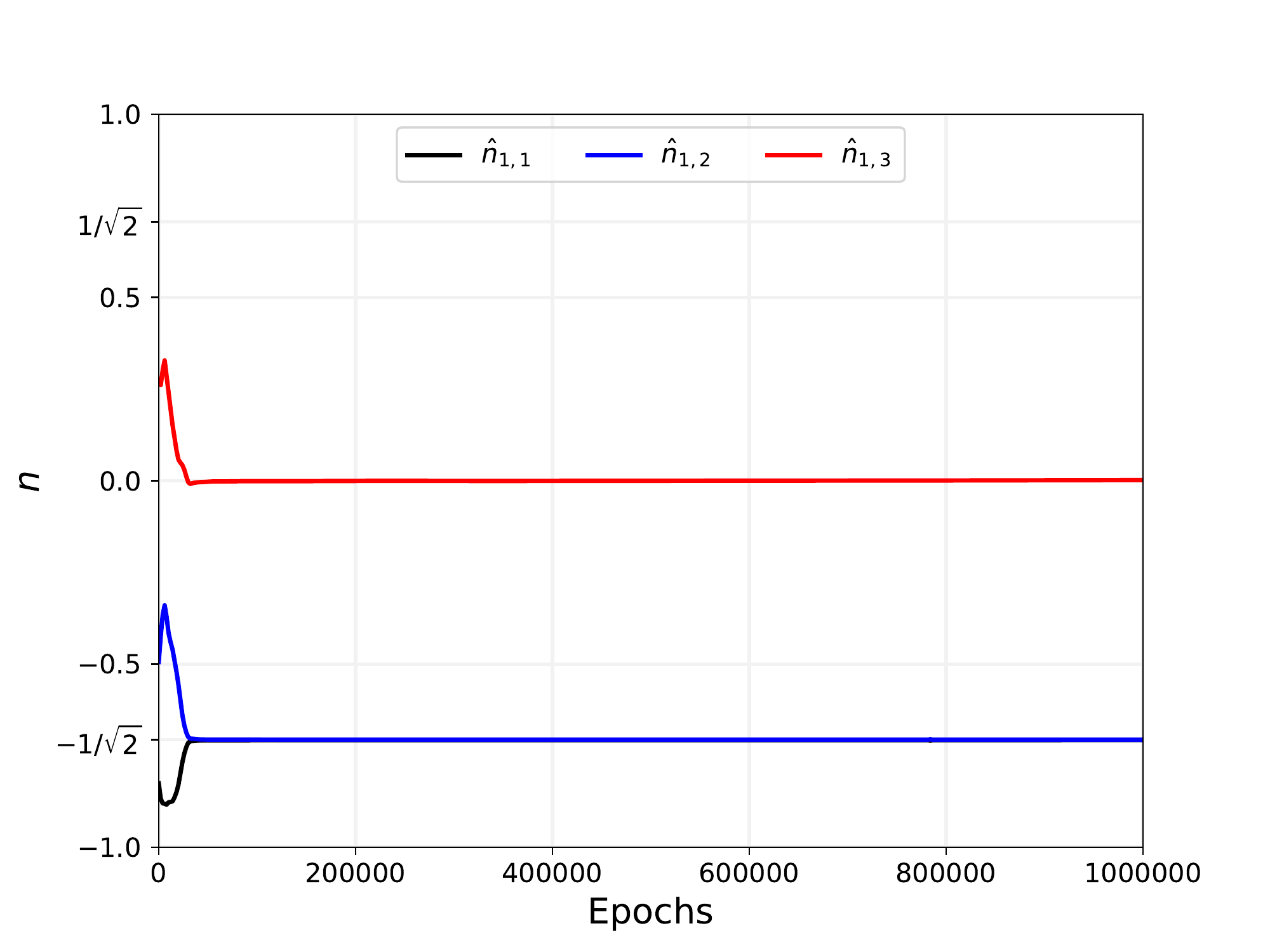}
\caption{}\label{fig:PrefDir1_over_epochs_ortho}
\end{subfigure}
\begin{subfigure}[b]{0.5\linewidth}
\includegraphics[scale=0.3]{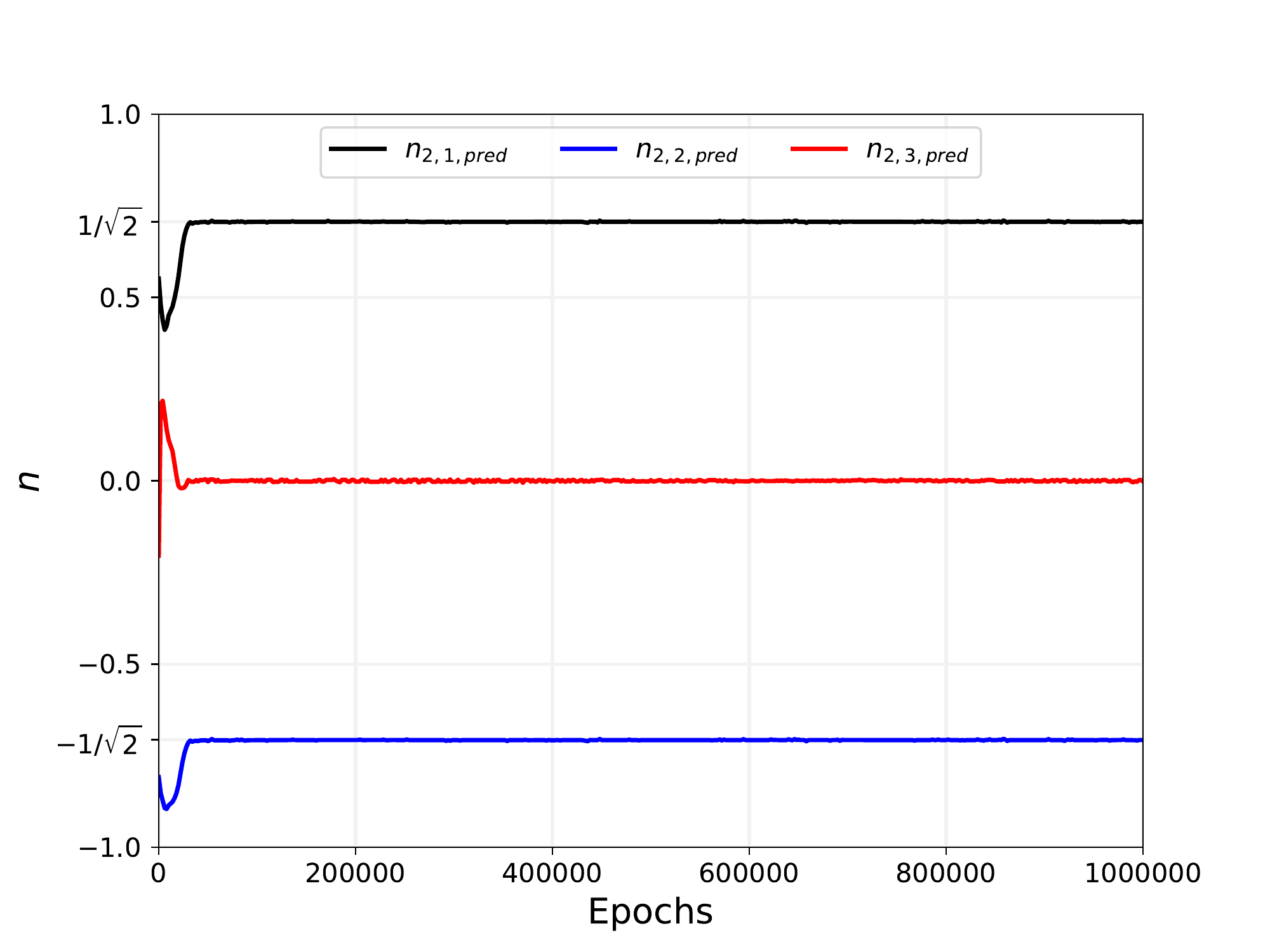}
\caption{}\label{fig:PrefDir2_over_epochs_ortho}
\end{subfigure}
\begin{center}
\begin{subfigure}[b]{0.5\linewidth}
\includegraphics[scale=0.3]{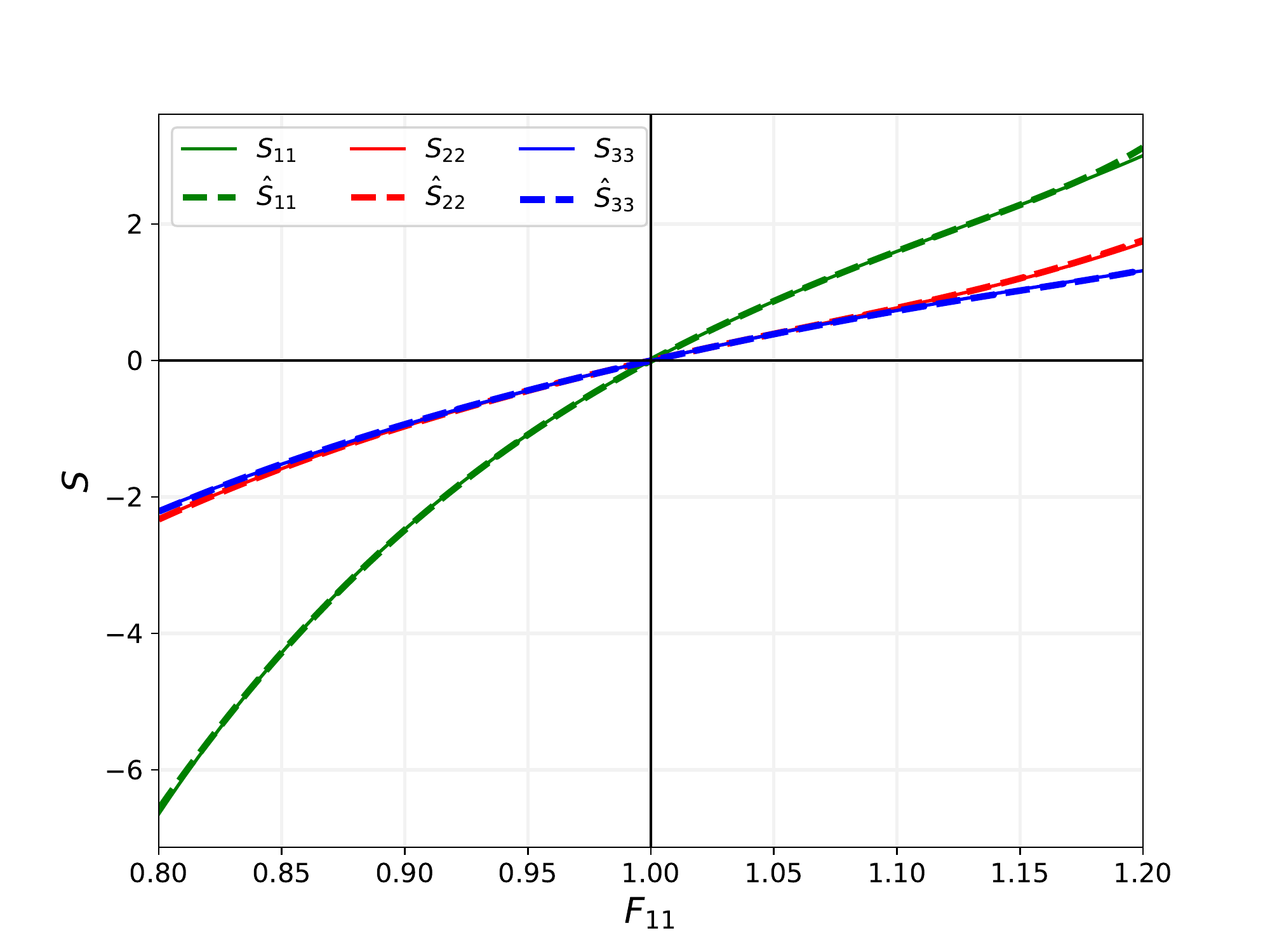}
\caption{}\label{fig:StressOrtho}
\end{subfigure}
\end{center}
\caption{Results for the \textbf{orthotropic} material model of \sref{sec::ModelVeri}. (a) Training loss over epochs, (b) anisotropic coefficients over epochs, (c-d) preferred directions. Known solutions of the preferred directions $\bm{n}_{1} = (\frac{1}{\sqrt{2}},\frac{1}{\sqrt{2}},0)$ and $\bm{n}_{2} = (\frac{1}{\sqrt{2}},-\frac{1}{\sqrt{2}},0)$.
and (d) true (solid lines) and predicted stresses (dashed lines) over $F_{11}$..
}
\end{figure}

Lastly, we check if the proposed approach is able to accurately recover the anisotropy defining coefficients and orientations of the orthotropic material model described in \sref{sec::ModelVeri}.
Here, we expect both of the trainable anisotropy-defining coefficients $\alpha_{1}$ and $\alpha_{2}$ to be non-zero.
The two preferred directions of the data source model are chosen to be the off-axis unit-vectors $\nb_{1} = (\frac{1}{\sqrt{2}},\frac{1}{\sqrt{2}},0)$ and $\nb_{2} = (\frac{1}{\sqrt{2}},-\frac{1}{\sqrt{2}},0)$.
\Fref{fig:loss_over_epochs_ortho} shows the training loss over the training process.
The evolution of the two anisotropic coefficients is plotted in \fref{fig:AniCoeffs_over_epochs_ortho}.
It can be seen that both coefficient values converge to non-zero values after $\approx 200,000$ epochs.
Figures \ref{fig:PrefDir1_over_epochs_ortho} and \ref{fig:PrefDir2_over_epochs_ortho} plot the components of the associated preferred directions $\bm{n}_{1}$ and $\bm{n}_{2}$.
They converge to an accurate match (up to a sign) of the ground truth $\nb_i$.
These results prove that the anisotropic properties of an orthotropic material can be correctly recovered.
To highlight the prediction quality in terms of stress outputs of the trained surrogate, the true and predicted $S_{11}$, $S_{22}$ and $S_{33}$ stress components are plotted in \fref{fig:StressOrtho} over $F_{11}$ for the uniaxial test case.

\subsection{Discovering anisotropy structure and orientation}
For the following two examples neither the type of the anisotropy nor the orientations are directly expressible as a subclass of the parameterized orthotropic stress representation of \eref{eq::MethodS}.
Applying our proposed framework on data of this kind will allow us to find best fitting anisotropy structures and orientations in terms of \eref{eq::MethodS}.
This can help us gain insight into ``effective" anisotropies and orientations that are hidden in the data.

\begin{figure}
\begin{subfigure}[b]{0.5\linewidth}
\includegraphics[scale=0.3]{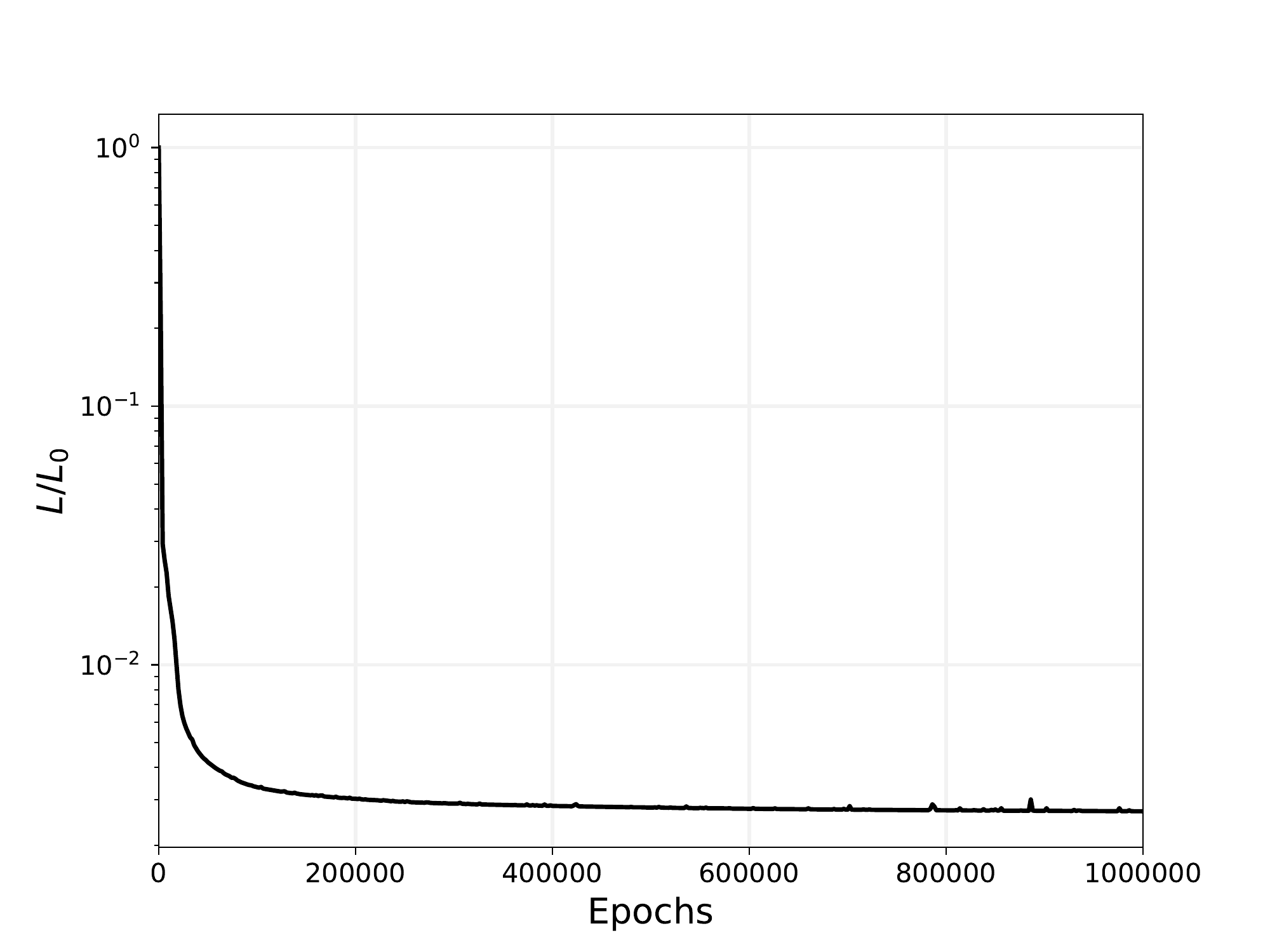}
\caption{Loss}\label{fig:loss_over_Fiber}
\end{subfigure}
\begin{subfigure}[b]{0.5\linewidth}
\includegraphics[scale=0.3]{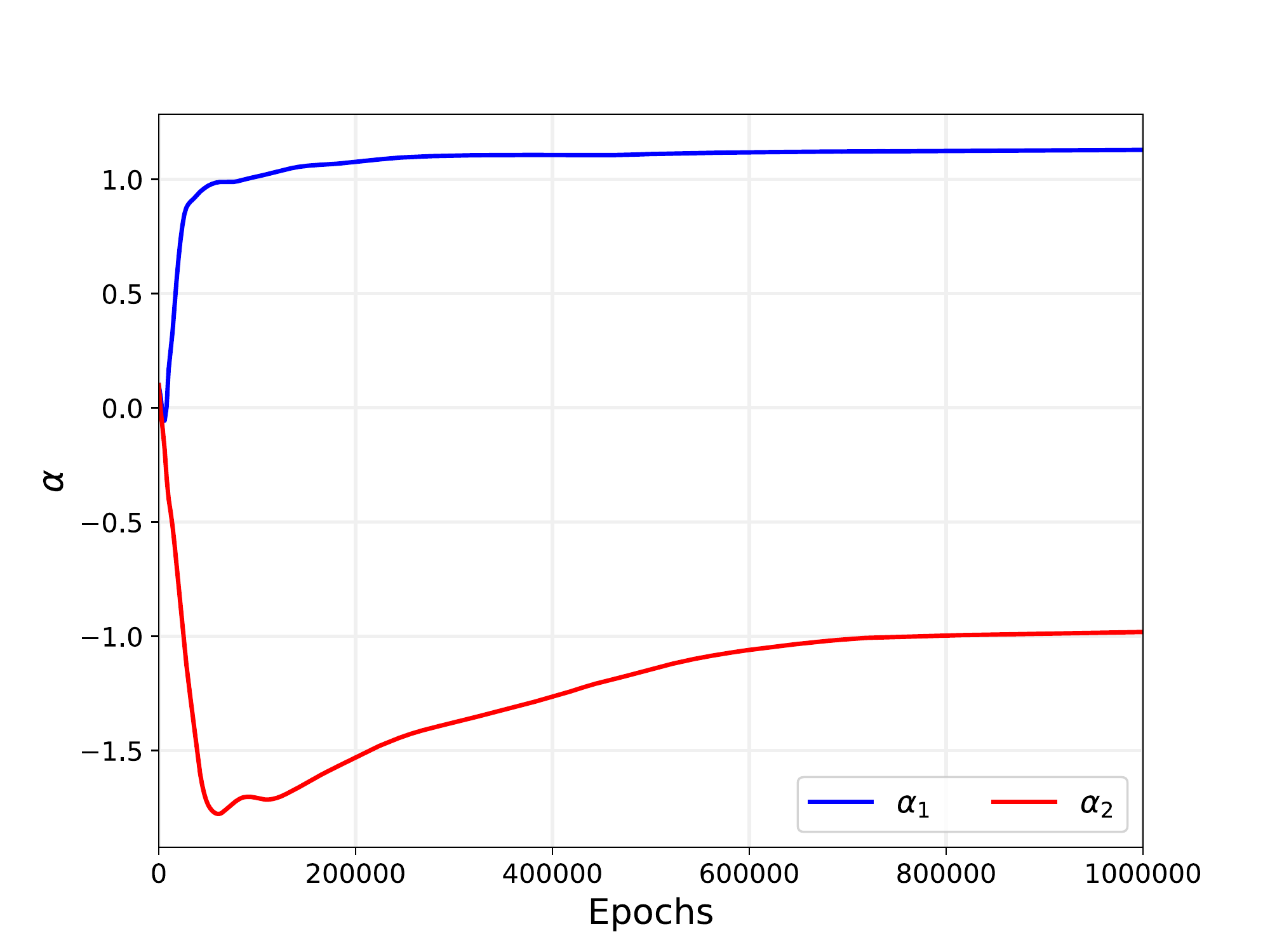}
\caption{Anisotropic coefficients}\label{fig:AniCoeffs_over_Fiber}
\end{subfigure}
\begin{subfigure}[b]{0.5\linewidth}
\includegraphics[scale=0.3]{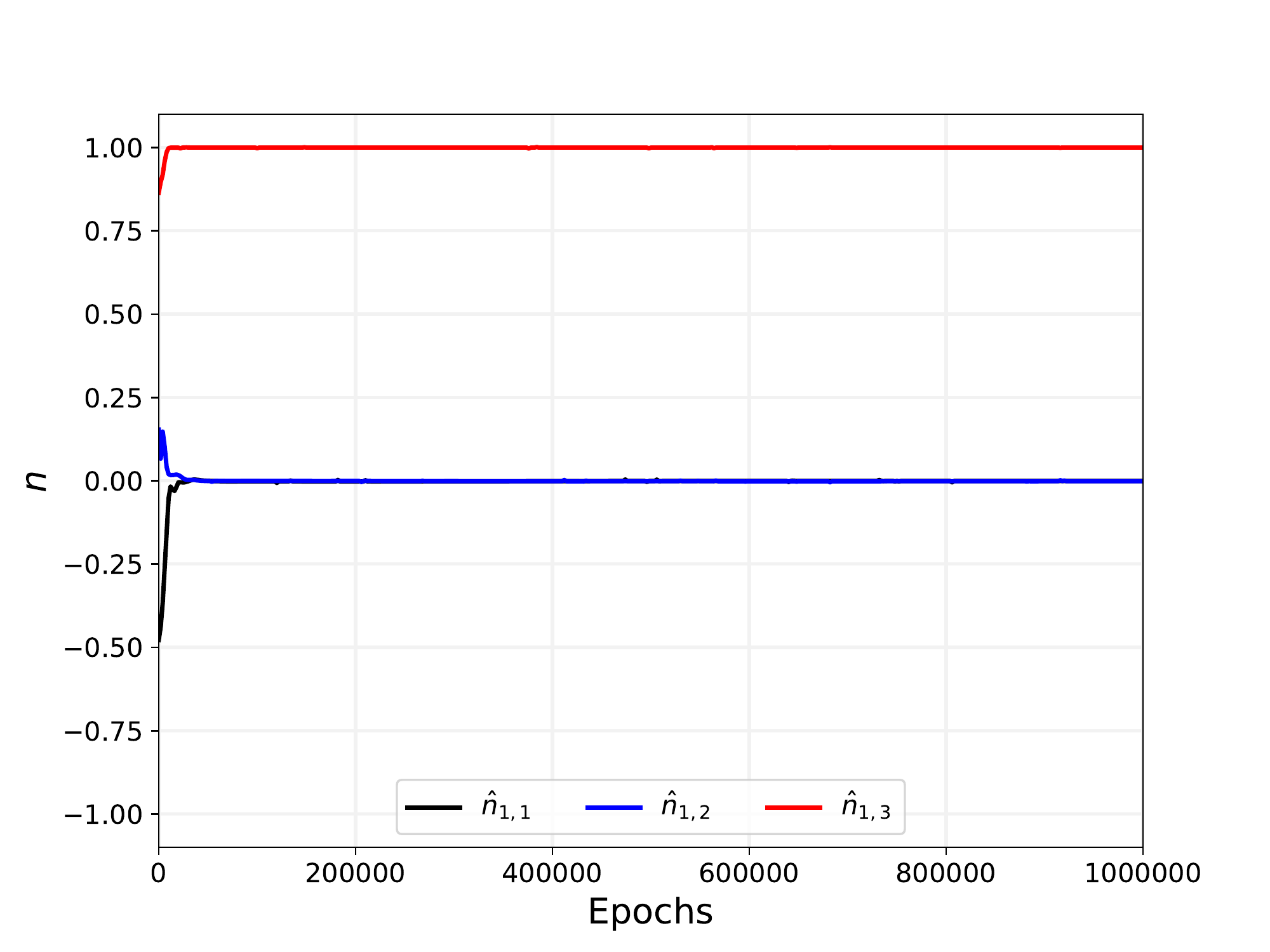}
\caption{Preferred direction 1}\label{fig:n1Fiber}
\end{subfigure}
\begin{subfigure}[b]{0.5\linewidth}
\includegraphics[scale=0.3]{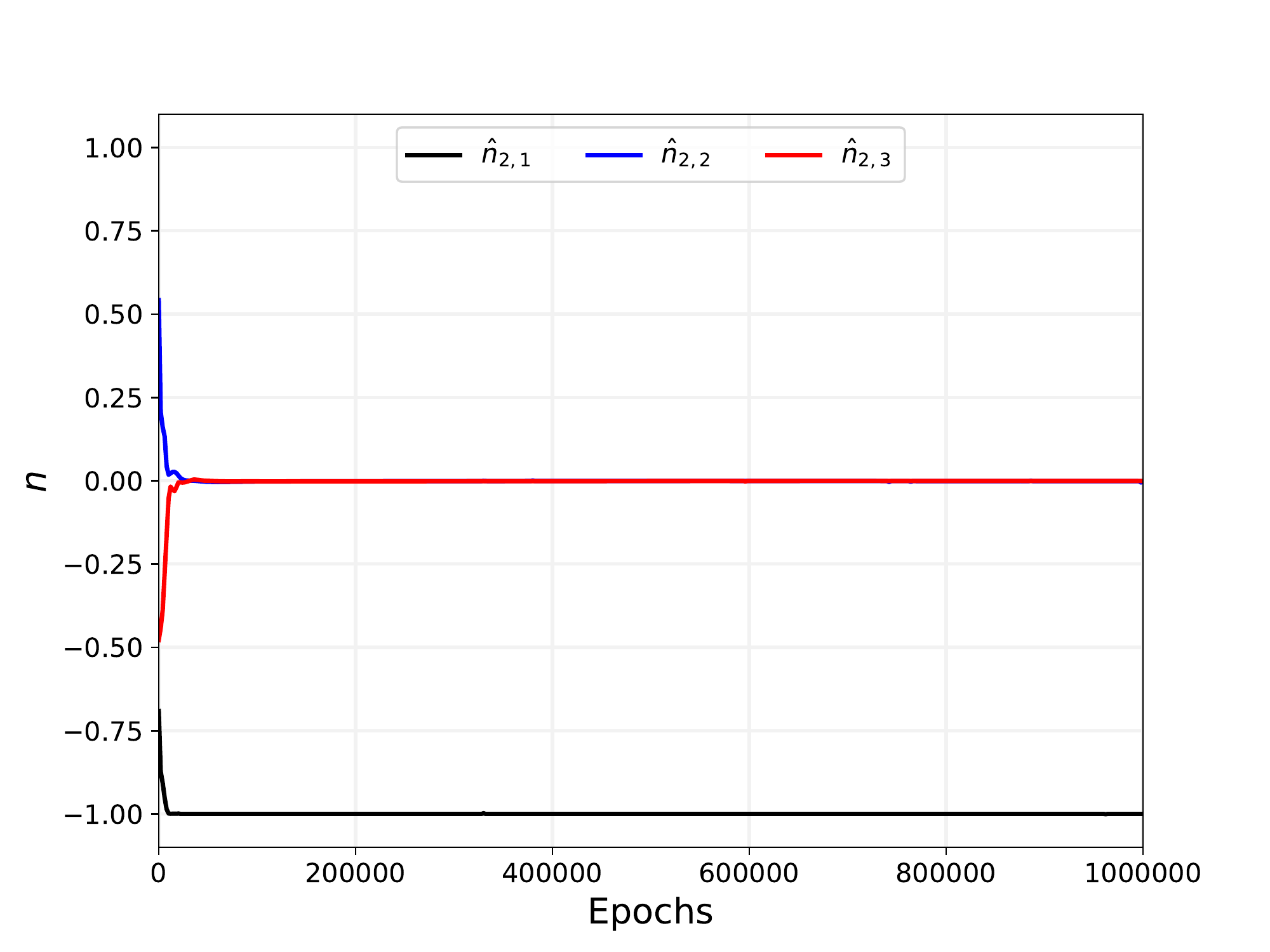}
\caption{Preferred direction 2}\label{fig:n2Fiber}
\end{subfigure}
\begin{center}
\begin{subfigure}[b]{0.5\linewidth}
\includegraphics[scale=0.3]{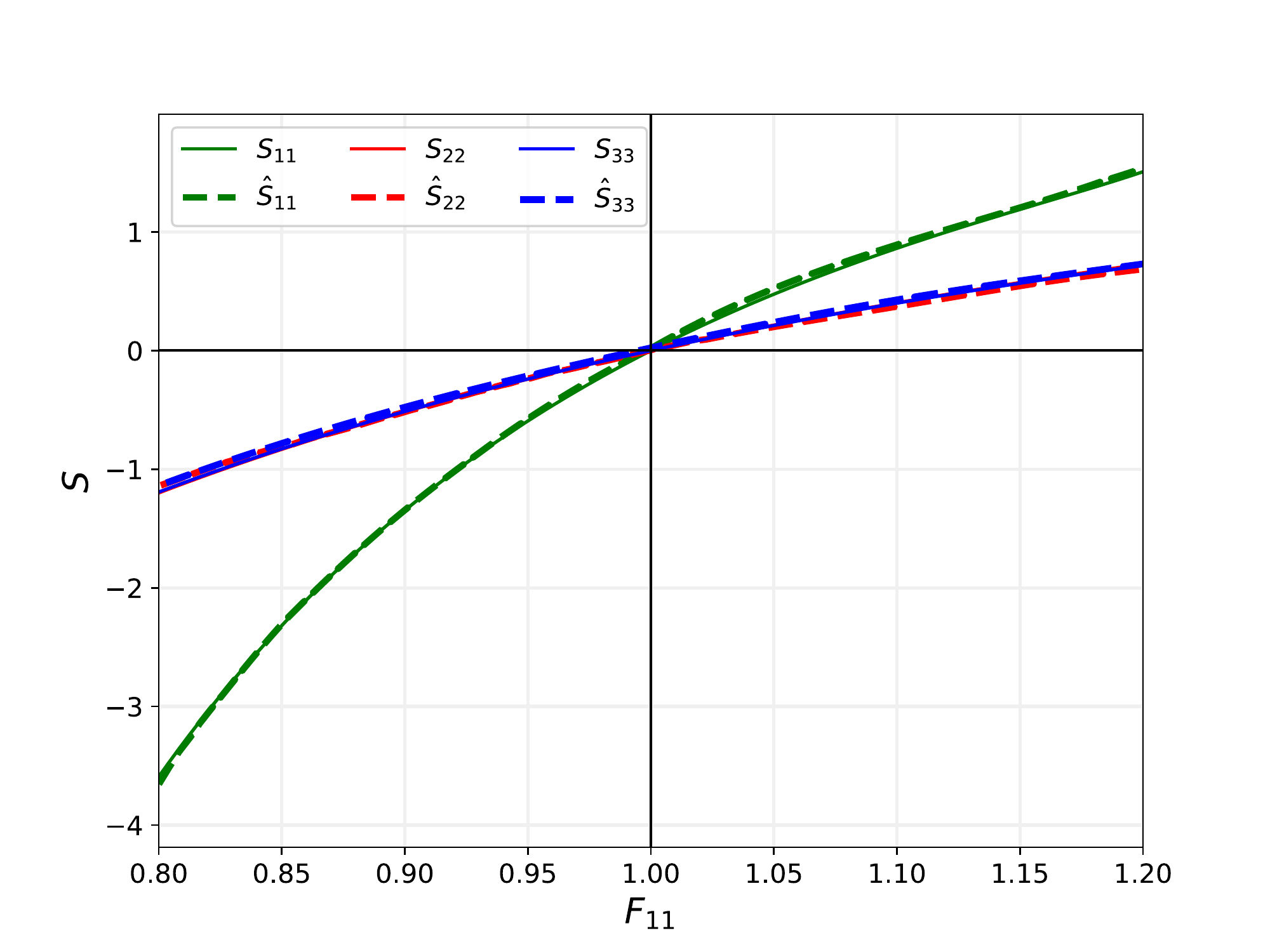}
\caption{Stress comparison}\label{fig:Stress_over_Fiber}
\end{subfigure}
\end{center}
\caption{Results for the \textbf{fiber anisotropy} material model of \sref{sec::ModelVali} . (a) Training loss over epochs, (b) anisotropic coefficients over epochs, (c-d) approximated preferred directions, (e) true (solid lines) and predicted stresses (dashed lines) over $F_{11}$.}
\label{fig::FibreResults}
\end{figure}

First, we discuss the results for data from the fiber anisotropy model presented in \sref{sec::ModelVali}.
For this case, the loss is plotted over the number of epochs \fref{fig:loss_over_Fiber}.
We can see that the training procedure is able to significantly decrease the initial loss.
The evolution of the anisotropic coefficients is shown in \fref{fig:AniCoeffs_over_Fiber}.
Interestingly, it appears that the trained model predicts the stress-strain data to have a structure that closely resembles that of an orthotropic material, i.e. both coefficients are decidedly non-zero.
This makes sense given the symmetries of the fiber distribution plotted in \fref{fig:rho}.
\Fref{fig:n1Fiber} and \ref{fig:n2Fiber} show the convergence of the two associated preferred directions of this type of anisotropy.
Lastly, the ground truth and predicted diagonal stress components of the second Piola-Kirchhoff stress as a function of $F_{11}$ are plotted in  \fref{fig:Stress_over_Fiber}.
As with the previous cases, it can be seen that the trained surrogate is able to accurately recreate the correct stress-strain relationship even on data that are not in the training set.

\begin{figure}
\begin{subfigure}[b]{0.5\linewidth}
\includegraphics[scale=0.3]{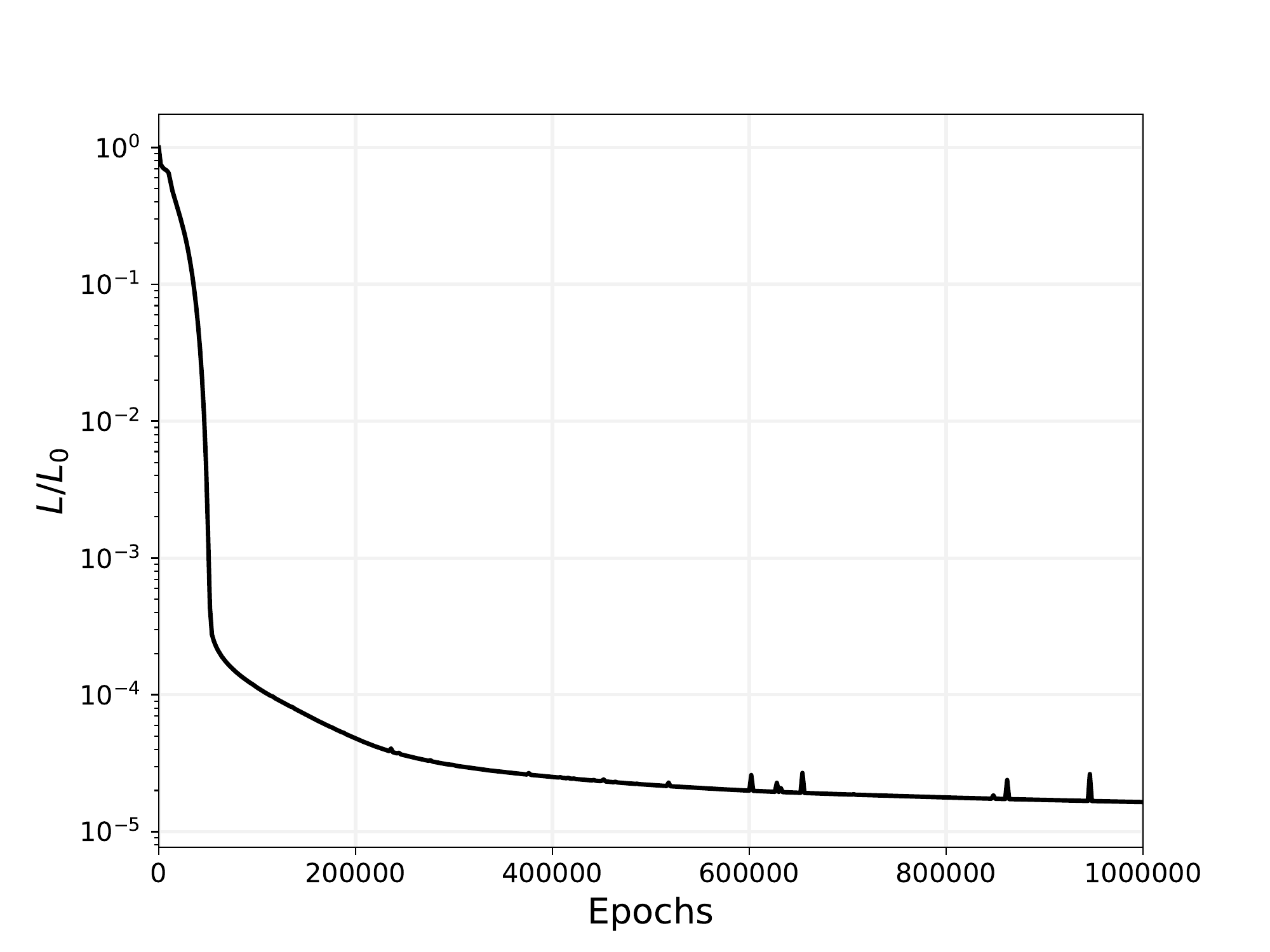}
\caption{Loss}\label{fig:loss_over_RVE}
\end{subfigure}
\begin{subfigure}[b]{0.5\linewidth}
\includegraphics[scale=0.3]{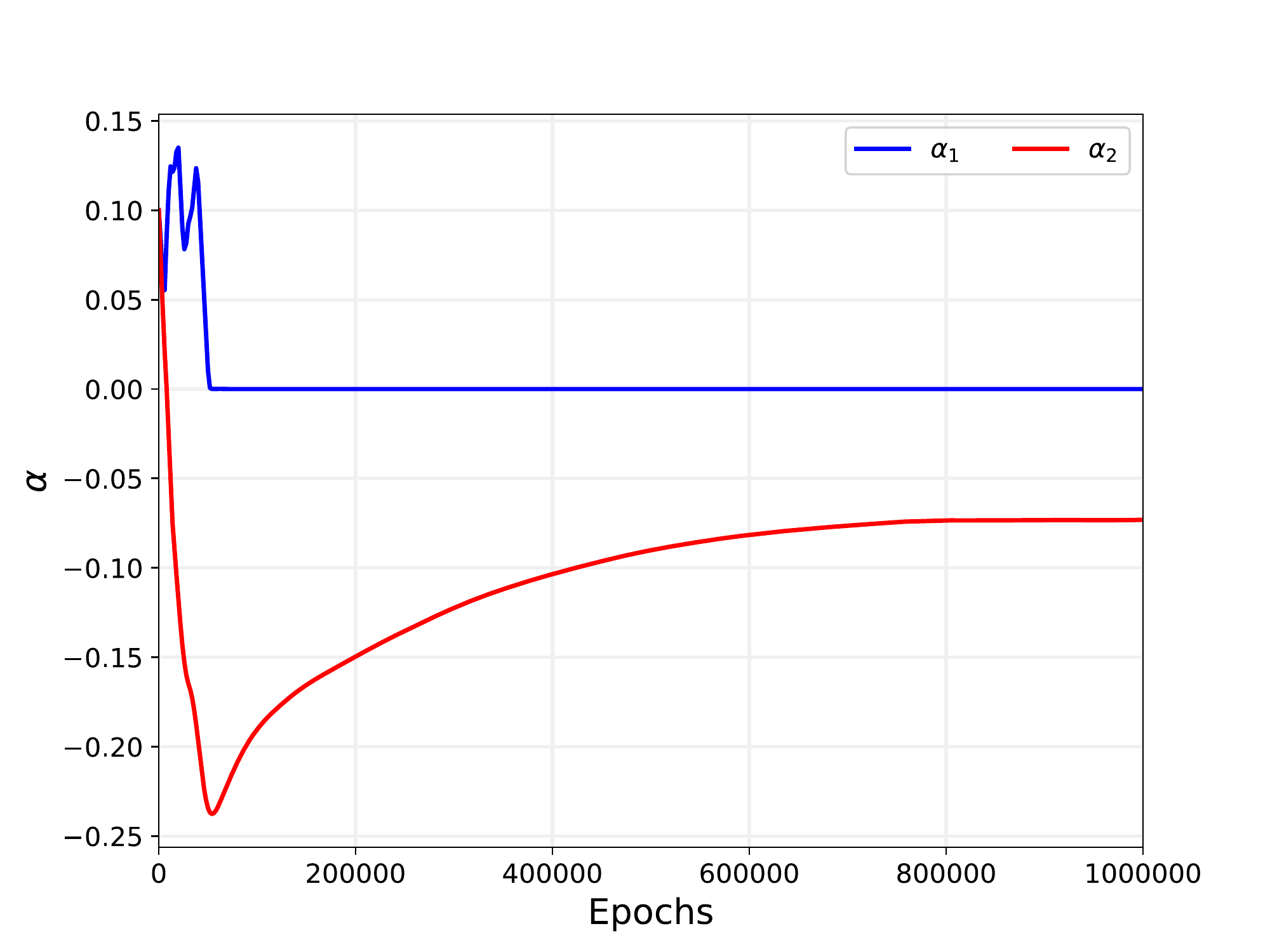}
\caption{Anisotropic coefficients}\label{fig:AniCoeffs_over_RVE}
\end{subfigure}
\begin{subfigure}[b]{0.5\linewidth}
\includegraphics[scale=0.3]{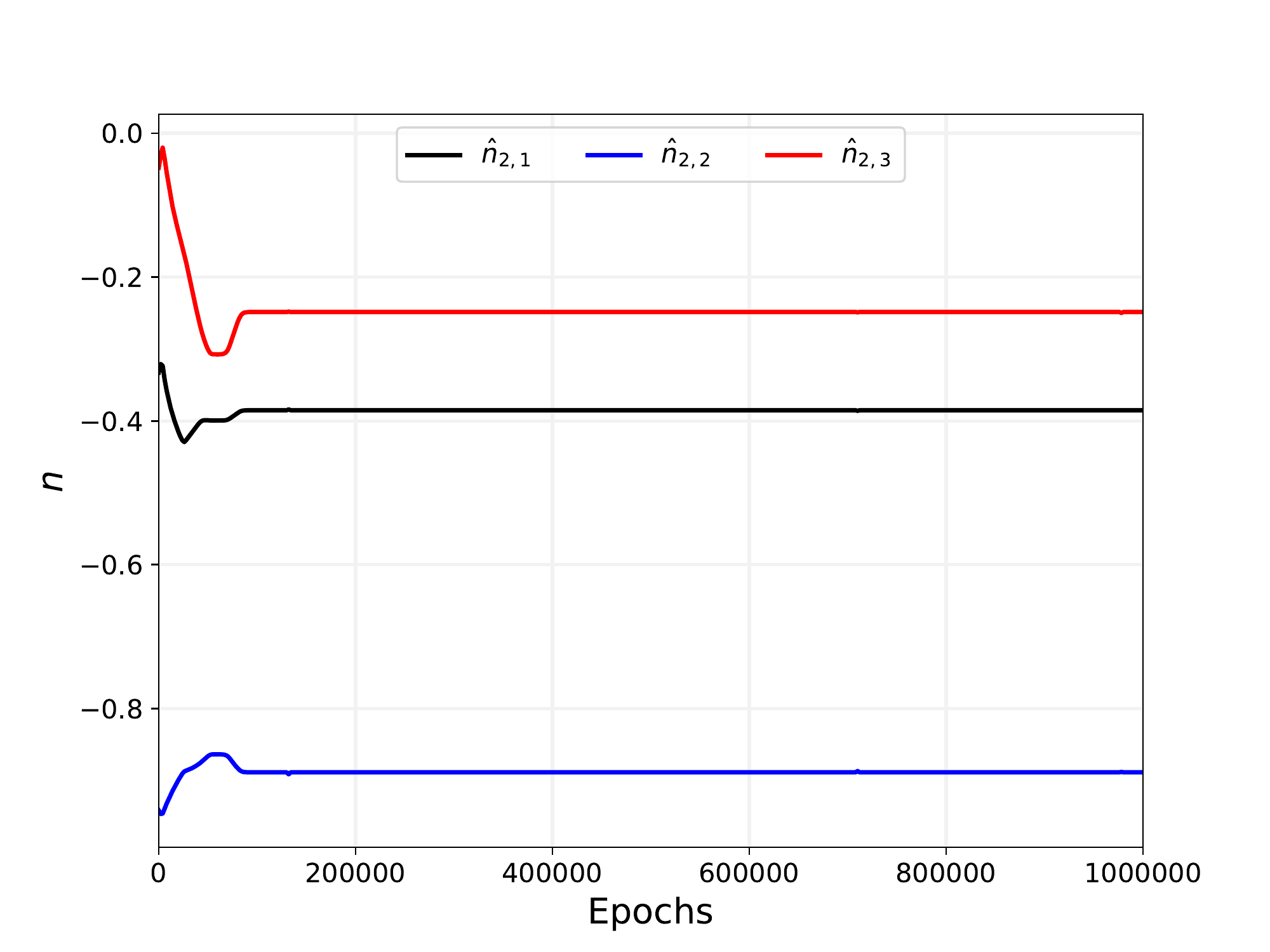}
\caption{Preferred direction}\label{fig:PrefDir_over_RVE}
\end{subfigure}
\begin{subfigure}[b]{0.5\linewidth}
\includegraphics[scale=0.3]{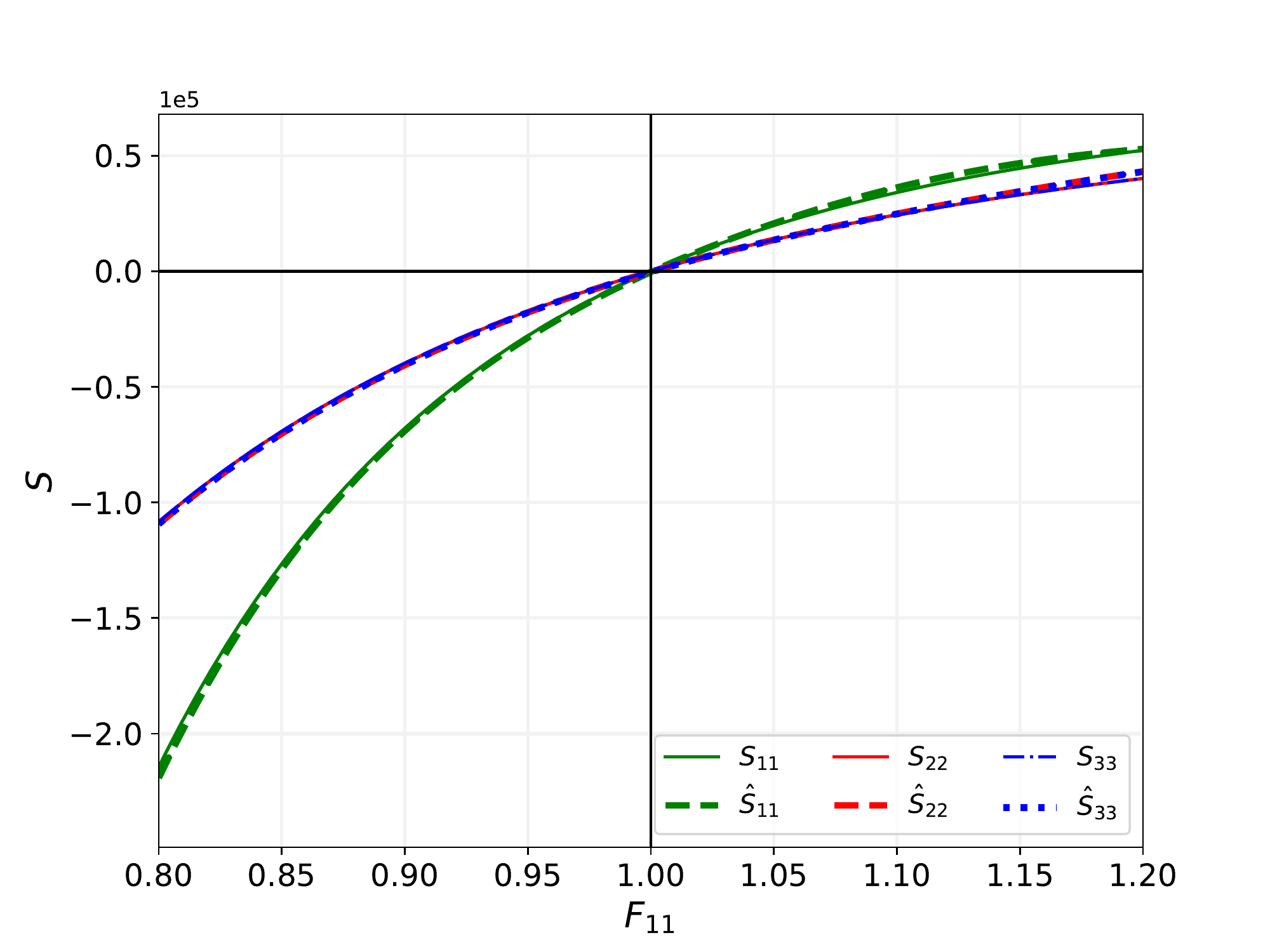}
\caption{Stress comparison}\label{fig:Stress_over_RVE}
\end{subfigure}
\caption{Results for the effective material model of the \textbf{microstructure with inclusions}, c.f. \sref{sec::ModelVali}. (a) Training loss over epochs, (b) anisotropic coefficients over epochs, (c) approximated preferred direction, (d) true (solid lines) and predicted stresses (dashed lines) over $F_{11}$.
}\label{fig::RVEResults}
\end{figure}

The final example involves the microstructure problem discussed in \sref{sec::ModelVali}.
The loss over the training process is shown in \fref{fig:loss_over_RVE} where we can see that the initial loss is significantly reduced.
\Fref{fig:AniCoeffs_over_RVE} plot the evolution of the two anisotropic coefficients.
The proposed framework identifies a representation with one preferred direction as the best fit.
This preferred direction is plotted in \fref{fig:PrefDir_over_RVE}.
In this case it is difficult to get an expectation of what the symmetries should be from the microstructure but we are are assured by the high accuracy the TBNN model achieves.
The stress response of the trained surrogate is compared to the ground truth test case results obtained from FE simulations in \fref{fig:Stress_over_RVE}.

\begin{figure}
\begin{subfigure}[b]{0.5\linewidth}
\includegraphics[scale=0.9]{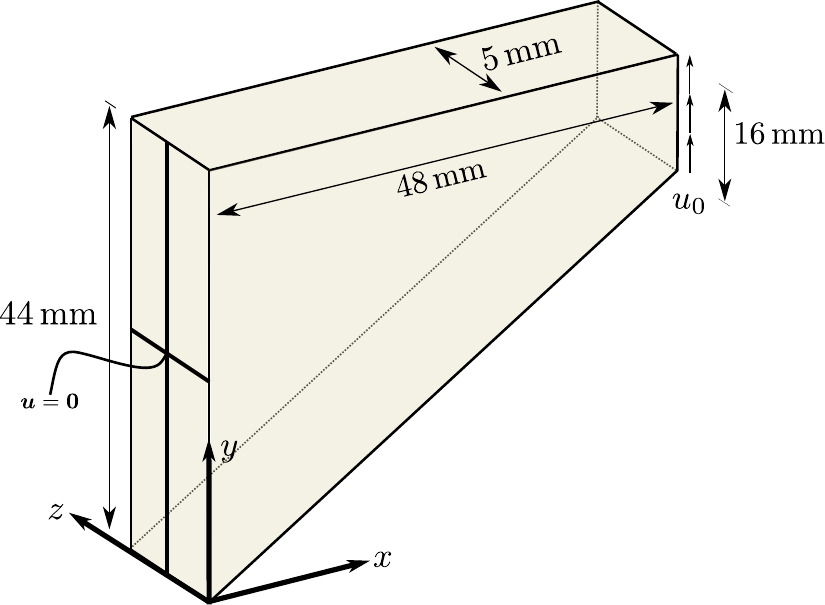}
\caption{Schematic}\label{fig:CooksMem}
\end{subfigure}
\begin{subfigure}[b]{0.5\linewidth}
\includegraphics[scale=0.22]{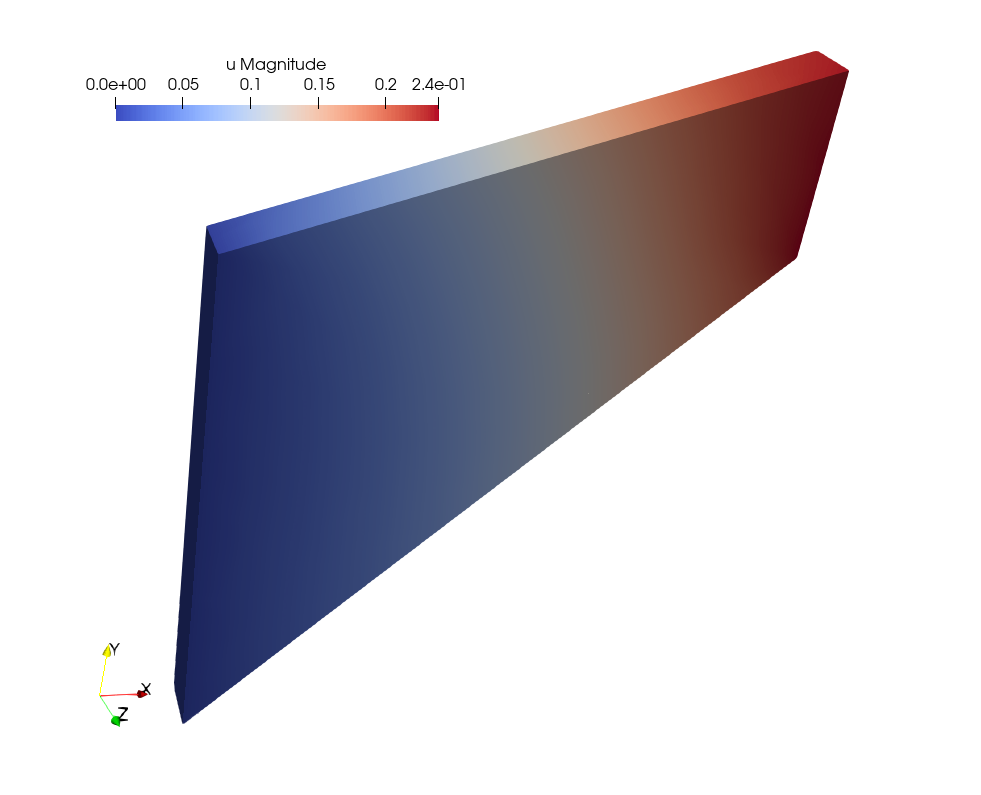}
\caption{Displacement}\label{fig:outputRVE}
\end{subfigure}
\caption{Embedding TBNN surrogates trained using effective responses of \textbf{microstructure with inclusions}, c.f. \sref{sec::ModelVali} into a Finite Element framework. (a) Structural problem: Cook's membrane, (b) displacement magnitude using the fitted surrogate as the material model.}\label{fig::RVEFEM}
\end{figure}

Lastly, we test the trained TBNN surrogate model by embedding it into the C++ FE code provided by  \cref{yaghoobi2019prisms}.
As a commonly applied benchmark for hyperelastic response \cite{schroder2021selection}, we employ the so-called Cook’s membrane problem, illustrated in \fref{fig:CooksMem}, where a hexahedral structural problem is pinned on the left and a vertical displacement of $u_{0} = 0.12 m$ is applied.
The Cook’s membrane problem is typically used as a test case for combined bending and shear response with moderate distortion.
As a proof-of-concept we show the magnitude of the displacement response in \fref{fig:outputRVE} which is in accordance with the solution obtained in \cref{fuhg2022local}.

\section{Conclusion} \label{sec:conclusion}

We proposed a tensor-basis neural network framework based on classical representation theorems and machine learning that can discover both the orientation and the type of anisotropy purely from stress-strain datasets, and, as a byproduct, provide a surrogate for the stress-strain response.
We have applied this approach to recover the exact symmetries corresponding to phenomenological constitutive laws that we used as test cases as well as the best fits to data from materials with explicit microstructure.
Expanding the anisotropic basis from the orthotropic set assumed in this work may yield benefits in certain cases, e.g. crystal symmetries \cite{parry1976elasticity,green1960large}; however, it appears that cross term basis generators, such as $\sym \nb_1 \otimes \nb_2$ were not needed for the wide range of material responses we explored.
Furthermore, the viability of the trained TBNN surrogate as a usable constitutive law was demonstrated by solving a commonly employed structural test problem with a FE code.
In additional results described in \aref{append:ConverStud}, we found that the method provides robust inference of symmetries over a wide range of dataset sizes.
We also developed an extension of the proposed TBNN architecture to generate polyconvex potentials,  which is discussed in \aref{append:InputConvex} and could prove useful in cases where data availability is an issue since the polyconvexity provides another constraint on the form of the representation.

In future work we will pursue the application of the method to experimental data, and tackle issues of measurement noise as well as limited control and observation of the stress response.
In this endeavor we envision the use of active learning techniques to guide experiments to rapidly discover anisotropies from measurements, c.f.  \cref{fuhg2021state,rocha2021fly}.
For more complex materials we may need to resort to global optimization techniques, such as simulated annealing and multi-start initialization, to overcome issues with multiple minima in the objective function.
We may also explore alternative methods to induce sparsity, such as Orthogonal Matching Pursuit \cite{tropp2007signal,pati1993orthogonal}.

\section*{Acknowledgments}

N. Bouklas gratefully acknowledges support by the Air Force Office of Scientific Research under award number FA9550-22-1-0075.
Sandia National Laboratories is a multimission laboratory managed and operated by National Technology and Engineering Solutions of Sandia, LLC., a wholly owned subsidiary of Honeywell International, Inc., for the U.S. Department of Energy's National Nuclear Security Administration under contract DE-NA0003525.
The views expressed in the article do not necessarily represent the views of the U.S. Department of Energy or the United States Government.



\appendix
\numberwithin{equation}{section}
\section{Tensor calculus identities} \label{app:ids}

A number of fundamental tensor calculus identities are useful for this work:
\begin{eqnarray}
\partial_\Cb \tr \Cb^n &=& n \left( \Cb^{n-1} \right)^T \\
\partial_\Cb \left(\tr \Cb\right)^n &=& n \left( \tr \Cb^{n-1} \right) \Ib \\
\partial_\Cb \det(\Cb) &=&  \det(\Cb) \, \Cb^{-T}
\end{eqnarray}
Derivatives of invariants follow these identities.
For the principal isotropic invariants:
\begin{eqnarray}
I_1 = \tr \Cb &&
\partial_\Cb I_1 =  \Ib \\
I_2 = 1/2 ( \tr^2 \Cb - \tr \Cb^2 ) &&
\partial_\Cb I_2=  \tr(\Cb) \Ib - \Cb \\
I_3 = \det \Cb  &&
\partial_\Cb I_3 =  I_3 \Cb^{-T}  \\
J = \det \Cb^{1/2} &&
\partial_\Cb J  
= \frac{J}{2} \Cb^{-T} \\
&&
\partial_\Cb \log J = \frac{1}{2} \Cb^{-T} \ ,
\end{eqnarray}
the rescaled invariants:
\begin{eqnarray}
\bar{I}_1 = I_1/J^{2/3} &&
\partial_\Cb \bar{I}_1
= J^{-2/3} \Ib - \frac{1}{3} \bar{I}_1 \Cb^{-T} \\
\bar{I}_2 = I_2/J^{4/3} &&
\partial_\Cb \bar{I}_2 =
J^{-4/3} ( \tr(\Cb) \Ib - \Cb )   - \frac{2}{3} \bar{I}_2 \Cb^{-T} \ ,
\end{eqnarray}
and the anisotropic invariants:
\begin{eqnarray}
I_4  = \tr \Cb \Nb &&
\partial_\Cb I_4 =  \Nb \\
I_5  = \tr \Cb^2 \Nb &&
\partial_\Cb I_5 =  \Nb \Cb^T + \Cb \Nb
\end{eqnarray}

\section{Influence of the number of training points}\label{append:ConverStud}

The results of \sref{sec:results} were obtained using $N=$2,500 sample points. In this section we investigate the influence of $N$ on the ability of the proposed framework to recover the anisotropy structure and orientation of the material models discussed in the context of {\it{verification}} in  \sref{sec::ModelVeri}.

Using the sampling method described in  \sref{sec::Sampling}, $N=$100, 1,000, and 10,000 stress-strain sample datasets are obtained for each of the three constitutive laws.
All hyperparameter, solver and initialization choices are equivalent to the ones discussed in  \sref{sec:method}.
\footnote{
For the low data cases, e.g. $N=$100, the 3$\times$30 neural network has more trainable parameters than data points that are available, i.e. we have an underdetermined problem.
However, neural networks have empirically been shown to train well and be surprisingly resistant to overfitting even in the small data regime \cite{olson2018modern,salman2019overfitting}. This is also observed in this work.}

\begin{figure}
\begin{subfigure}[b]{0.5\linewidth}
\includegraphics[scale=0.3]{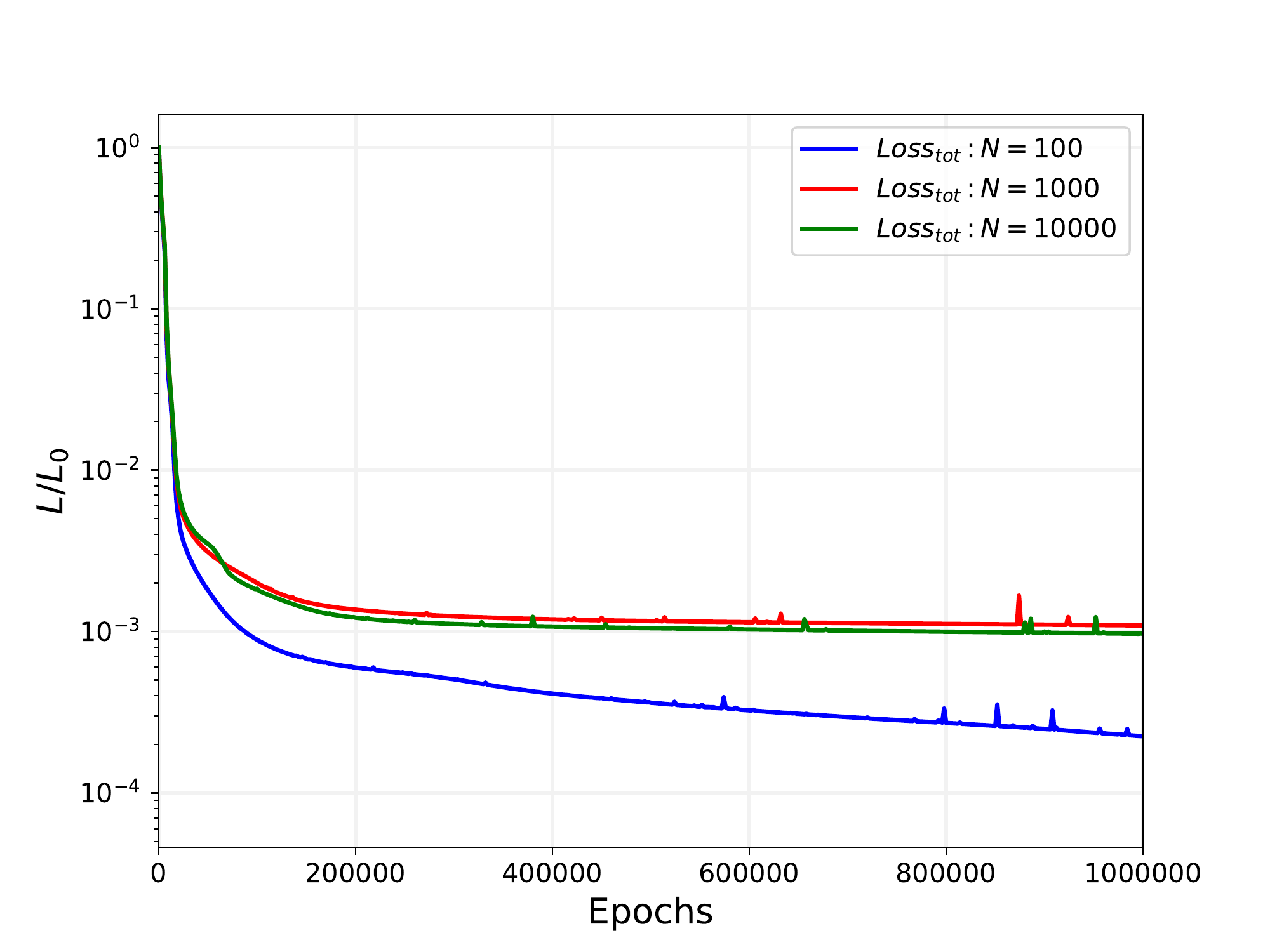}
\caption{Loss}\label{fig:ConvIsoA}
\end{subfigure}
\begin{subfigure}[b]{0.5\linewidth}
\includegraphics[scale=0.3]{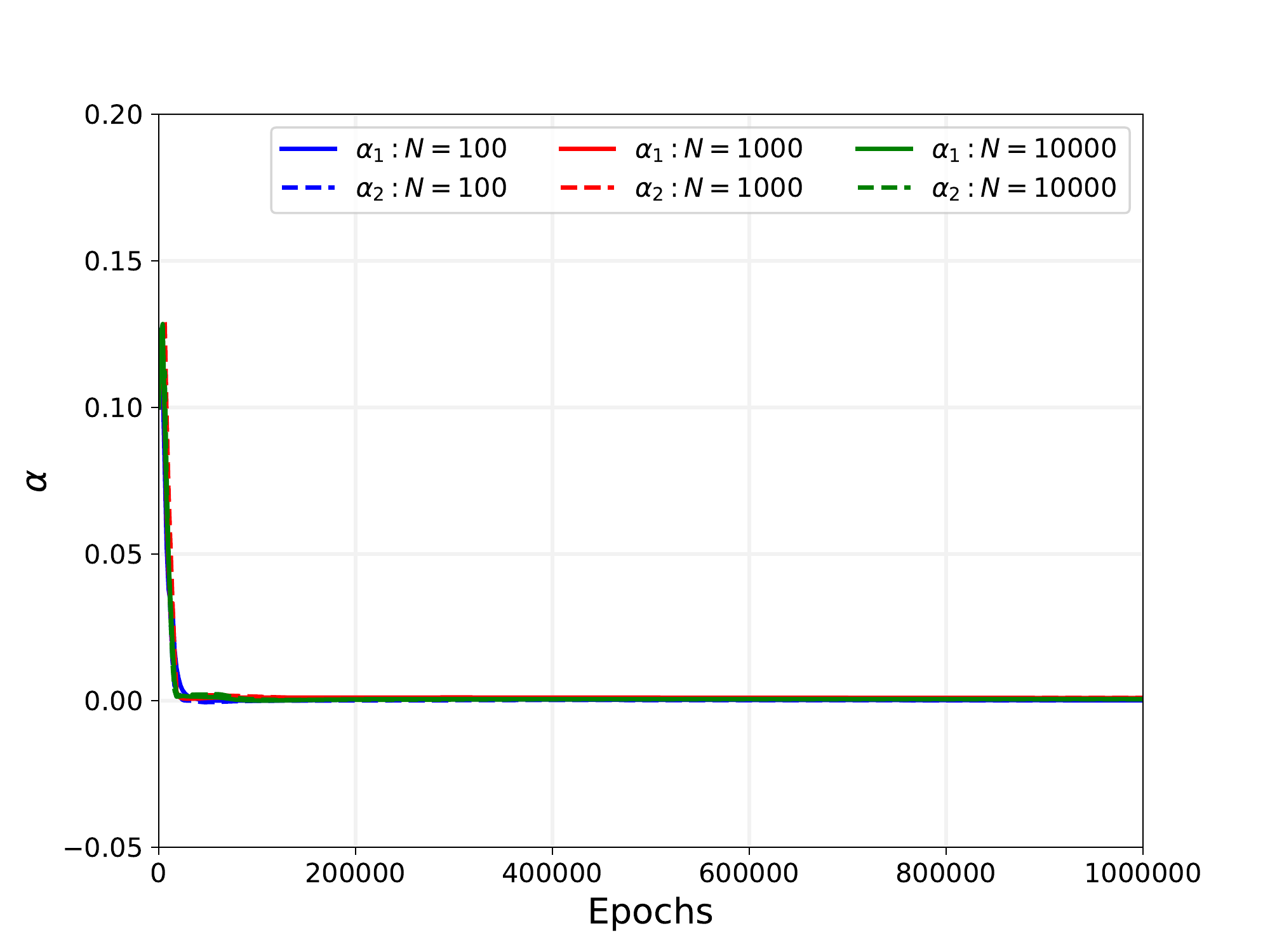}
\caption{Anisotropic coefficients}\label{fig:ConvIsoB}
\end{subfigure}
\caption{Results for the \textbf{isotropic} material model of  \sref{sec::ModelVeri} for three different training dataset sizes. (a) Training loss over epochs, (b) anisotropic coefficients over the training process.}\label{fig::ConvIso}
\end{figure}

We first consider the isotropic material model discussed in \sref{sec::ModelVeri}. For this case we expect the anisotropic coefficients  ($\alpha_{1}, \alpha_{2}$)  to go toward zero.
\Fref{fig::ConvIso} plots the training loss as well as the values of the coefficients over the training process for all three dataset cases. It can be seen that the trained model is able to significantly reduce the initial training error. Furthermore the two $\alpha$-values are zero after training for all three cases, indicating that the proposed approach is able to accurately identify that the data was generated from an isotropic material model even when only $100$ training points are available.
Note the fact that the loss is continuing to decrease for the low data case, $N=100$, provides some indication that the model might be overfitting; however, for the purposes of discovering the anisotropy, i.e. the inverse problem, this is essentially irrelevant.

\begin{figure}
\begin{subfigure}[b]{0.5\linewidth}
\includegraphics[scale=0.3]{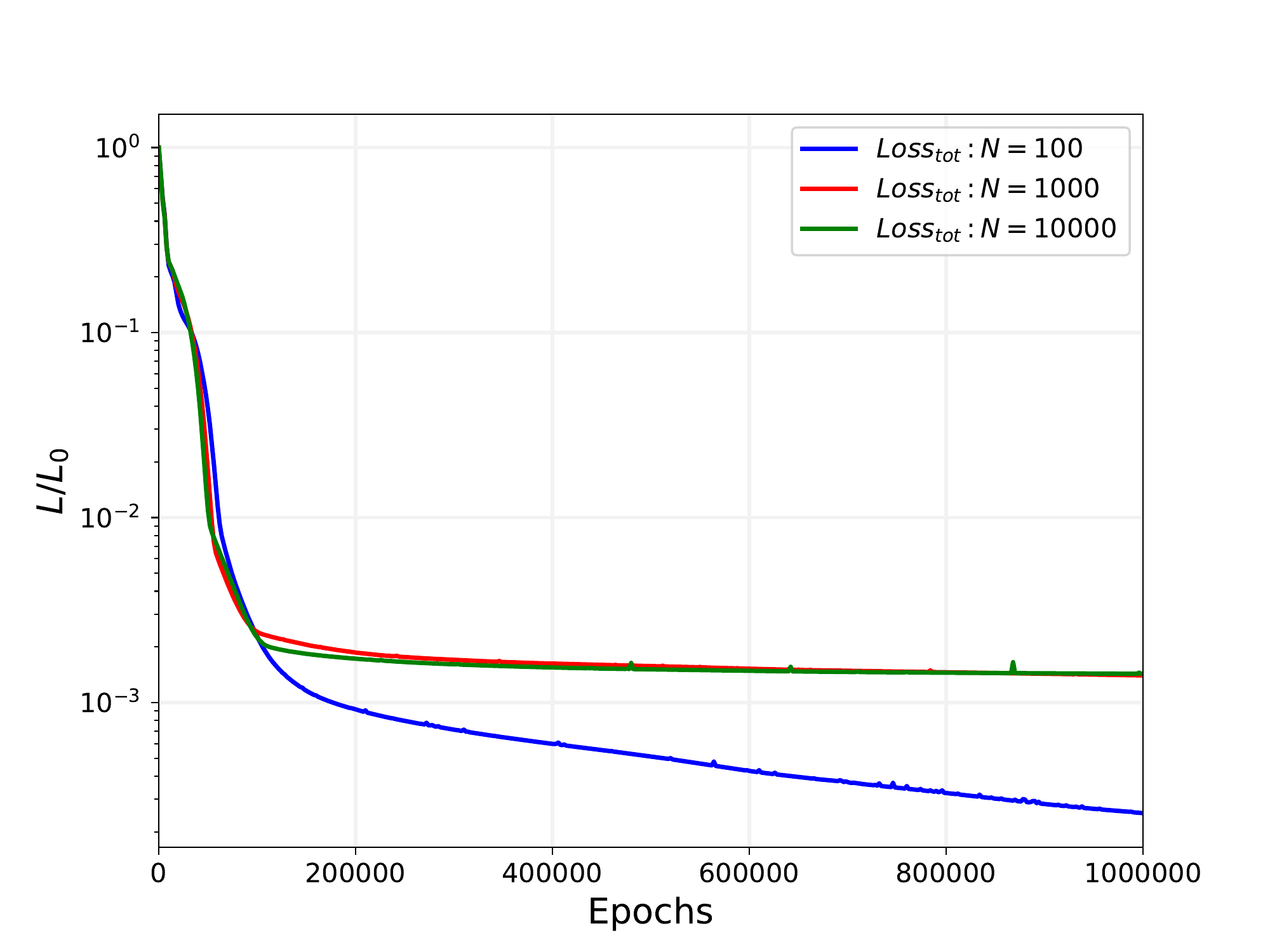}
\caption{Loss} 
\end{subfigure}
\begin{subfigure}[b]{0.5\linewidth}
\includegraphics[scale=0.3]{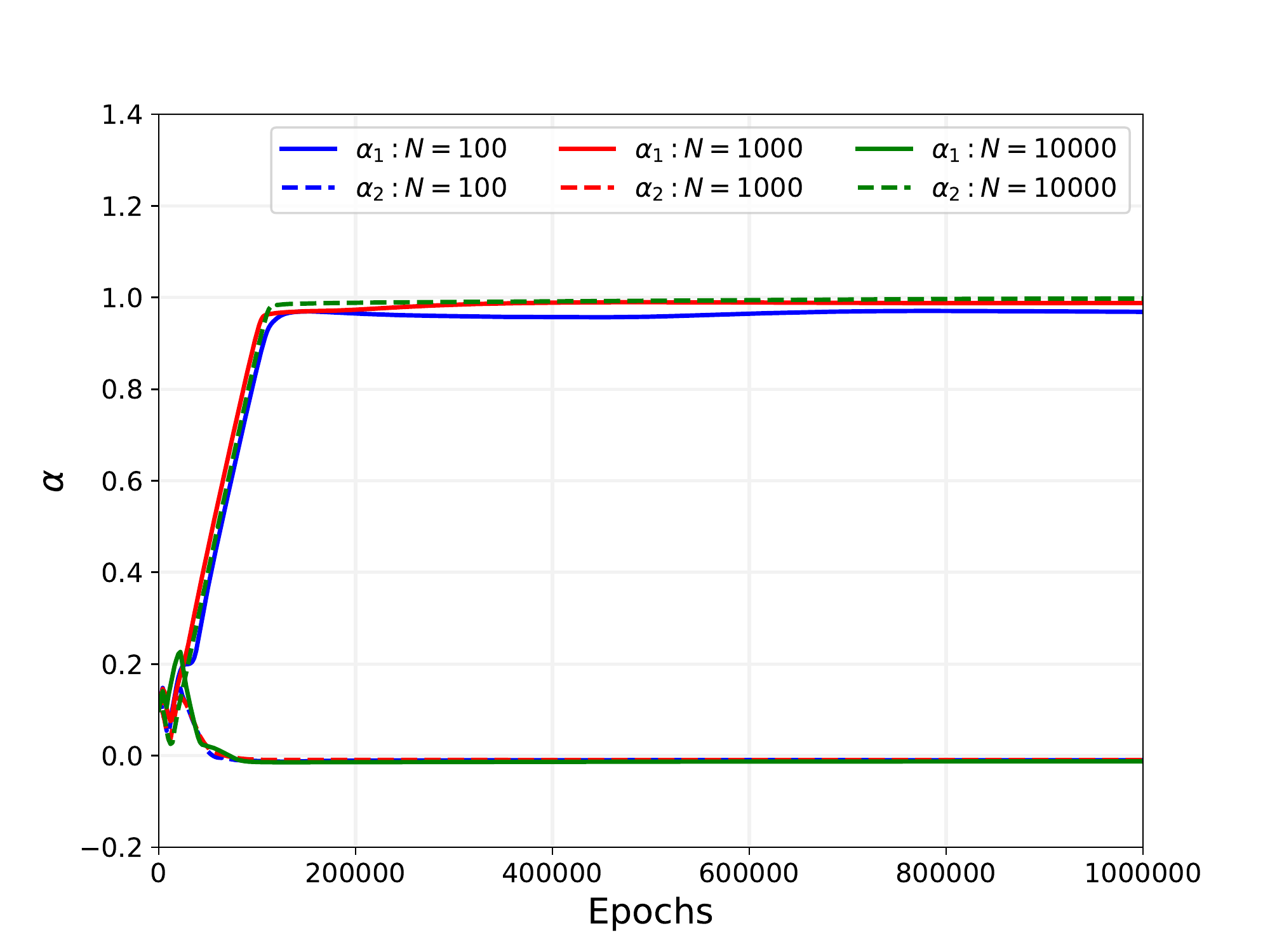}
\caption{Anisotropic coefficients} 
\end{subfigure}
\caption{Results for the \textbf{transversely isotropic} material model of \sref{sec::ModelVeri} for three different training dataset sizes. (a) Training loss over epochs, (b) anisotropic coefficients over the training process.}\label{fig::ConvTransLossAni}
\end{figure}

\begin{figure}
\begin{subfigure}[b]{0.5\linewidth}
\includegraphics[scale=0.3]{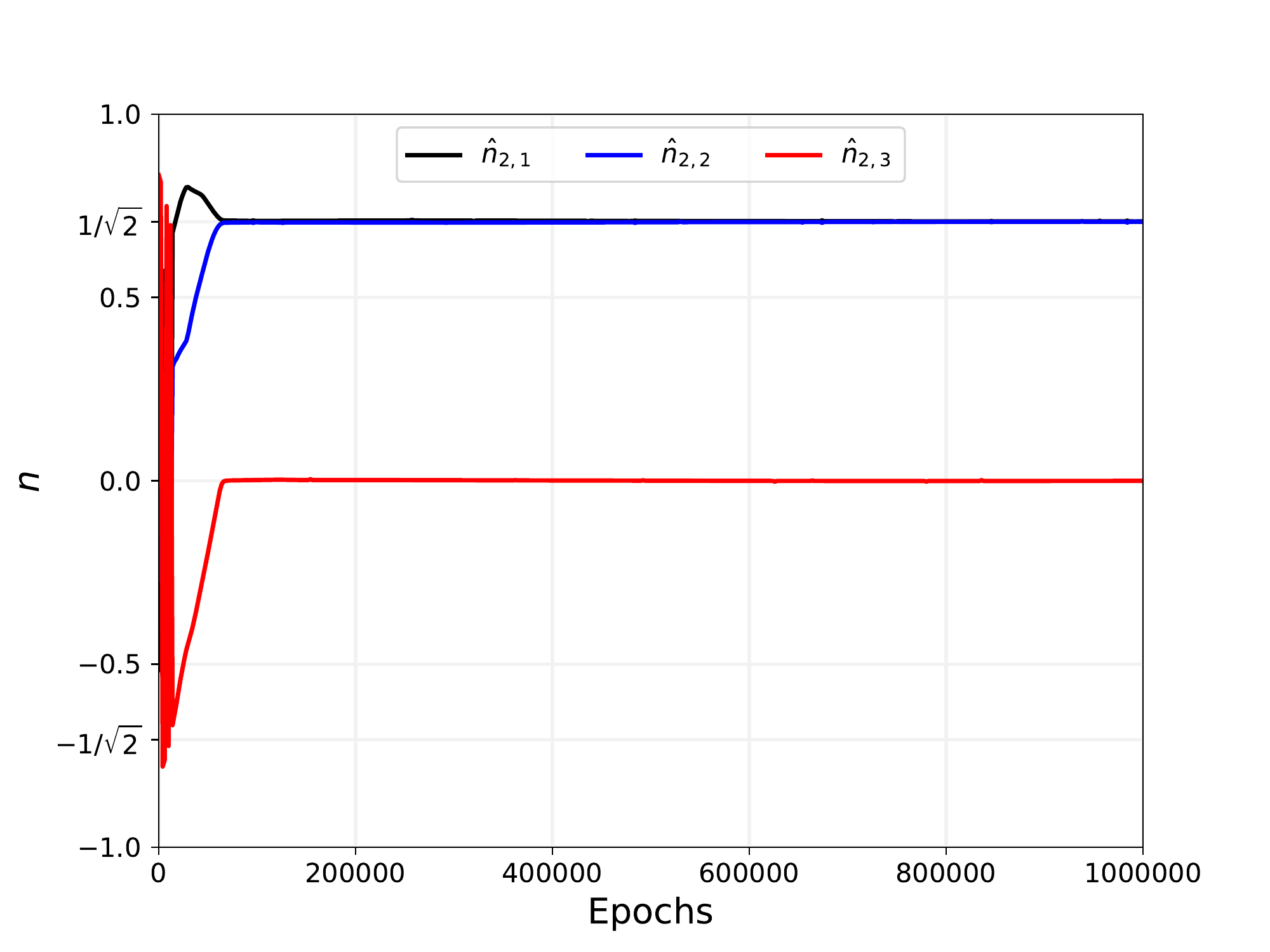}
\caption{$\nb_{1}$ for $N = 100$} 
\end{subfigure}
\begin{subfigure}[b]{0.5\linewidth}
\includegraphics[scale=0.3]{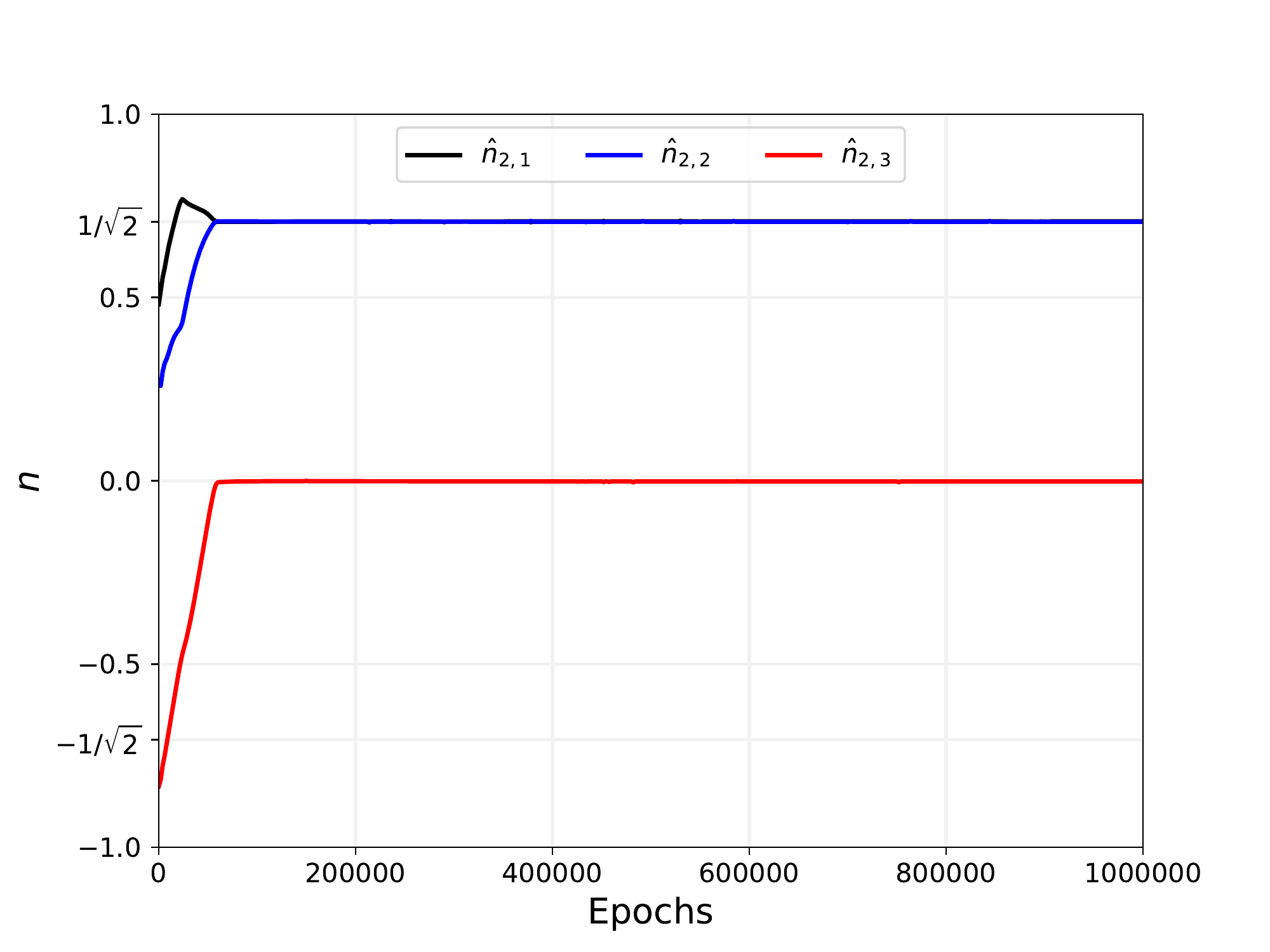}
\caption{$\nb_{1}$ for $N = 1000$} 
\end{subfigure}
\begin{center}
\begin{subfigure}[b]{0.5\linewidth}
\includegraphics[scale=0.3]{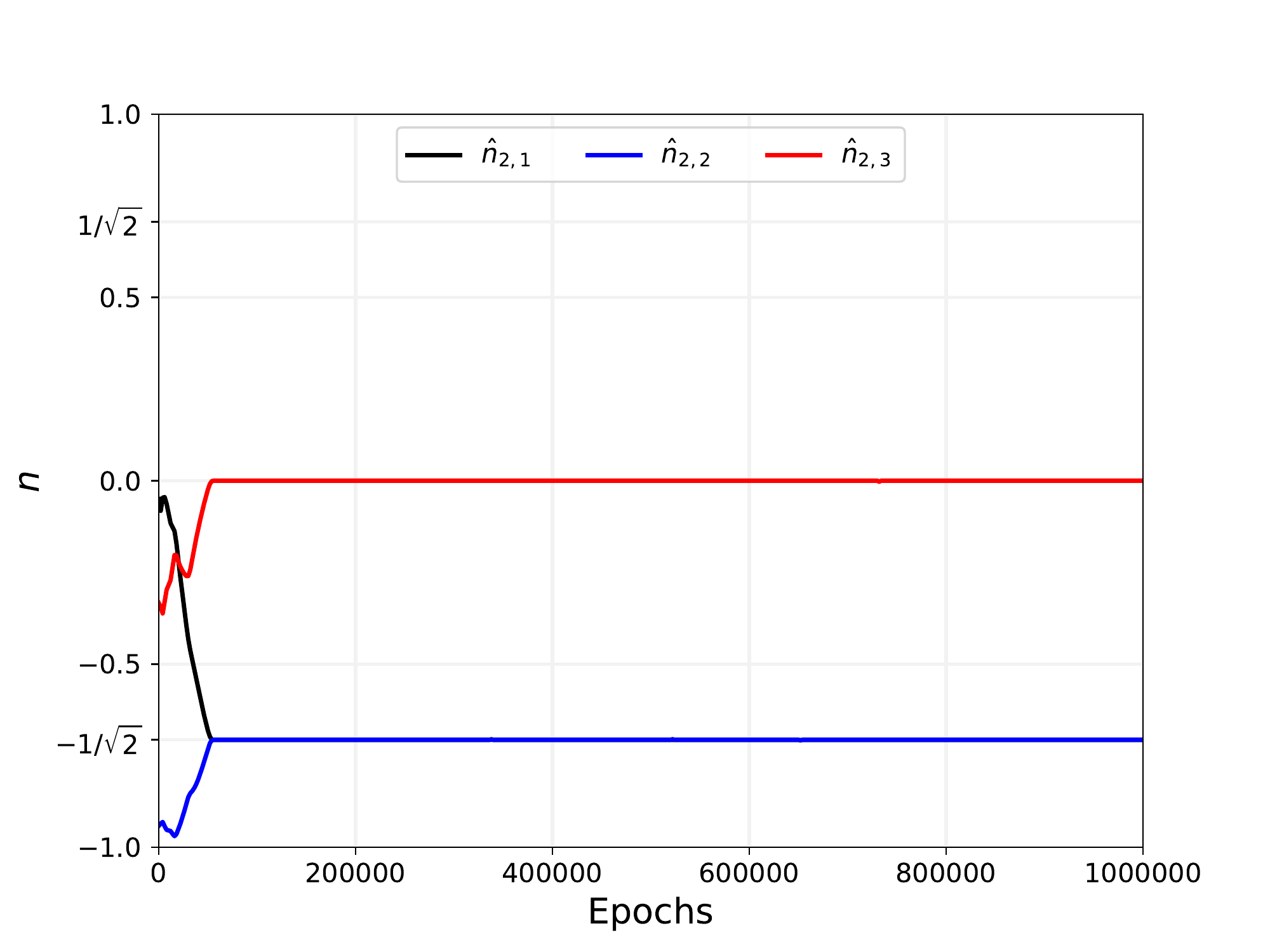}
\caption{$\nb_{2}$ for $N = 10000$} 
\end{subfigure}
\end{center}
\caption{Recovered preferred directions where $\alpha \neq 0$ for the \textbf{transversely isotropic} material model of \sref{sec::ModelVeri} for three different training dataset sizes. Ground truth: $\nb = (1/\sqrt{2}, 1/\sqrt{2}, 0)$. (a) $N=100$, (b) $N=1,000$, (c) $N=10,000$.}\label{fig::ConvTransPrefDire}
\end{figure}

Next, we study the influence of the training dataset size for the transversely isotropic hyperelastic law presented in \sref{sec::ModelVeri}. In contrast to the first example we expect one anisotropic coefficient to be non-zero and the corresponding preferred direction to be equivalent to a factor of $\nb = (1/\sqrt{2}, 1/\sqrt{2}, 0)$.
For this example, the evolution of the training loss and the coefficients are shown \Fref{fig::ConvTransLossAni}.
We can see that the proposed framework is able to recover the correct form of anisotropy (only one non-zero coefficient) for the small, medium and large datasets.
Additionally, the respective preferred directions are accurately fit at the end of the training process as seen in \fref{fig::ConvTransPrefDire}.
We can see again that even with only $100$ training points the type of the anisotropy and the preferred orientations are correctly recovered.

\begin{figure}
\begin{subfigure}[b]{0.5\linewidth}
\includegraphics[scale=0.3]{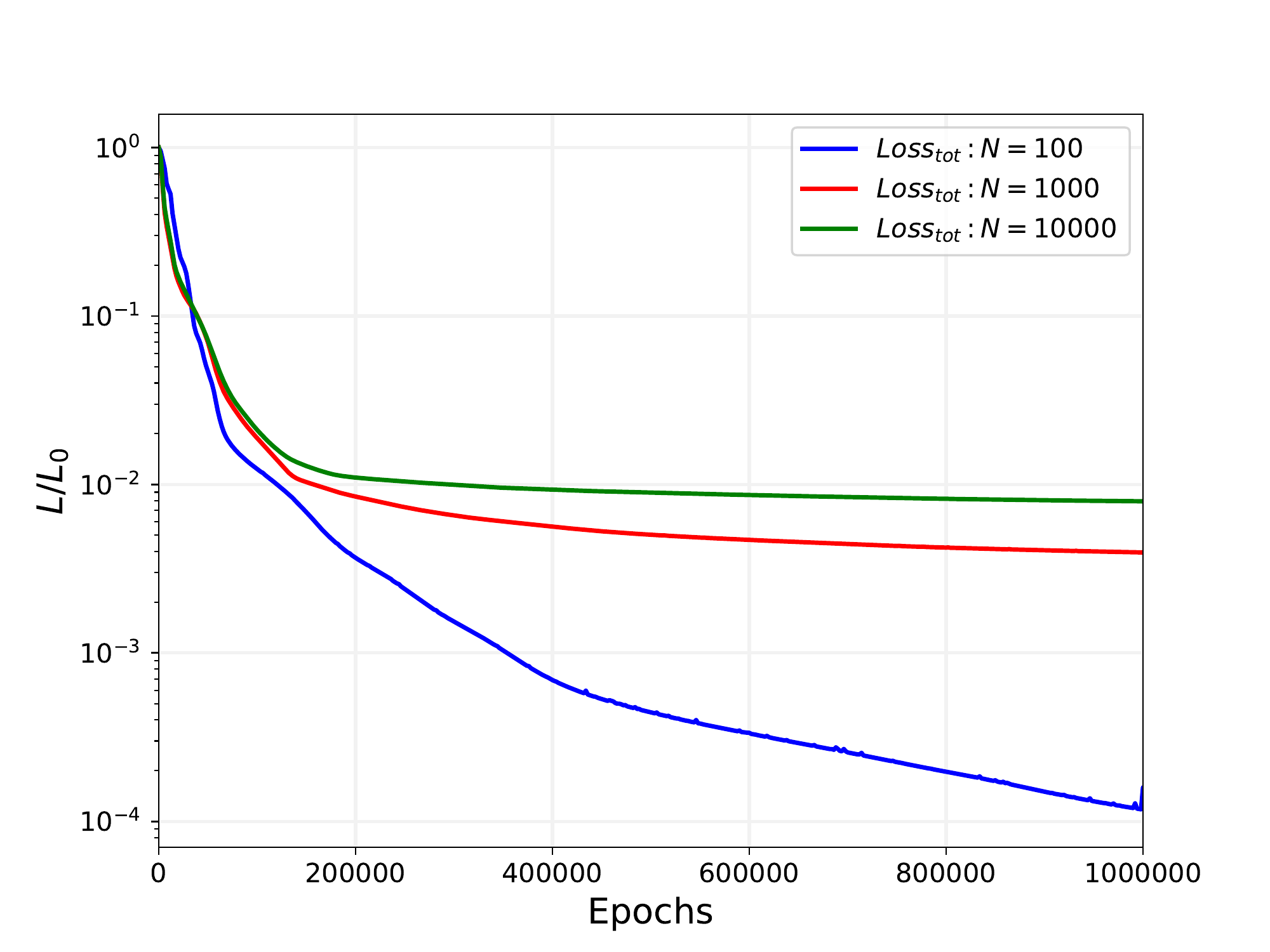}
\caption{Loss} 
\end{subfigure}
\begin{subfigure}[b]{0.5\linewidth}
\includegraphics[scale=0.3]{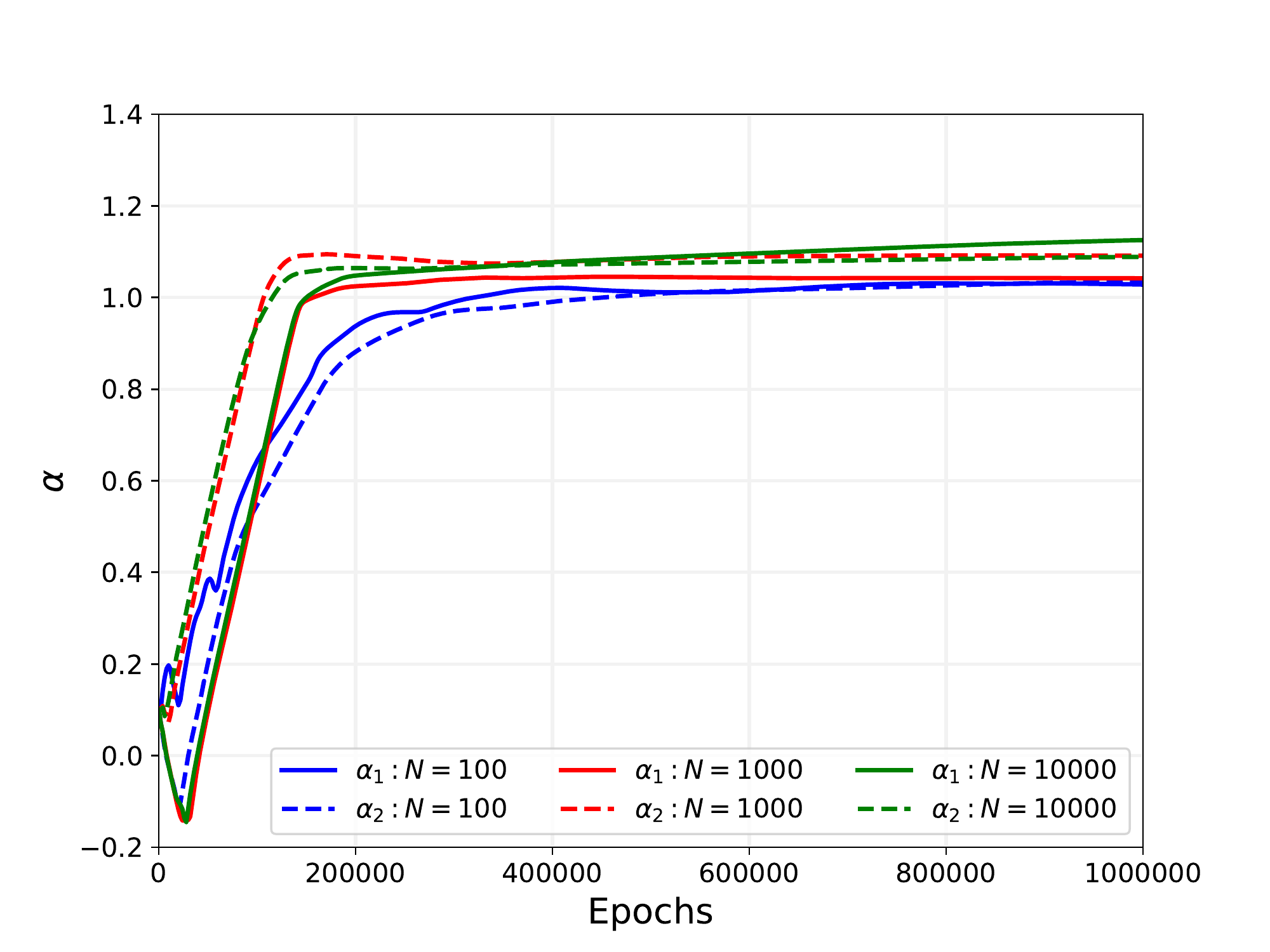}
\caption{Anisotropic coefficients}\label{fig:ConvOrthoLossAniB}
\end{subfigure}
\caption{Results for the \textbf{orthotropic} material model of \sref{sec::ModelVeri}  for three different training dataset sizes. (a) Training loss over epochs, (b) anisotropic coefficients over the training process.}\label{fig::ConvOrthoLossAni}
\end{figure}

\begin{figure}
\begin{subfigure}[b]{0.5\linewidth}
\includegraphics[scale=0.3]{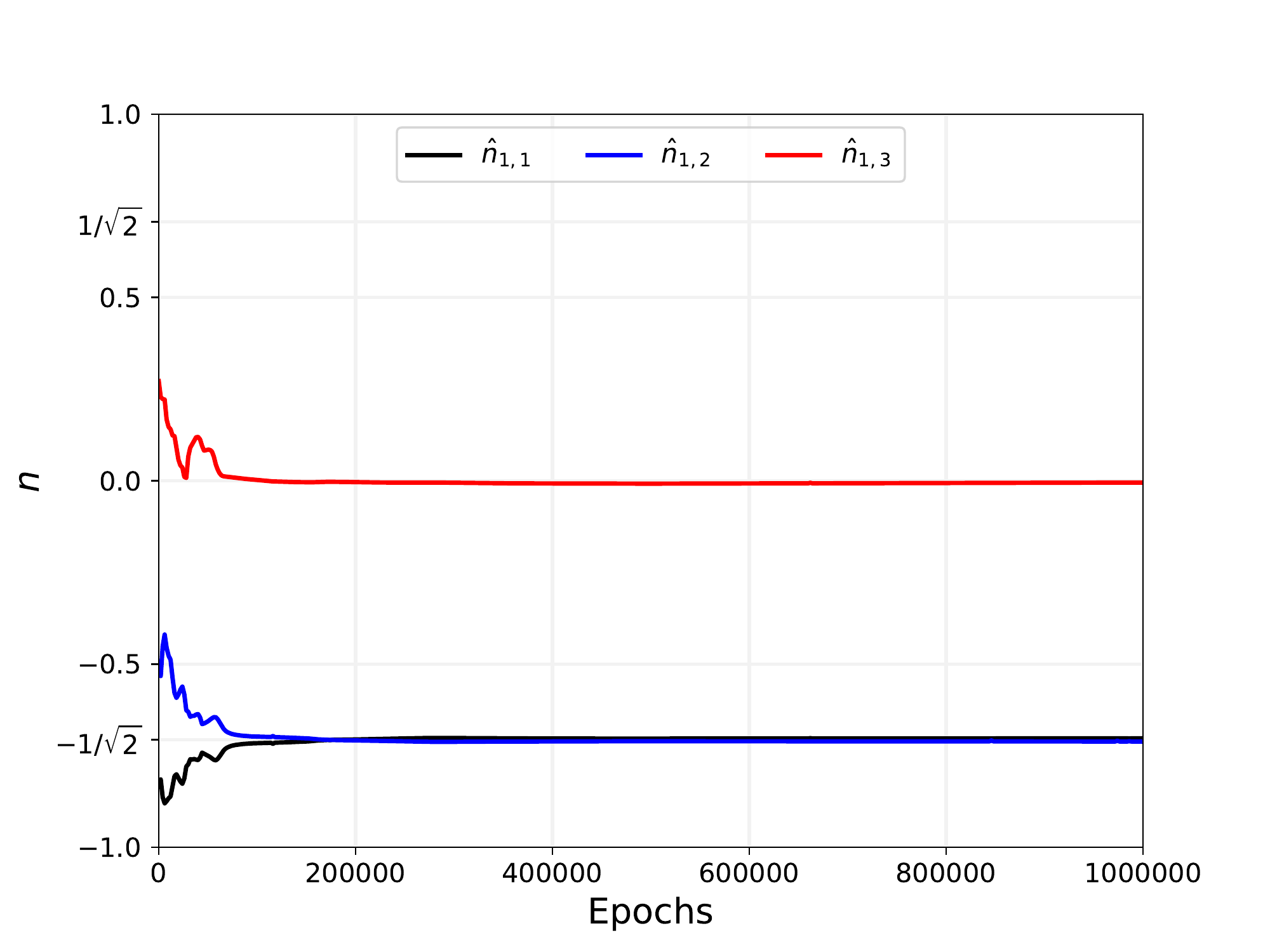}
\caption{$\nb_{1}$ for $N = 100$} 
\end{subfigure}
\begin{subfigure}[b]{0.5\linewidth}
\includegraphics[scale=0.3]{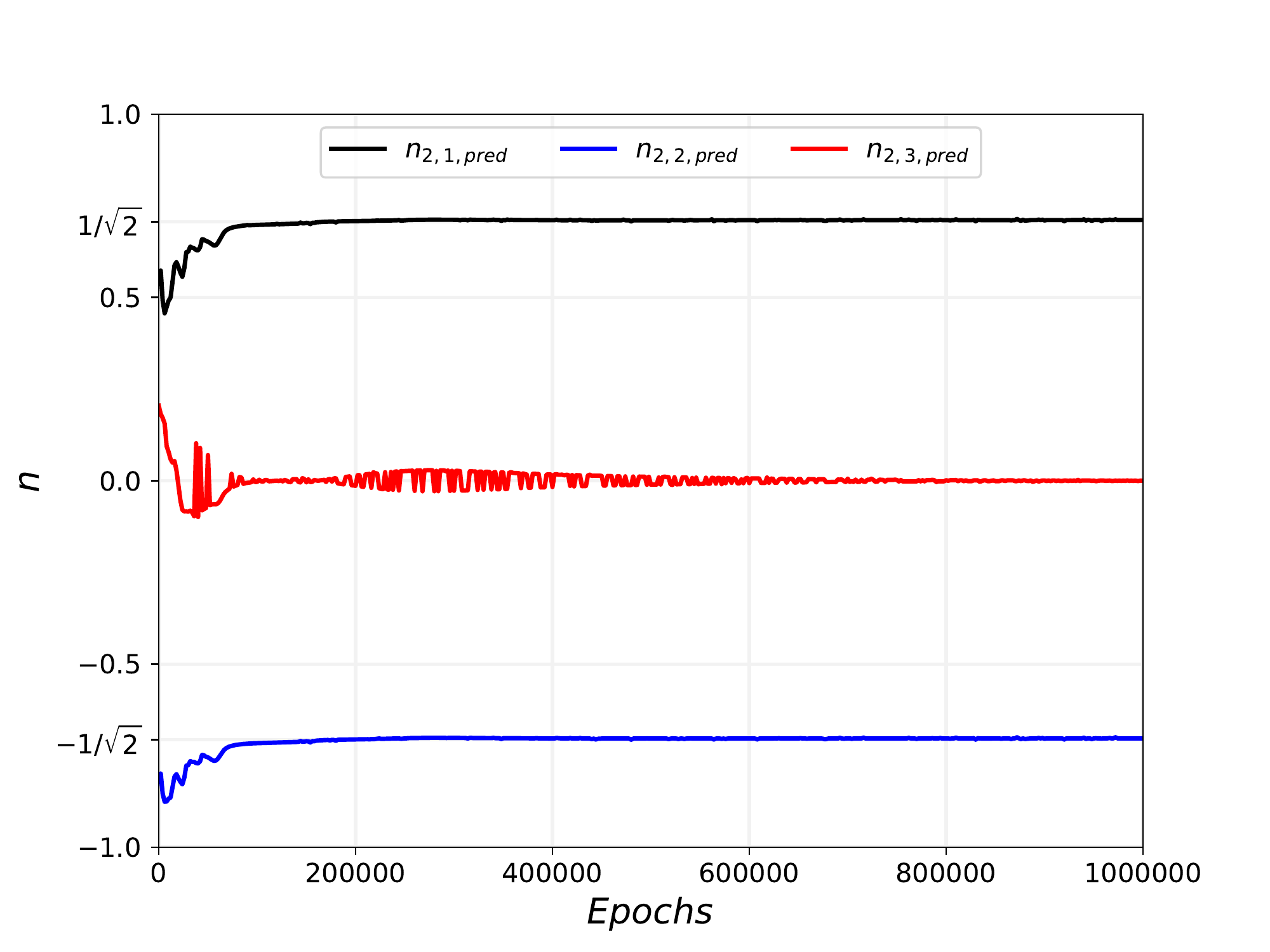}
\caption{$\nb_{2}$ for $N = 100$} 
\end{subfigure}
\begin{subfigure}[b]{0.5\linewidth}
\includegraphics[scale=0.3]{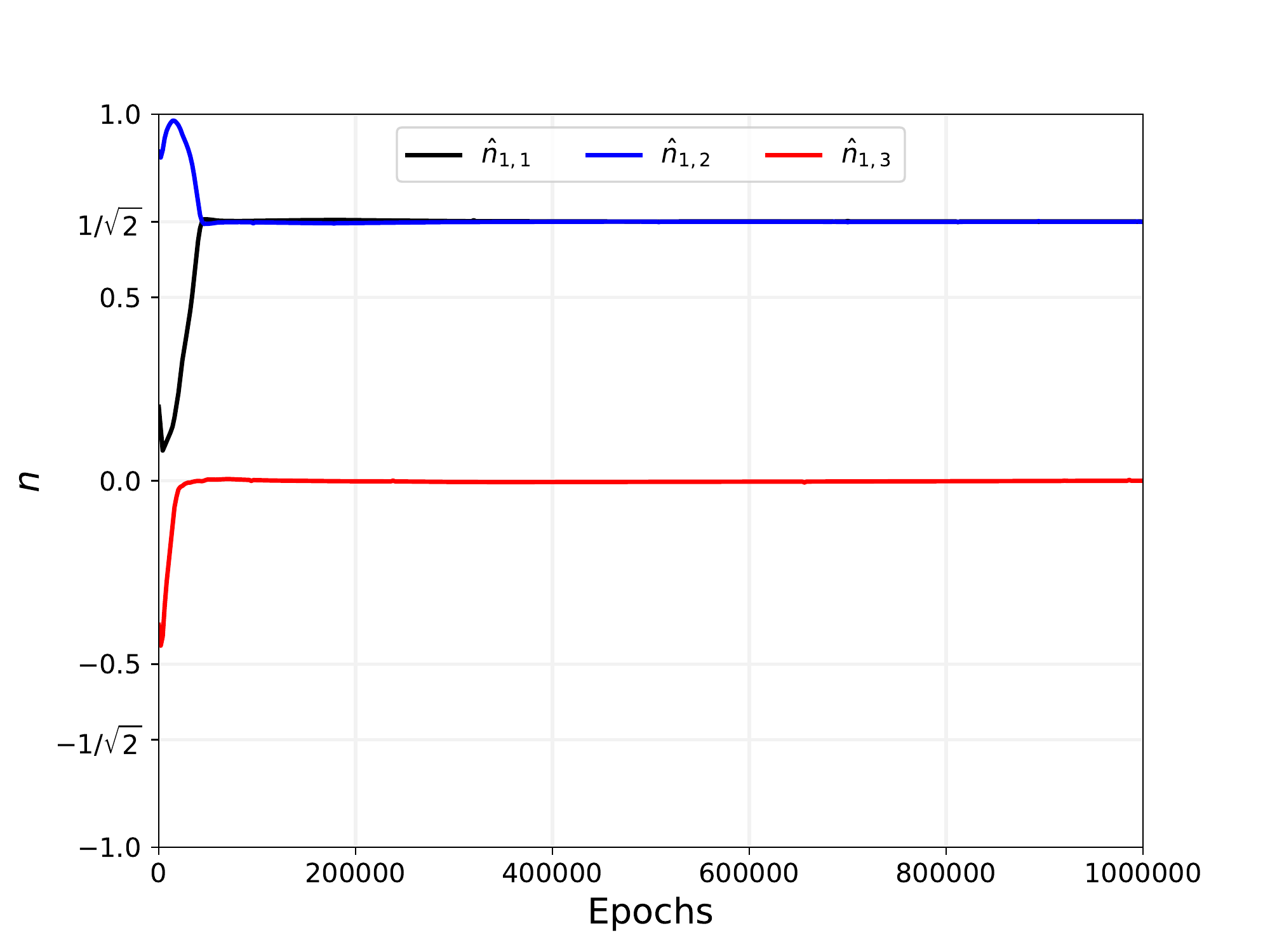}
\caption{$\nb_{1}$ for $N = 1000$} 
\end{subfigure}
\begin{subfigure}[b]{0.5\linewidth}
\includegraphics[scale=0.3]{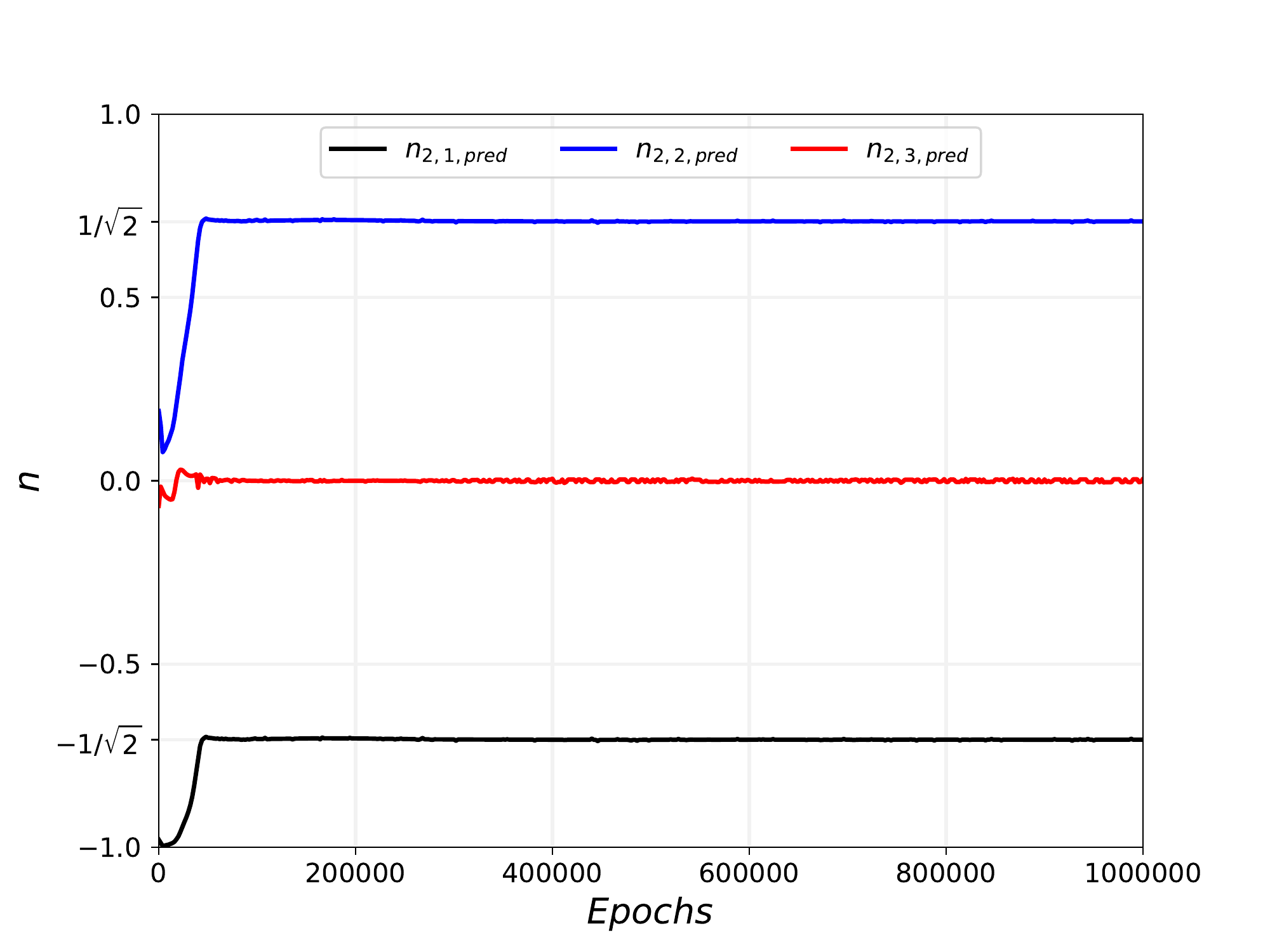}
\caption{$\nb_{2}$ for $N = 1000$} 
\end{subfigure}
\begin{subfigure}[b]{0.5\linewidth}
\includegraphics[scale=0.3]{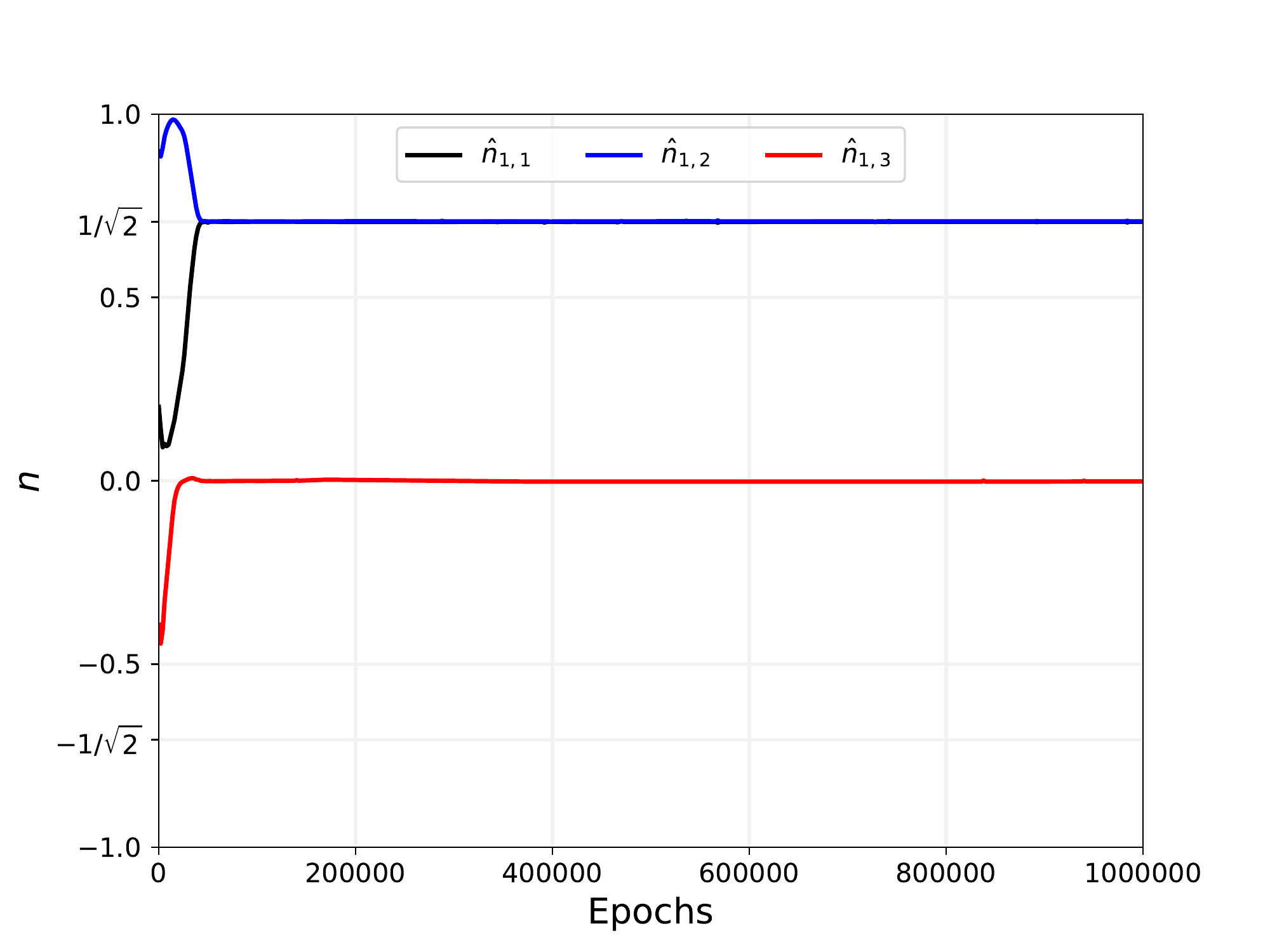}
\caption{$n_{1}$ for $N = 10000$}
\end{subfigure}
\begin{subfigure}[b]{0.5\linewidth}
\includegraphics[scale=0.3]{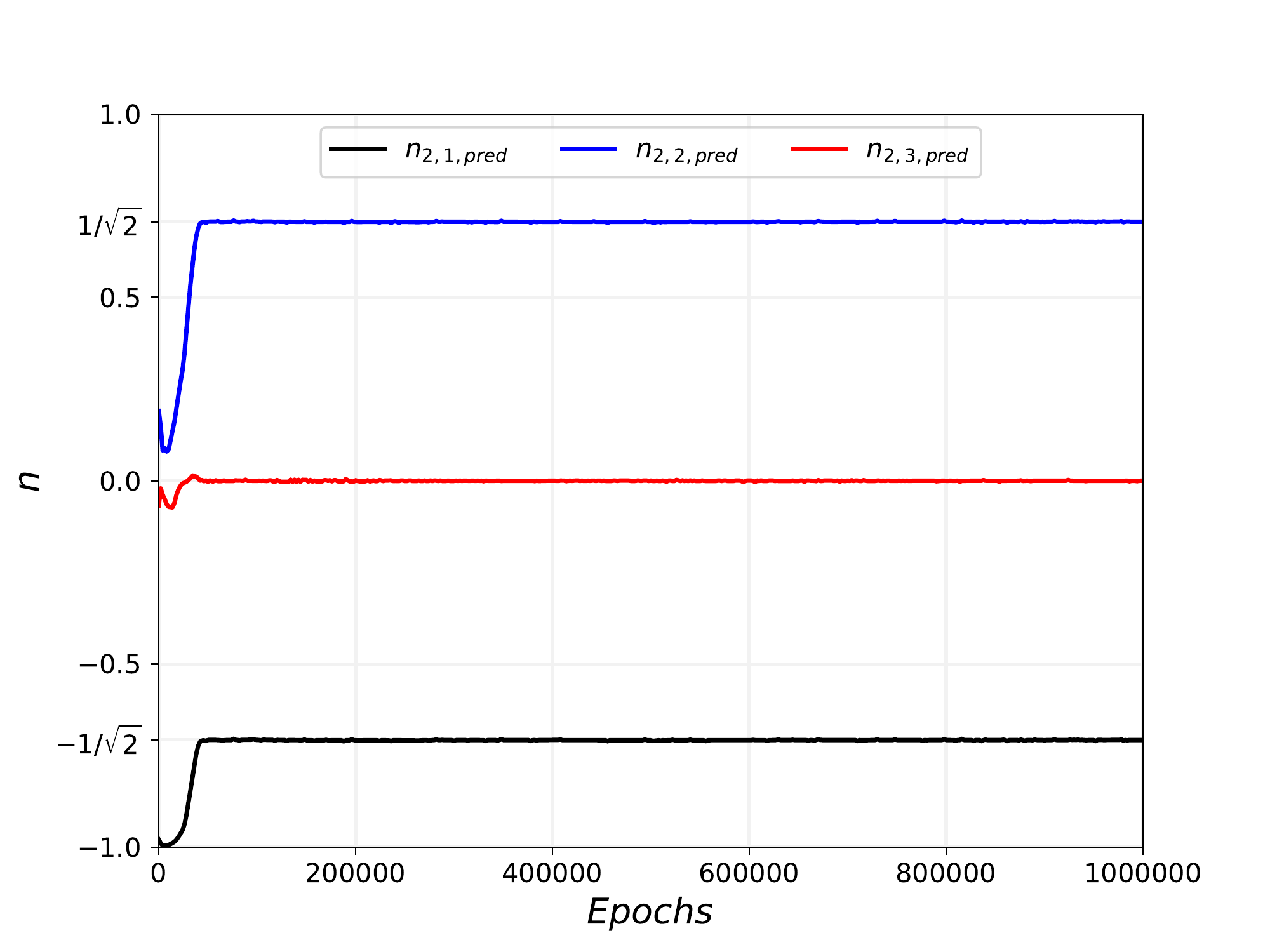}
\caption{$n_{2}$ for $N = 10000$}
\end{subfigure}
\caption{Recovered preferred directions over the training process for the \textbf{orthotropic} material model of \sref{sec::ModelVeri} for three different training dataset sizes. Ground truth: $\nb_{1} = (1/\sqrt{2}, 1/\sqrt{2}, 0)$ and $\nb_{2} = (1/\sqrt{2}, -1/\sqrt{2}, 0)$. (a,b) $N=100$, (c,d) $N=1,000$, (e,f) $N=10,000$.}\label{fig::ConvOrthoPrefDir}
\end{figure}

Lastly, we investigate the ability of the proposed framework to recover orthotropic material behavior based on these three datasets. As an example we employ the orthotropic hyperelastic law discussed in  \sref{sec::ModelVeri}. For this material law, both anisotropic coefficients should be non-zero and the two corresponding directions should be equivalent to $\nb_{1} = (1/\sqrt{2}, 1/\sqrt{2}, 0)$ and $\nb_{2} = (1/\sqrt{2}, -1/\sqrt{2}, 0)$.
We can see in \fref{fig:ConvOrthoLossAniB} that the coefficient values of all three cases are decidedly non-zero at the end of the training process meaning that the correct type of anisotropy is found.
\Fref{fig::ConvOrthoPrefDir} shows that all three dataset sizes allow for a correct identification of the  anisotropy orientations.

Overall, we can see that the proposed approach shows a surprising resilience to decreasing the size of the stress-strain dataset. It appears that even $100$ training points are enough to accurately recover the three major classes of anisotropy studied in this work.

\section{Input-convex neural networks enforcing polyconvexity}\label{append:InputConvex}
The standard TBNN formulation aims to find the best (strain energy) potential that fits the data, refer to \sref{sec:method}, here we explore a variant that embeds polyconvexity of the potential.
In finite elasticity theory,
Ball \cite{ball1976convexity} proved that minimizers for existing variational functionals
exist if the energy potential is polyconvex.
The polyconvexity condition is fulfilled if and only if a function $\mathcal{P}\, :\, \mathbb{R}^{3 \times 3} \times \mathbb{R}^{3 \times 3} \times \mathbb{R} \rightarrow \mathbb{R} $ exists such that
\begin{equation}
\Psi(\Fb) = \mathcal{P}(\Fb, \text{Cof} \,\Fb, \det \Fb)
\end{equation}
where $\mathcal{P}$ is convex with regards to all the arguments.
Here, $\text{Cof}  \,\Fb$ denotes the cofactor operator, which results in the transpose of the adjugate of $\Fb$.
The growth condition on the strain energy function is a necessary part of the proof that polyconvexity guarantees the existence of a solution of a boundary value problem \cite{hartmann2003polyconvexity}. In a simplified form it can be expressed as
\begin{equation}\label{eq::Growth}
\Psi(\Fb) \rightarrow \infty, \qquad \text{as} \quad \det \Fb \rightarrow 0 ^{+}.
\end{equation}
The polyconvexity condition (as well as objectivity and material symmetry conditions) can be fulfilled if  $\mathcal{P}$ is a convex function of a set of invariants which are polyconvex functions of $(\Fb, \text{Cof} \,\Fb, \det \Fb)$ \cite{steigmann2003frame}.
For orthotropic materials such a set is given by \cite{schroder2005variational}
\begin{equation}\label{eq::PolyInvSet}
\Ic_\text{ortho, poly} = \{\tr \Cb, \tr \text{Cof} \, \Cb, \det \Cb,  \tr \Cb \Nb_1, \tr \Cb^2 \Nb_1
\tr \Cb \Nb_2, \tr \Cb^2 \Nb_2
\} \ .
\end{equation}
Note $\Cb^2$ was used for the anisotropic generators as in \cref{schroder2005variational}, and shown to be equivalent to the use of $\text{Cof}\, \Cb$ in \cref{steigmann2003frame}.

In \sref{sec:method} a classical feedforward neural network architecture is applied to establish the TBNN framework. However, these are not able to robustly establish a polyconvex potential, since the network can not guarantee convexity of the output energy with regards to the input invariants.
Recently \cite{amos2017input}, input convex neural networks (ICNN) have been proposed which, due to their specific architecture, ensure convexity of the output with regards to the input of the network.

\paragraph{Input convex neural networks}
In input convex neural networks the standard update formula (cf. \eref{eq::NNStand}) of neural networks can be rewritten to
\begin{equation}
\bm{z}_{i+1} = a_i (\bm{W}_{i}^{z} \bm{z}_{i} + \bm{W}_{i}^{x} \bm{x} + \bm{b}_{i}),
\end{equation}
where $\bm{W}_{0}^{z} = \bm{0}$, $\bm{z}_{0} = \bm{0}$, $\bm{z}_{k} = \Phi$ and $ \lbrace \bm{W}_{1:k-1}^{z}, \bm{W}_{0:k-1}^{x}, \bm{b}_{0:k-1} \rbrace$ denote the set trainable parameters.
In order to guarantee convexity with these networks all weights $\lbrace \bm{W}_{i}^{z} \rbrace_{i=1}^{k-1}$ need to be non-negative and the activation functions $g_{i}$ are required to be non-decreasing and convex.

\begin{figure}
\begin{subfigure}[b]{0.5\linewidth}
\includegraphics[scale=0.3]{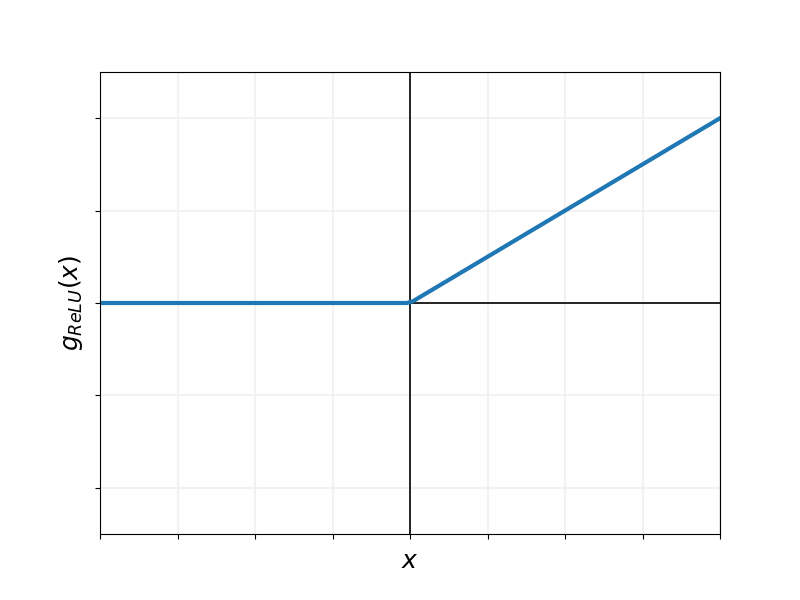}
\caption{ReLU} 
\end{subfigure}
\begin{subfigure}[b]{0.5\linewidth}
\includegraphics[scale=0.3]{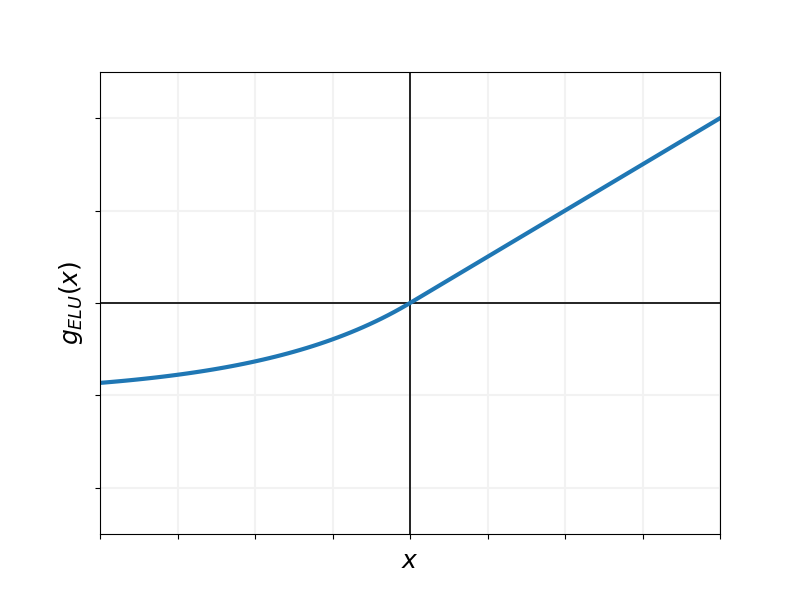}
\caption{ELU} 
\end{subfigure}
\caption{Comparison of two activation functions used with input convex neural networks.}\label{fig::ReluvsELU}
\end{figure}

In \cref{amos2017input}  ``passthrough" layers, i.e. the input $\bm{x}$ is directly connected to the hidden and output layers, are established to increase the expressiveness of the network architecture.
\begin{figure}
\begin{center}
\includegraphics[scale=0.6]{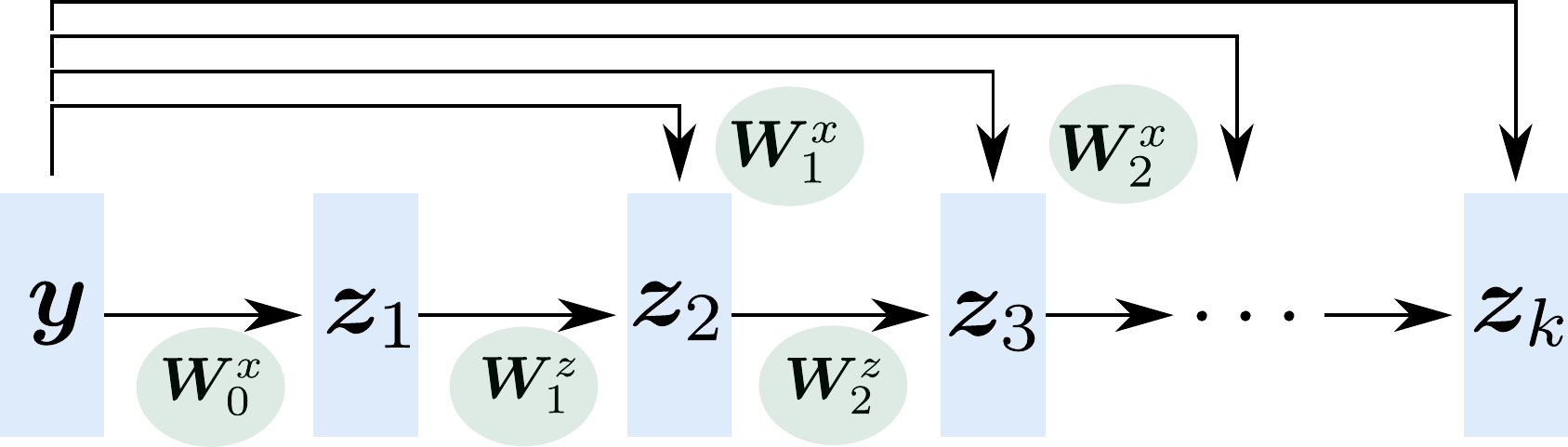}
\caption{Schematic representation of the input convex neural network architecture.}\label{fig::ICNNRep}
\end{center}
\end{figure}
Initially, Amos \etal \cite{amos2017input} used the Rectified Linear Unit (ReLU) as the activation function of choice to fulfill the requirements on $a_{i}$. Recently, the Exponential Linear Units (ELU) has been proposed as an alternative \cite{sivaprasad2021curious} whose first derivative is smooth (in contrast to ReLU). The shape of both types of activation functions can be seen in \fref{fig::ReluvsELU}.
A schematic representation of the input convex neural network architecture is shown in \fref{fig::ICNNRep}.
In solid mechanics applications, ICNNs have already been applied to fitting yield functions \cite{fuhg2022machine,fuhg:hal-03619186} as well as strain energy functions \cite{klein2022polyconvex}.

\paragraph{Polyconvex tensor basis neural network and training specifications}
Using input convex neural networks allows for the output potential to be convex with regards to the invariant inputs. However, they do not fulfill the growth condition as described in \eref{eq::Growth}. Following \cref{klein2022polyconvex} this requirement can be established by adding an analytical term to the output of the network
\begin{equation}
\Psi = \Phi + ( \det \Fb + \frac{1}{\det \Fb} - 2 )^{2}.
\end{equation}
For a general orthotropic material defined using the invariant set of \eref{eq::PolyInvSet}
the stress response function then yields
\begin{eqnarray}\label{eq::StressPoly}
\Sb &=&2 \left[ ( \partial_{I_1}  \Psi +   I_{1} \partial_{I_2}  \Psi) \, \Ib - ( \partial_{I_2} \Psi ) \, \Cb + ( \partial_{I_3} \Psi )\,  \text{cof}\, \Cb \right. \\
&+&
\left.
\alpha_{1} \left( ( \partial_{I_4}  \Psi ) \, \Nb_1
+ ( \partial_{I_5}  \Psi ) \, [  \Cb  \Nb_1 + \Nb_1 \Cb ] \right)
+  \alpha_{2} \left(
( \partial_{I_6}  \Psi ) \, \Nb_2
+ ( \partial_{I_7}  \Psi ) \, [  \Cb  \Nb_2 + \Nb_2 \Cb ] \right) \right].
\nonumber
\end{eqnarray}
In order to solve the inverse problem, i.e. to find the type of symmetry and the orientations, two trainable parameters $\alpha_{1}$ and $\alpha_{2}$ are established that allow to promote sparsity in the anisotropic part (equivalently to \eref{eq::MethodS}).
Given stress-strain data, the loss for the TBNN formulation that guarantees a polyconvex potential can then be established with the same regularization formulated in \sref{sec:method}
\begin{equation}
L = \| \Sb - \hat{\Sb} \|_2^2 + \varepsilon  \left( \left| \alpha_{1} \right| + \left| \alpha_{2} \right|  \right) \ ,
\end{equation}
where $\varepsilon$ is a L1 penalty parameter.
In the following we used $\epsilon=1 \times 10^{-4}$.

We investigated the efficacy of the polyconvex-TBNN in recovering the correct anisotropy type and orientation of the hyperelastic models (isotropic, transversely isotropic and orthotropic) discussed in the context of {\it{verification}} in \sref{sec::ModelVeri}.
Similarly to the results of the standard feed-forward architecture (cf. \sref{sec:results}) $N=2,500$ sample points are utilized for all following results.

The ICNN network architecture was implemented in Pytorch \cite{NEURIPS20199015}. The network consist of 3 hidden layers resulting in roughly $3,700$ trainable parameters. All other hyperparameter, solver and initialization choices are equivalent to the ones discussed in \sref{sec:method}.

\begin{figure}
\begin{subfigure}[b]{0.5\linewidth}
\includegraphics[scale=0.3]{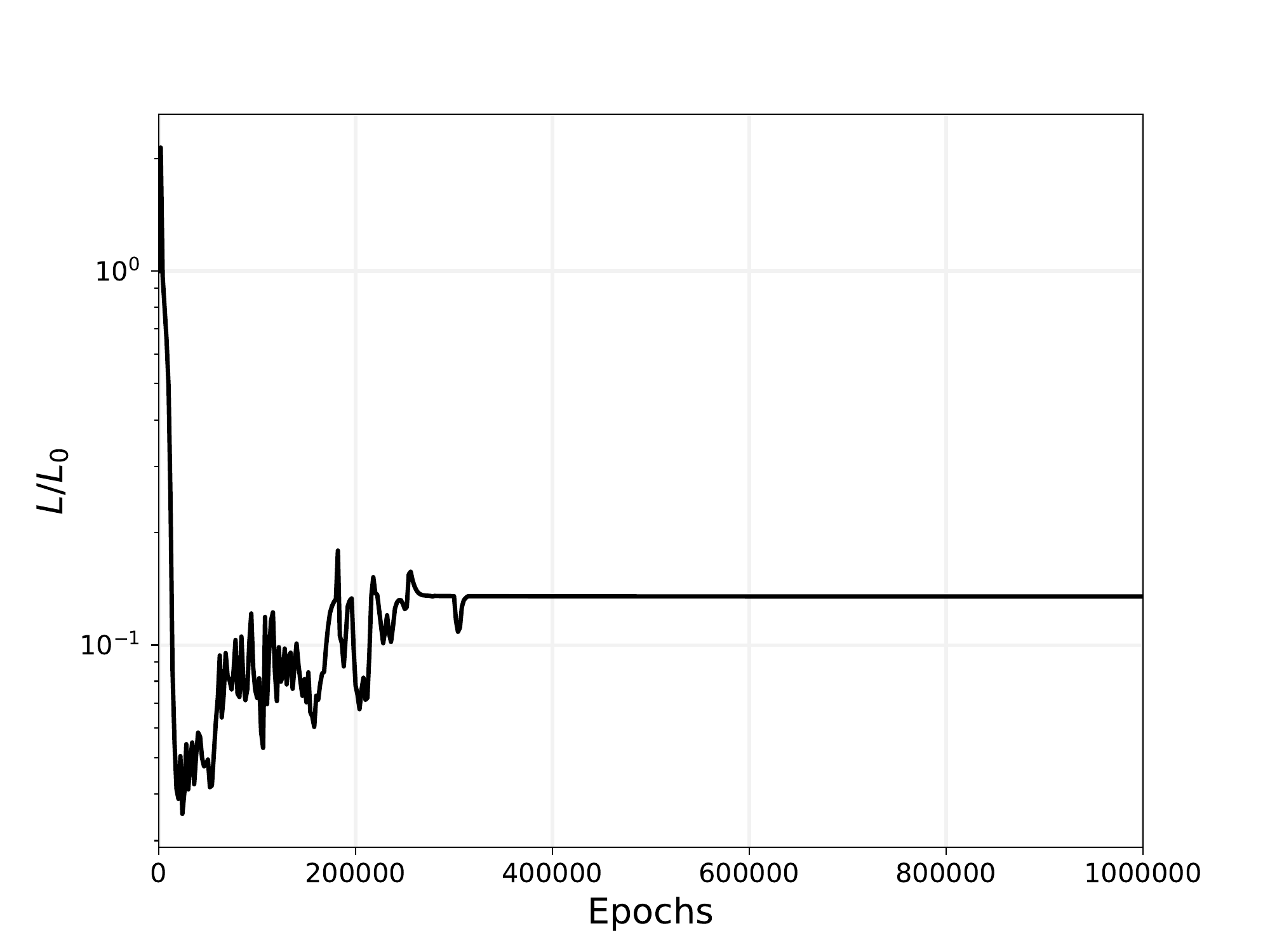}
\caption{Iso. with ReLU} 
\end{subfigure}
\begin{subfigure}[b]{0.5\linewidth}
\includegraphics[scale=0.3]{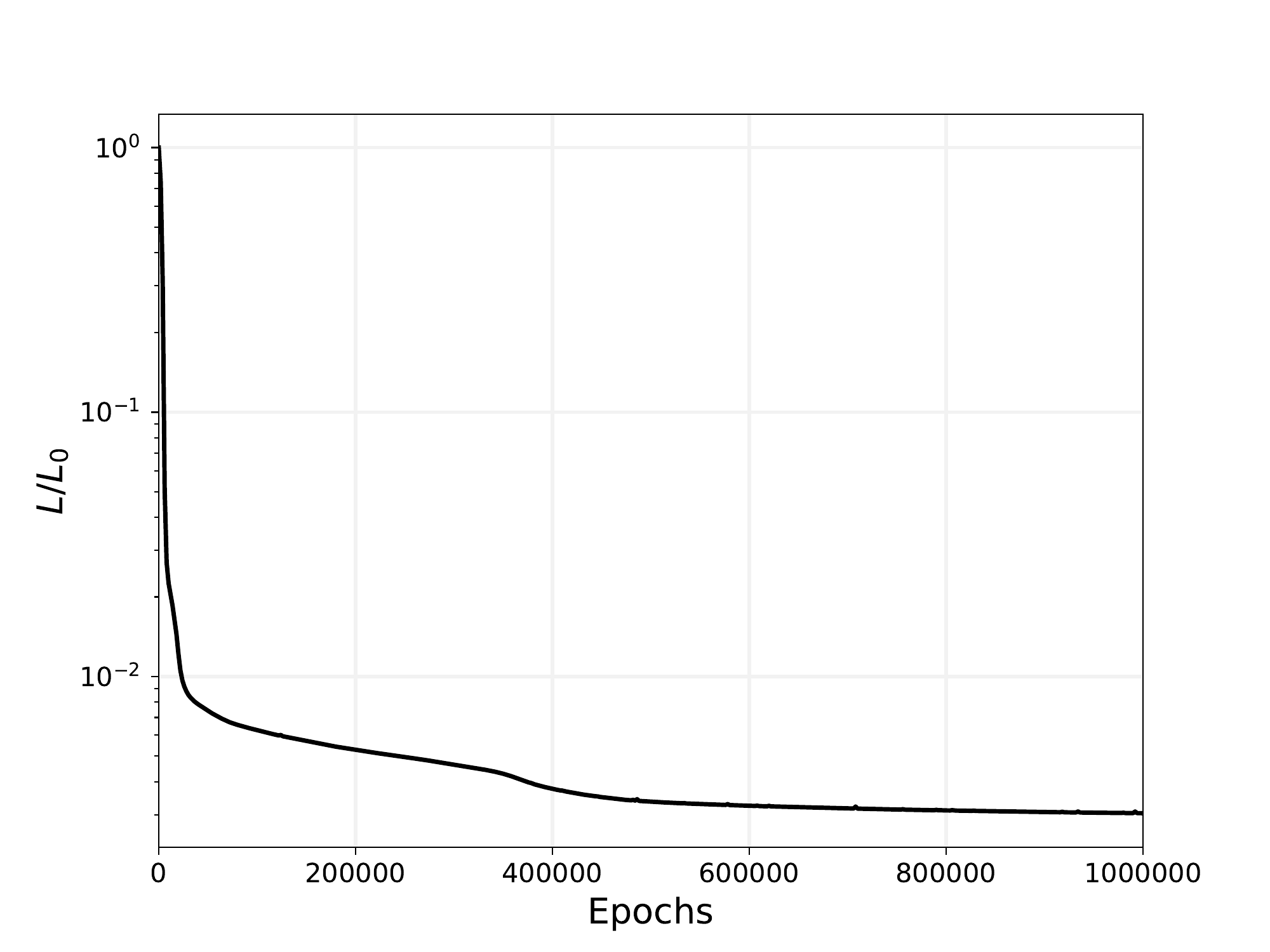}
\caption{Iso. with ELU} 
\end{subfigure}
\begin{subfigure}[b]{0.5\linewidth}
\includegraphics[scale=0.3]{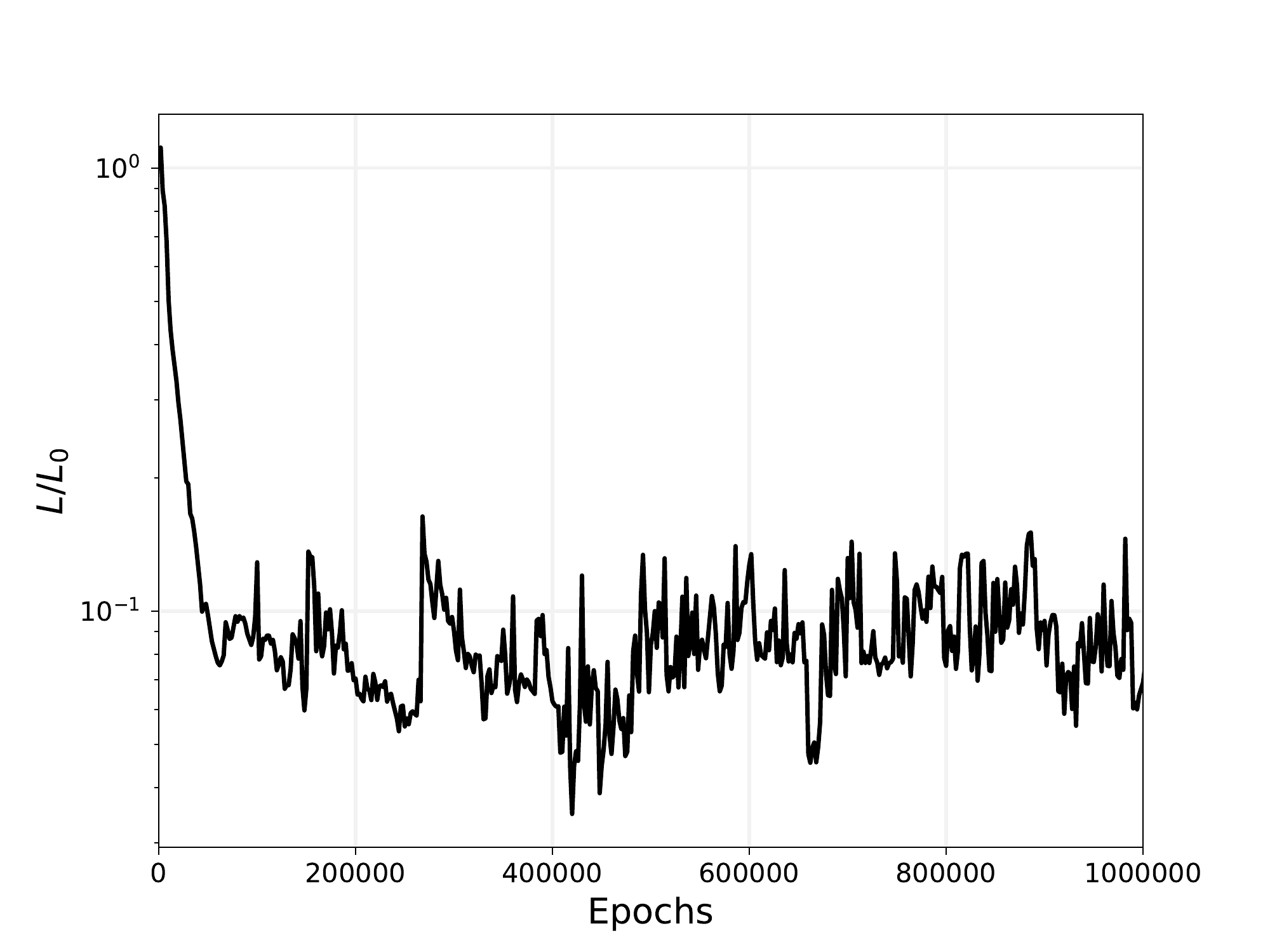}
\caption{Trans. iso. with ReLU} 
\end{subfigure}
\begin{subfigure}[b]{0.5\linewidth}
\includegraphics[scale=0.3]{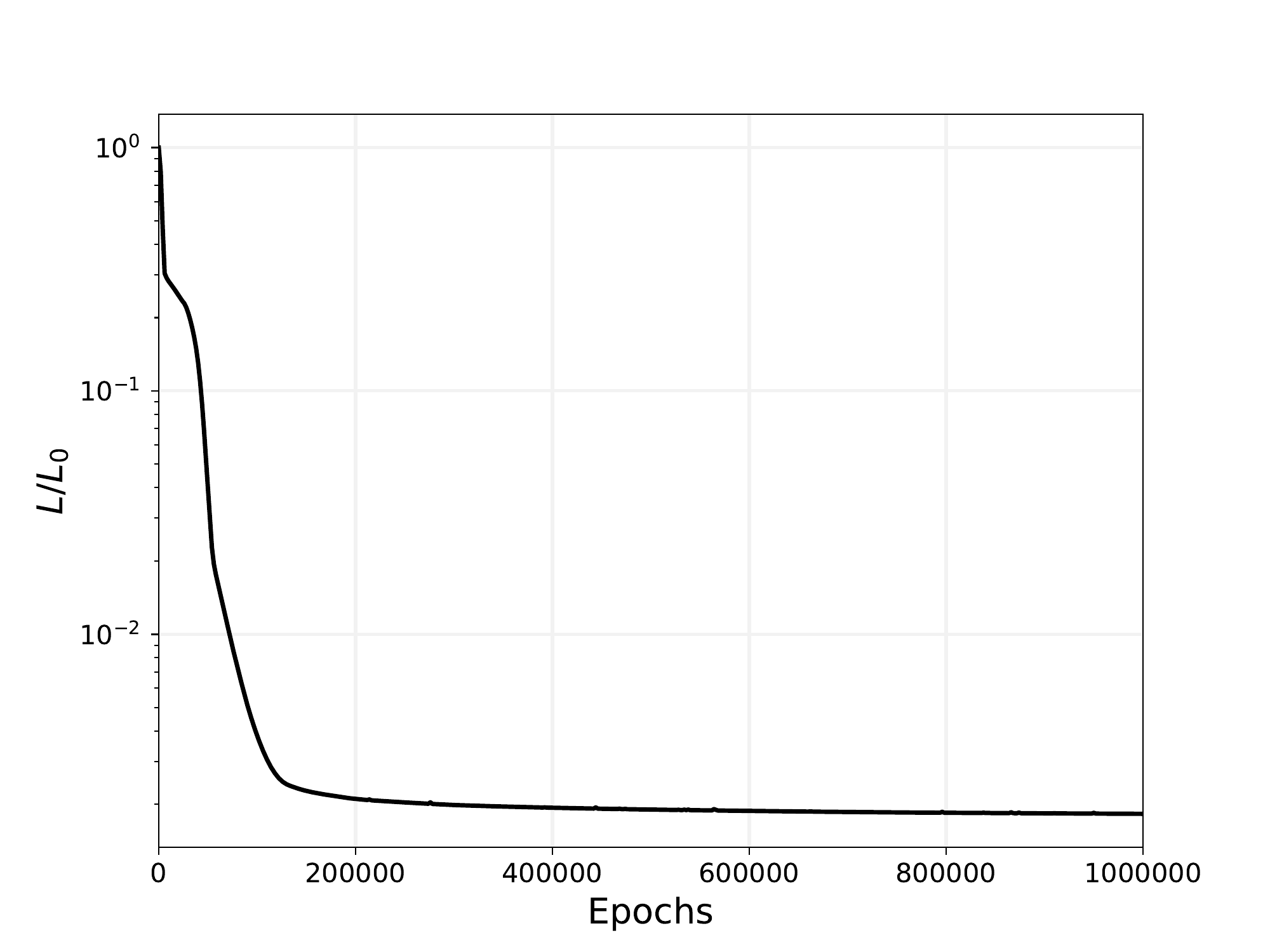}
\caption{Trans. iso. with ELU} 
\end{subfigure}
\begin{subfigure}[b]{0.5\linewidth}
\includegraphics[scale=0.3]{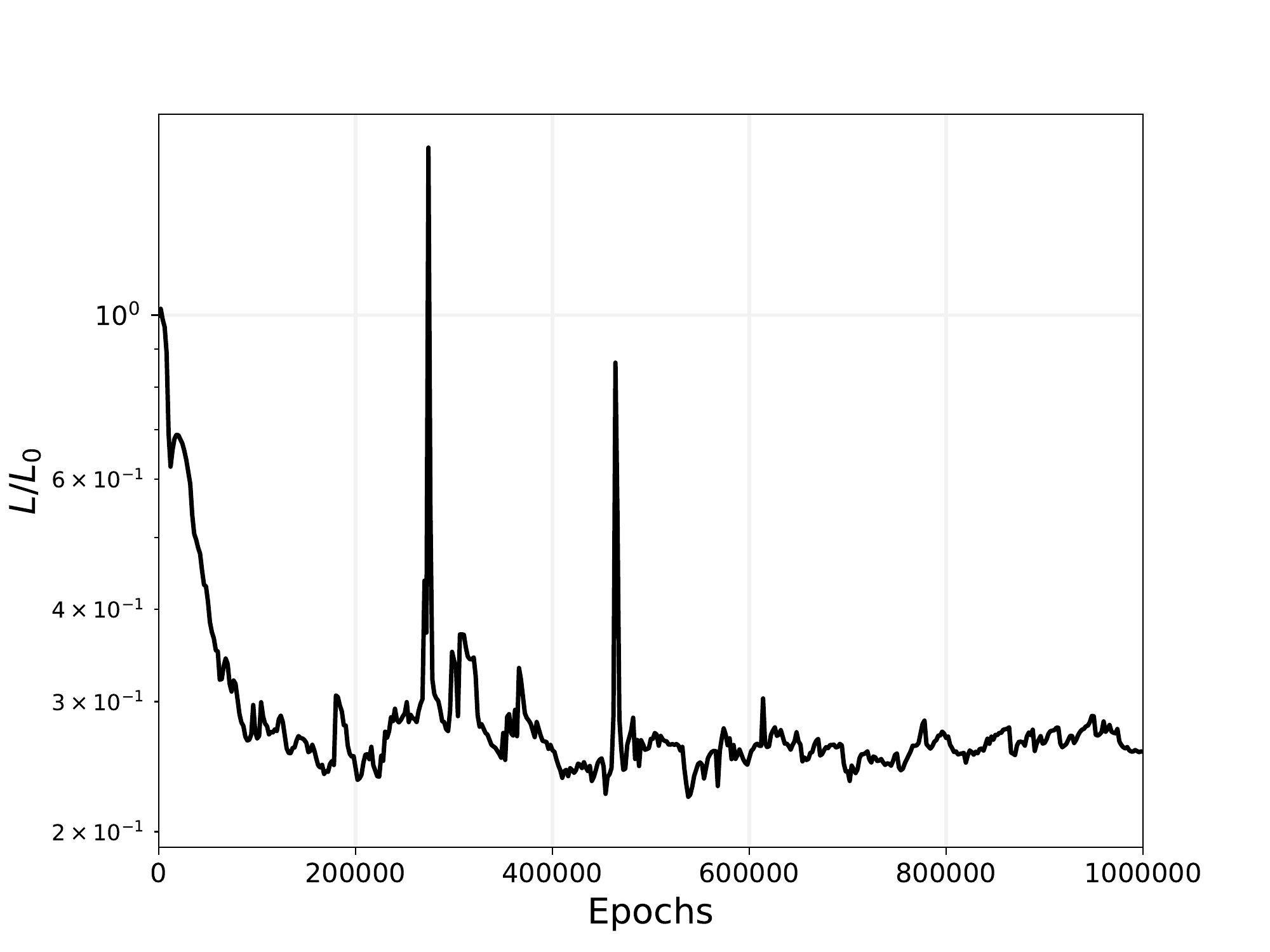}
\caption{Ortho. with ReLU} 
\end{subfigure}
\begin{subfigure}[b]{0.5\linewidth}
\includegraphics[scale=0.3]{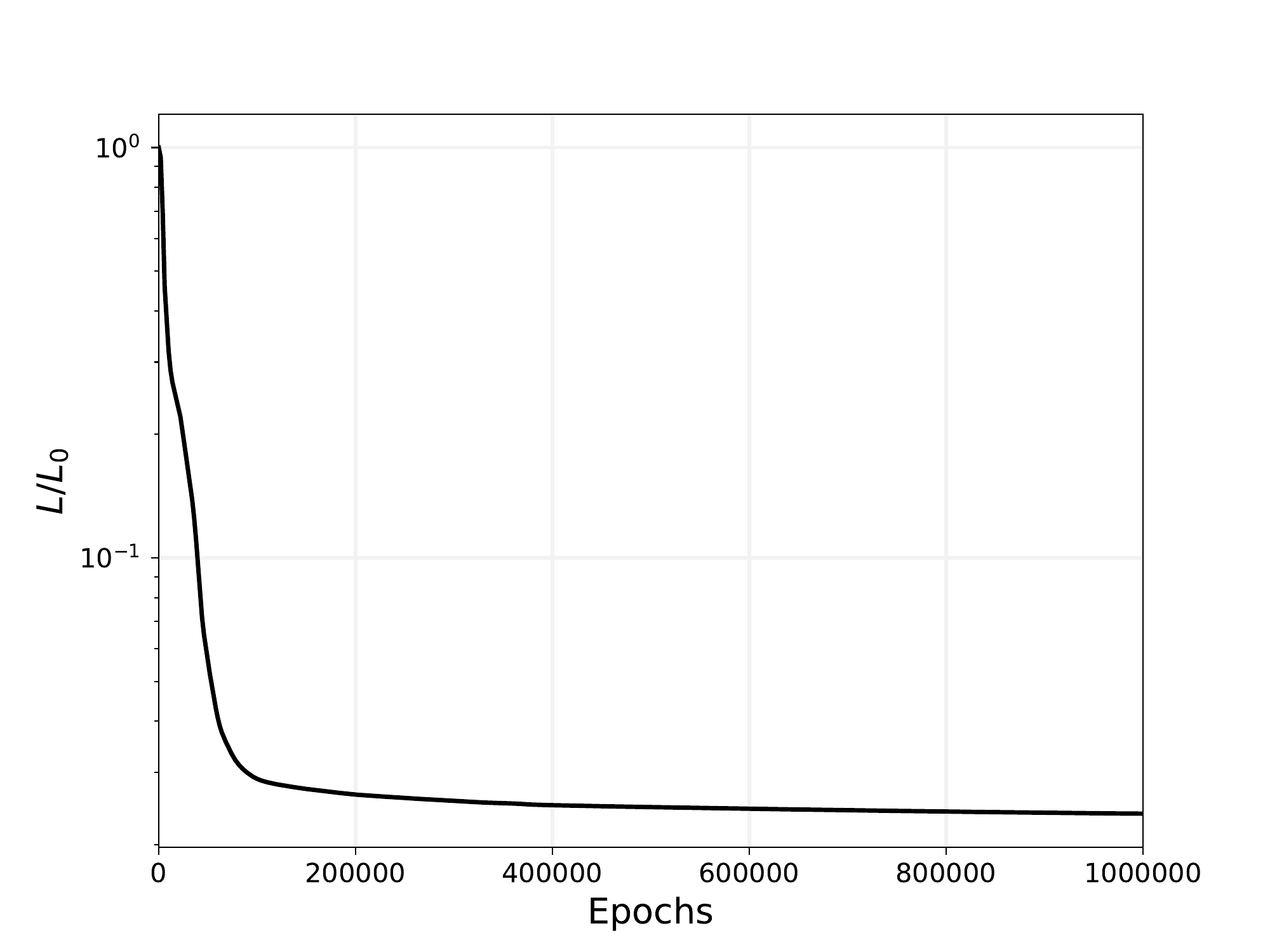}
\caption{Ortho. with ELU} 
\end{subfigure}
\caption{Comparison of training losses for the polyconvex TBNN with ReLU and ELU activation functions for the material models of  \sref{sec::ModelVeri}. (a,b) Isotropic model with ReLU and ELU activation functions, (c,d) transversely isotropic model with ReLU and ELU activation functions, (e,f) orthotropic model with ReLU and ELU activation functions. }\label{fig::ResultsReluVsElu}
\end{figure}

We start with a study of how the choice of the activation function affects the performance of the ICNN architecture. As mentioned the authors of the original ICNN paper only apply ReLU activation functions, c.f. \cref{amos2017input}. As seen in  \fref{fig::ReluvsELU} its first derivative is non-smooth. This appears problematic considering that, in the presented approach, the stress data is fitted through the derivative of the neural network output with regards to their inputs. Hence, a non-smooth prediction of the second Piola-Kirchhoff stress is likely.
To empirically investigate this, \fref{fig::ResultsReluVsElu} plots the training losses for all the hyperelastic model types for both ReLU and ELU activation functions over the number of epochs respectively. We can see that (presumably) due to ReLU being a $C^{0}$ function the convergence of the loss is particularly unstable, whereas for ELU the error reduces smoothly and ends up significantly lower.
Hence, when employing TBNNs with polyconvex potentials we recommend to use ELU activation functions which will also be employed in the following.

\begin{figure}
\centering
\includegraphics[scale=0.3]{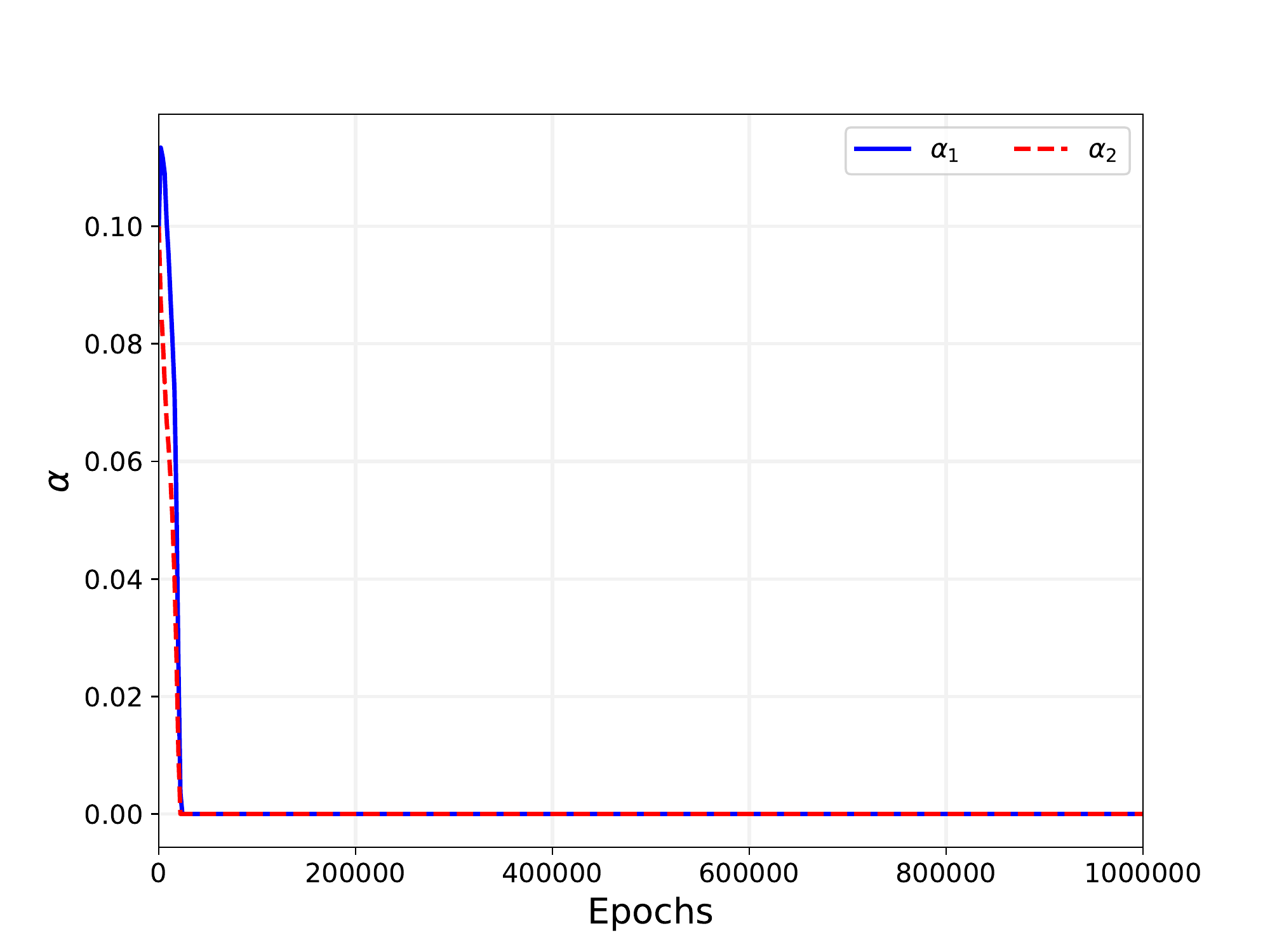}
\caption{Recovered anisotropic coefficients for the polyconvex TBNN with ELU activation function for the \textbf{isotropic} material model of \sref{sec::ModelVeri}
}\label{fig::IsoPolyElu}
\end{figure}

We proceed by studying if the proposed polyconvex approach is able to accurately recover an isotropic material model when starting from the stress representation of \eref{eq::StressPoly}. For this form of material symmetry we expect the anisotropic coefficients ($\alpha_{1}$ and $\alpha_{2}$) to vanish.
\Fref{fig::IsoPolyElu} plots the evolution of the coefficients over the training process. It can be seen that the proposed model is able to recover this isotropic behavior since the coefficients are zero at the end of the training process.

\begin{figure}
\begin{subfigure}[b]{0.5\linewidth}
\includegraphics[scale=0.3]{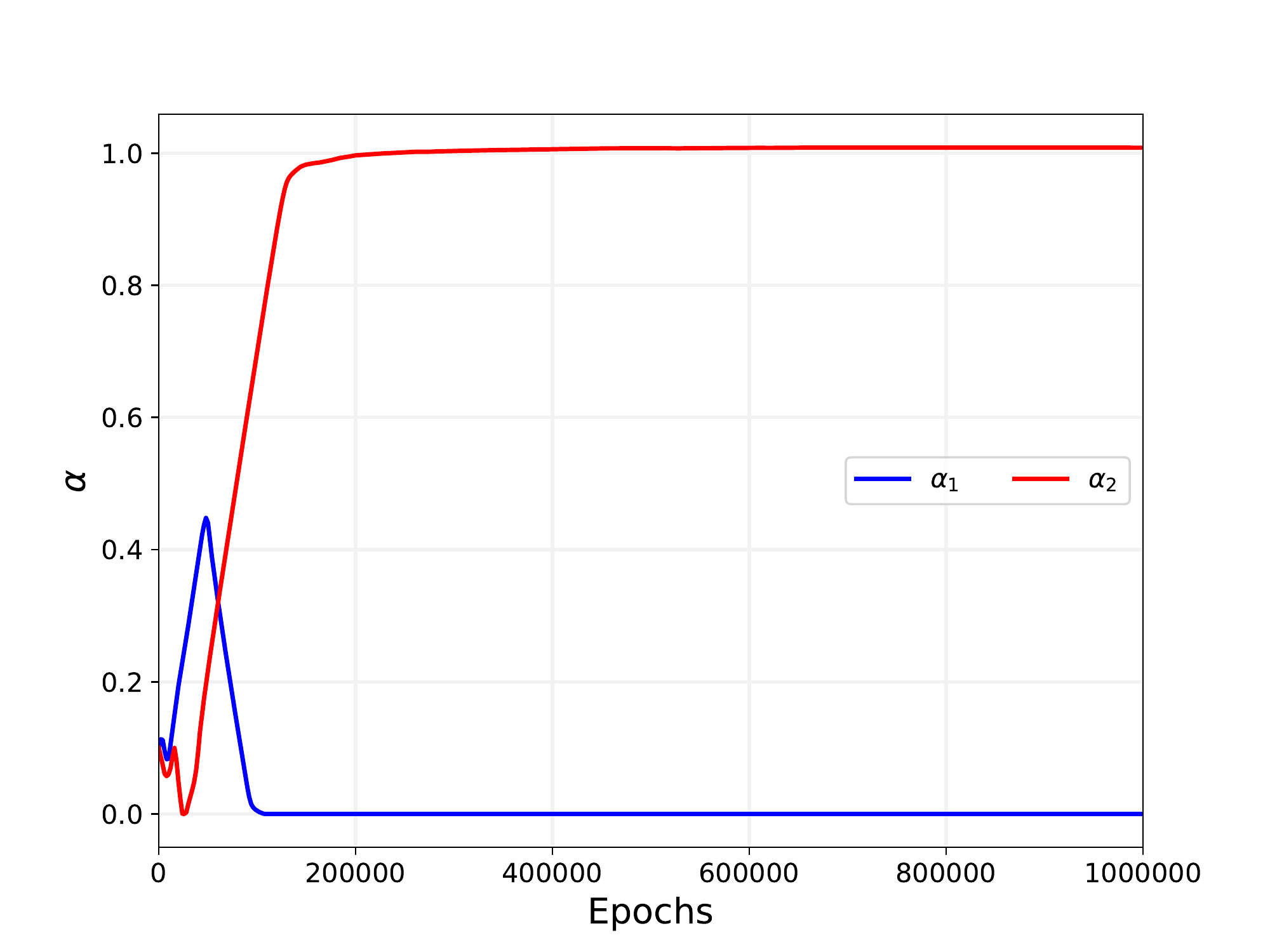}
\caption{Anisotropic coefficients}\label{fig:TransPolyEluA}
\end{subfigure}
\begin{subfigure}[b]{0.5\linewidth}
\includegraphics[scale=0.3]{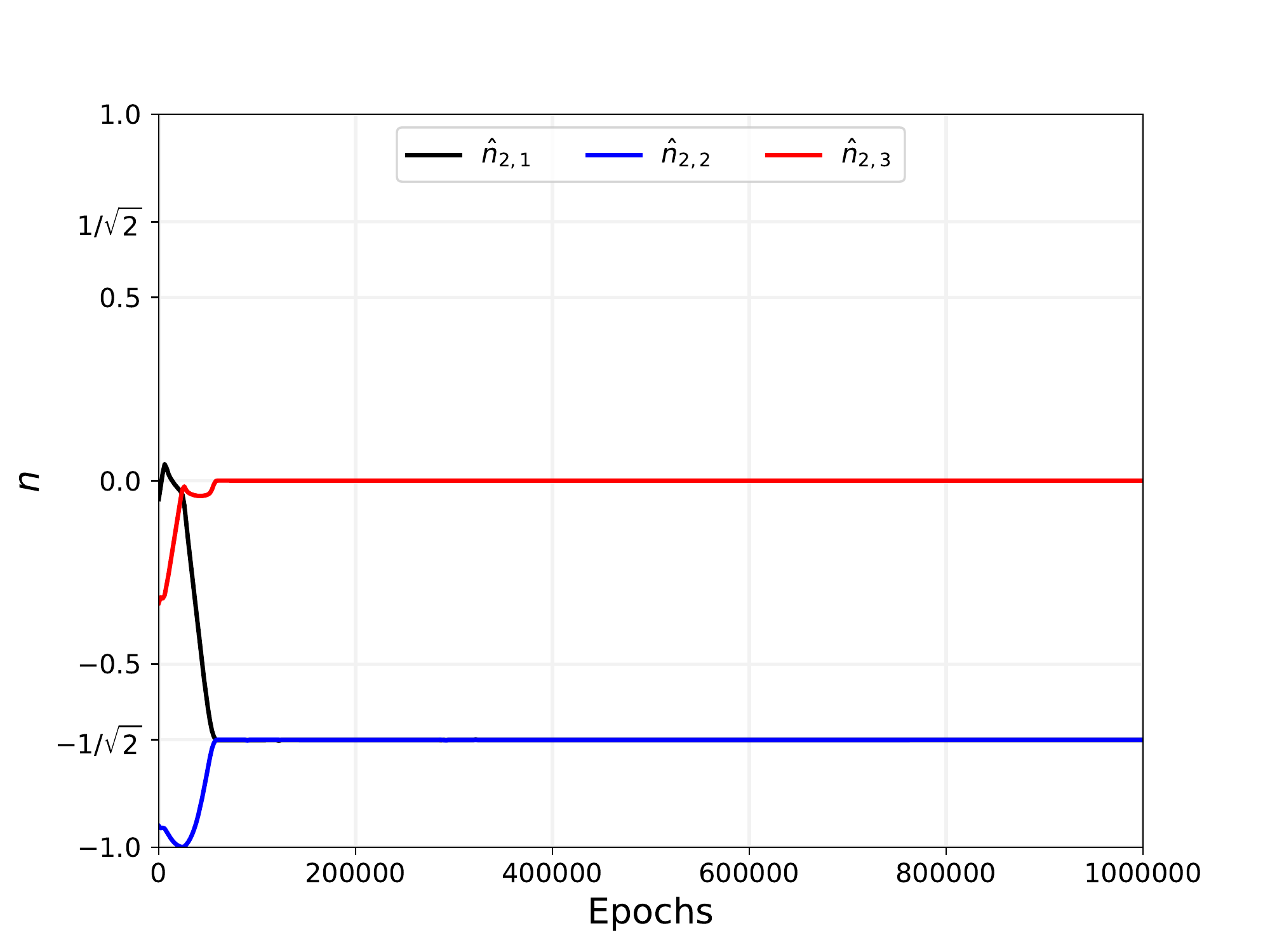}
\caption{Preferred direction}\label{fig:TransPolyEluB}
\end{subfigure}
\caption{Results for the polyconvex TBNN with ELU activation function for the \textbf{transversely isotropic} material model of \sref{sec::ModelVeri}. Ground truth: $\nb=[\frac{1}{\sqrt{2}},\frac{1}{\sqrt{2}},0]^{T}$. (a) Recovered anisotropic coefficients, (b) Recovered preferred direction for direction of $\alpha_{2}$.}\label{fig::TransPolyElu}
\end{figure}

Equivalently, we investigate if the polyconvex framework is able to correctly identify the structure and orientation of the transversely isotropic hyperelastic law presented in \sref{sec::ModelVeri}.
The ground truth direction is assumed to be $\nb=(\frac{1}{\sqrt{2}},\frac{1}{\sqrt{2}},0)$.
The evolution of the anisotropic coefficients are shown in \fref{fig:TransPolyEluA}. We can see that the proposed framework correctly recovers the right degree of anisotropy by having only one remaining coefficient by the end of training. The corresponding direction of this coefficient over the training process is plotted in \fref{fig:TransPolyEluB} which matches with the ground truth after the training is finished.

\begin{figure}
\begin{center}
\begin{subfigure}[b]{0.5\linewidth}
\includegraphics[scale=0.3]{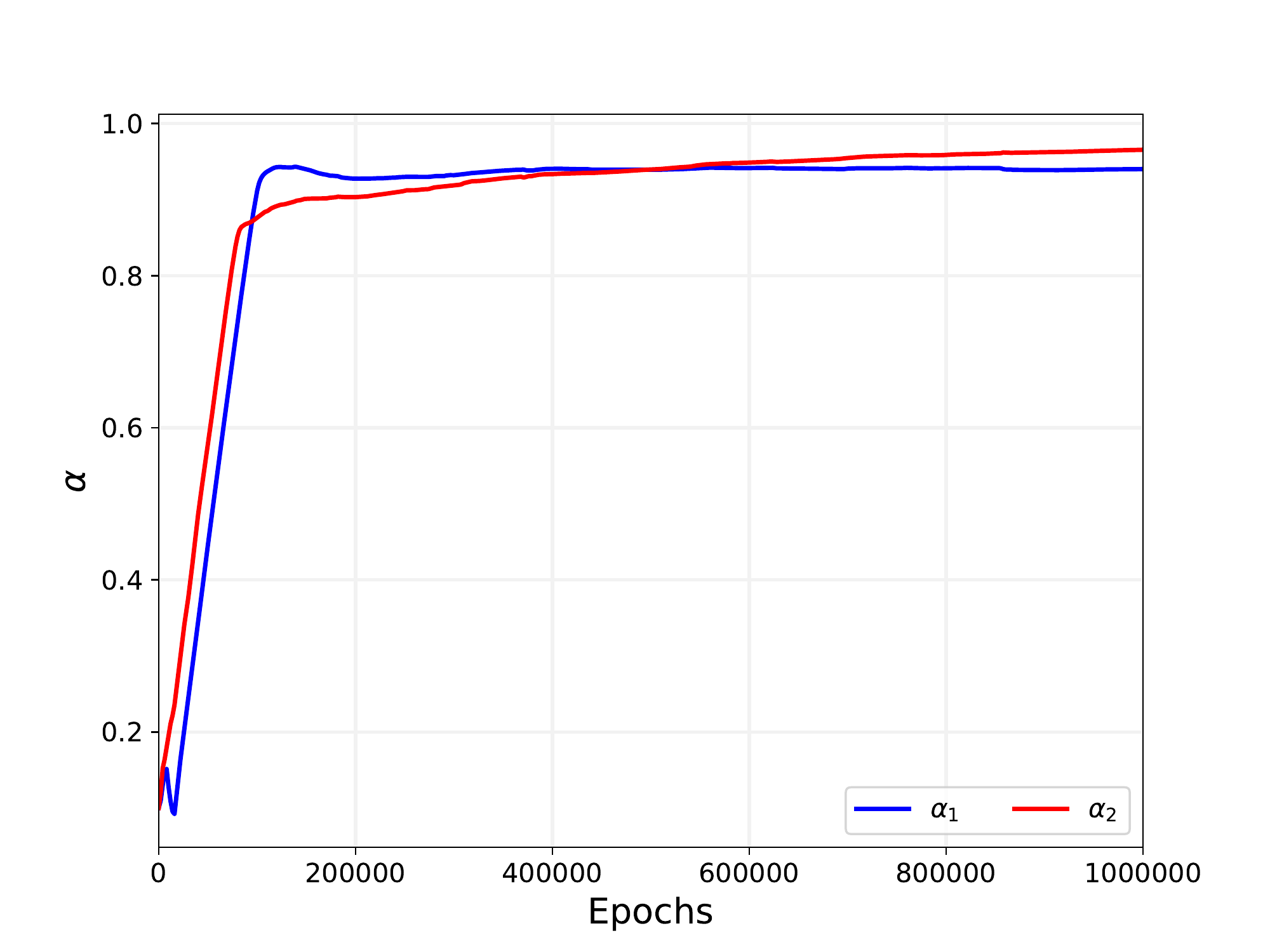}
\caption{Anisotropic coefficients}\label{fig:OrthoPolyEluA}
\end{subfigure}
\end{center}

\begin{subfigure}[b]{0.5\linewidth}
\includegraphics[scale=0.3]{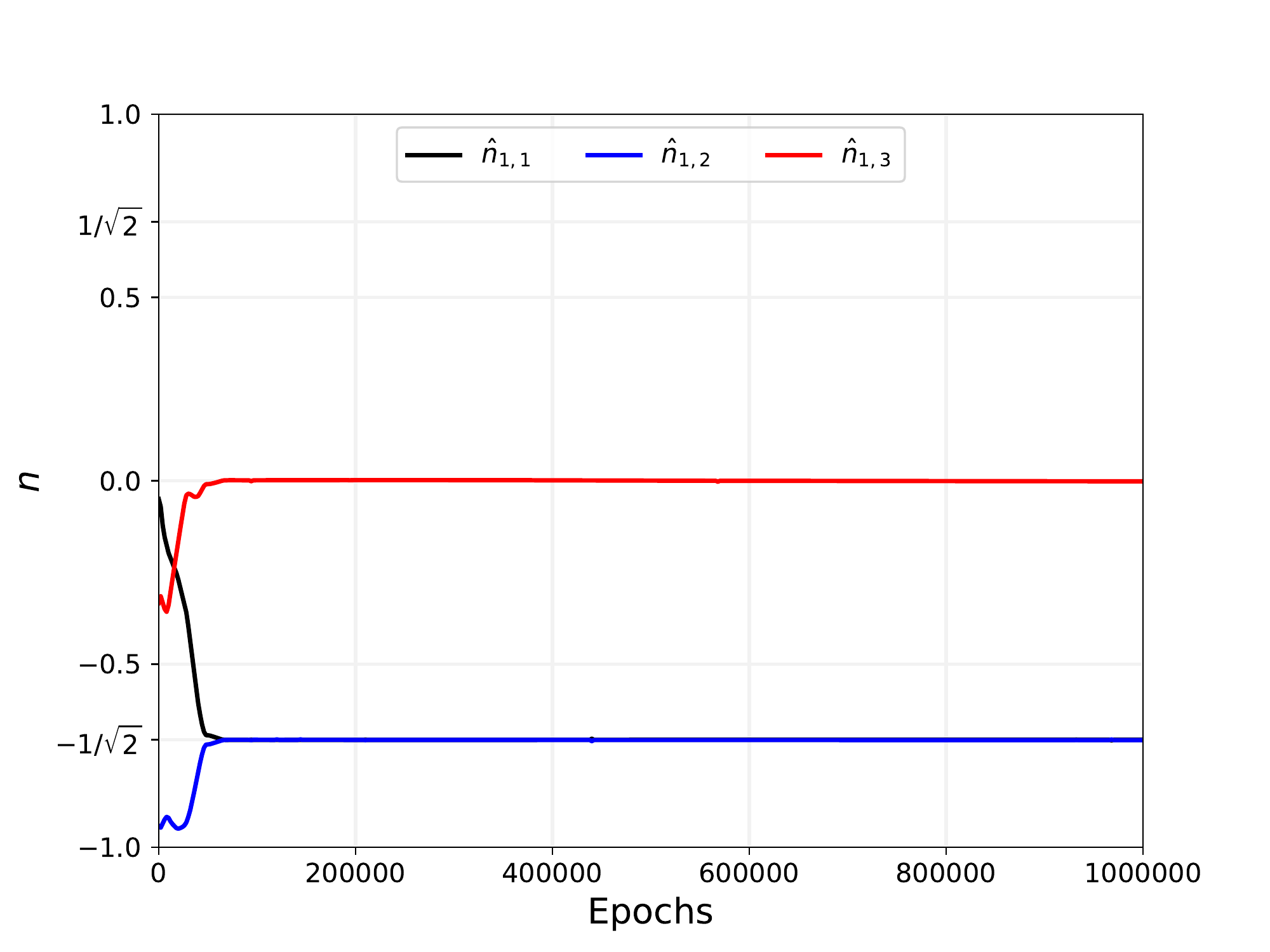}
\caption{Preferred direction 1}\label{fig:OrthoPolyEluB}
\end{subfigure}
\begin{subfigure}[b]{0.5\linewidth}
\includegraphics[scale=0.3]{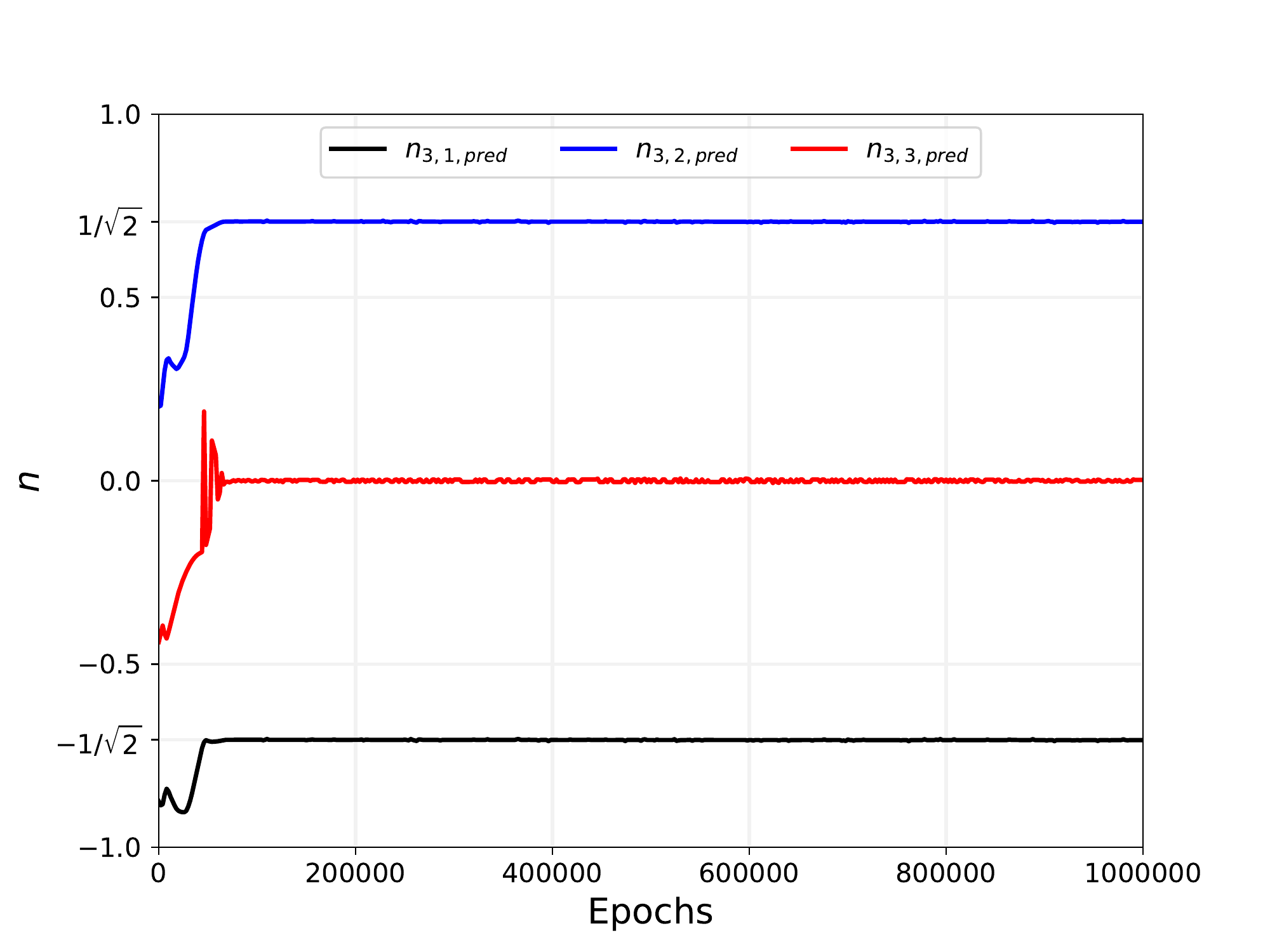}
\caption{Preferred direction 2}\label{fig:OrthoPolyEluC}
\end{subfigure}
\caption{Results for the polyconvex TBNN with ELU activation function for the \textbf{orthotropic} material model of \sref{sec::ModelVeri}.
Ground truth: $\bm{n}_{1} = [\frac{1}{\sqrt{2}},\frac{1}{\sqrt{2}},0]^{T}$ and $\bm{n}_{2} = [-\frac{1}{\sqrt{2}},\frac{1}{\sqrt{2}},0]^{T}$. (a) Recovered anisotropic coefficients, (b,c) Recovered preferred direction.}\label{fig::OrthoPolyElu}
\end{figure}

Lastly, we test the polyconvex potential on its ability to correctly recover properties of orthotropic materials. For this we use the orthotropic hyperelastic law provided in \sref{sec::ModelVeri} with the two ground truth directions chosen as $\bm{n}_{1} = (\frac{1}{\sqrt{2}},\frac{1}{\sqrt{2}},0)$ and $\bm{n}_{2} = (-\frac{1}{\sqrt{2}},\frac{1}{\sqrt{2}},0)$.
From the evolution of the anisotropic coefficients we can see that the model is able to correctly classify the data as being from an orthotropic material (c.f. \fref{fig:OrthoPolyEluA}). Furthermore, the corresponding orientations also match the ground truth (Figs. \ref{fig:OrthoPolyEluB} and \ref{fig:OrthoPolyEluC}).

Overall, for the presented cases, the polyconvex TBNN architecture is able to accurately recover anisotropic structures and orientations from hyperelastic stress-strain datasets.
The input-convex and the standard TBNNs achieved comparable accuracy; however, the built-in polyconvexity gives us more confidence in the predictions of this architecture.
Enforcing additional physical principles and model requirements (such as polyconvexity in this case), has proven as a successful approach to allow machine learning approaches to be efficient in low-data regimes and also in generating models that can efficiently generalize \cite{ling2016machine,fuhg2021physics}.
A counterpoint is the standard TBNN presented in \sref{sec:method} is relatively straightforward to implement with standard layers available in Pytorch \cite{NEURIPS20199015} and the like.

\end{document}